**First image-guided treatment of a mouse tumor with radioactive ion beams**

Daria Boscolo[1,§], Giulio Lovatti[2,§], Olga Sokol[1,§], Tamara Vitacchio[1], Martina Moglioni[1], Francesco Evangelista[2], Emma Haettner[1], Walter Tinganelli[1], Christian Graeff[1,3], Uli Weber[1,4], Christoph Schuy[1], Munetaka Nitta[2], Daria Kostyleva[1], Sivaji Purushothaman[1], Peter G. Thirolf[2], Andreas Bückner[1], Jonathan Bortfeldt[2], Christoph Scheidenberger[1,5,6], Katia Parodi[2,#] and Marco Durante[1,7,8,#,*]

1. GSI Helmholtzzentrum für Schwerionenforschung, Darmstadt, Germany
2. Department of Medical Physics, Ludwig-Maximilians-Universität München (LMU), Munich, Germany.
3. Department of Electrical Engineering and Information Technology (ETIT), Technische Universität Darmstadt, Darmstadt, Germany
4. Life Science Engineering Faculty, Technische Hochschule Mittelhessen, Gießen, Germany
5. Institute of Physics, Justus-Liebig-Universität Gießen, Gießen, Germany
6. Helmholtz Research Academy Hesse for FAIR (HFHF), GSI Helmholtz Center for Heavy Ion Research, Campus Giessen, 35392 Giessen, Germany
7. Institute of Condensed Matter Physics, Technische Universität Darmstadt, Darmstadt, Germany
8. Department of Physics „Ettore Pancini“, University Federico II, Naples, Italy

§: these authors contributed equally

#: these authors contributed equally

*Corresponding author: m.durante@gsi.de

**Abstract**

Radioactive ion beams (RIB) are a key focus of current research in nuclear physics. Already long ago it was proposed that they could have applications in cancer therapy. In fact, while charged particle therapy offers unique advantages that make it one of the most effective radiotherapy techniques, it is highly sensitive to uncertainties in the beam range. RIB are ideal for image-guided particle therapy, as isotopes that undergo $\beta^+$-decay can be visualized using positron emission tomography (PET). This allows spatial localization of the particle distribution in vivo, which can be correlated with the expected dose deposition for online beam range verification. We successfully treated a mouse osteosarcoma using a radioactive $^{11}$C-ion beam. The tumor was located in the neck, close to the spinal cord, where even few mm deviations in beam range could lead to unintended dose deposition in the spine and radiation-induced myelopathy. We managed to completely control the tumor with the highest dose (20 Gy) while avoiding paralysis. Low-grade neurological side effects were correlated to the activity measured by PET in the spine. The biological washout of the activity from the tumor volume was dependent on the dose, indicating a potential component of vascular damage at high doses. This experiment marks the first instance of tumor treatment using RIB and paves the way for future clinical applications.

## Main

Nuclear physics methods have been instrumental in improving cancer radiotherapy using accelerated charged particles (protons or heavier ions). Charged nuclei have indeed a favorable depth-dose distribution in the human body thanks to the Bragg peak[1]. Therapy with accelerated $^{12}$C-ions is currently ongoing in 17 centers worldwide[2] and, even if more expensive than proton therapy, adds biological advantages to the physical benefit of the Bragg peak[3]. Particle therapy is, however, much more sensitive to uncertainties in the beam range than conventional X-rays, exactly because of the high dose deposited in the Bragg peak[4,5]. Several techniques are available to monitor the beam range exploiting the nuclear interactions of the ions in the tissue[6], including PET[7]. PET in carbon ion therapy exploits $\beta^+$-emitting isotopes, such as $^{11}$C and $^{10}$C, produced by the nuclear fragmentation of the therapeutic stable $^{12}$C beam in the patient's body. The method was extensively tested during the C-ion therapy pilot trial at GSI in Darmstadt (Germany)[8], then at HIT in Heidelberg (Germany)[9] and more recently at CNAO in Pavia (Italy)[10] and MIMM in Wuwei (China)[11]. However, the counting rate from projectile fragments is low, the activity peak is shifted with respect to the Bragg peak because the particle range of the isotopic fragments depends on their mass (e.g. the range of $^{11}$C is approximately 91% of the range of $^{12}$C at the same velocity), and the image analysis has been performed off-line. Therefore, PET in $^{12}$C-ion therapy remains marginal and could not reduce the range uncertainty as desired (<1 mm)[7].

Most of these problems can be overcome by using RIB rather than stable beams for therapy. RIB are generally acknowledged as the main tool to address the most important modern questions in nuclear physics, as they allow the study of nuclei at extreme conditions[12–14]. In cancer radiotherapy, RIB have the same biological effectiveness of the corresponding stable ions beams[15,16] but increase the PET signal-to-noise ratio by approximately an order of magnitude, reduce the shift between the activity and dose peaks, and mitigate the washout image blur with short-lived isotopes (e.g.$^{10}$C) and online acquisition[17,18]. The reduced uncertainty in range allows a shrinkage of the tumor margins around the clinical target volume (CTV), and this is expected to reduce toxicity for both serial or parallel organs-at-risk (OAR)[19]. Attempts to use RIB in cancer therapy started already in the 80s during the heavy ion therapy pilot project at the Lawrence Berkeley Laboratory (CA, USA)[20], but they were always hampered by the low intensities of the secondary beams produced by fragmentation of the primary ions used for therapy (for an historical review see ref.[21]). Modern high-intensity accelerators that can produce RIB with intensity sufficient for therapeutic treatments[22] can be used to test PET-guided heavy ion treatments. One of these facilities is GSI/FAIR in Darmstadt[23], where we started the BARB (Biomedical Applications of Radioactive Ion Beams) project whose goal was to perform the first in vivo tumor treatment with RIB[17].

Within BARB, we have already reported the RIB imaging resolution in phantoms[24,25] and transported the beam from the fragment separator (FRS) to the medical vault (Cave M, where animal experiments are possible) at the GSI accelerator facility[26] (Extended Data Fig. 1). In Cave M we have then installed the portable small-animal SIRMIO in-beam PET scanner[27], built by the Ludwig-Maximilian-University (LMU) group in Munich for online range verification in pre-clinical particle therapy experiments. PET physics inherently requires

~30–60 seconds to accumulate sufficient statistics from radioactive decays for meaningful image updates. While not truly instantaneous, the data processing itself occurs on the millisecond scale, pushing the boundaries of in-beam PET monitoring. The SIRMIO PET scanner is based on 56 scintillator blocks of pixelated LYSO crystals. The crystals inside each detector block are arranged to provide a pyramidal-step shape to optimize the geometrical coverage in a spherical configuration[28]. Inside the detector it is possible to accommodate an anesthetized mouse in vertical position, by using a custom 3D-printed holder, for simultaneous irradiation and online PET imaging. The mouse model used in this study is a syngeneic LM8 osteosarcoma[29] implanted in the C3H mouse neck. Osteosarcoma is a very radioresistant tumor[30] and for this reason it is a typical candidate for treatment with accelerated $^{12}$C-ions[31]. Fig. 1A shows μCT images of the tumor growth after injection in the C3H mouse and the actual visible tumor in the neck. Fig. 1B shows the contouring of the individual gross tumor volumes (GTV) of the different mice used in the experiments. By summing up all the tumor profiles and smoothing the resulting outline, we have contoured a generalized CTV applied to all mice in this study (Fig. 1C, see Methods for details). The proximity of the CTV to the spinal cord makes high precision and guidance during treatment delivery necessary to avoid radiation myelopathy[32–34], a severe late effect of radiotherapy caused by white matter injury that can lead to motor deficits and paralysis[35]. Measured endpoints were tumor growth, spinal cord toxicity, and washout rate of the radioactive signal from the tumor. We elected to use $^{11}$C projectiles even if our previous experiments[24,25] show that the highest range resolving power can be achieved with short-lived isotopes such as $^{10}$C or $^{15}$O. We preferred to use carbon, which is already used in many clinical facilities, rather than oxygen. Moreover, the intensity of isotopes that have lost two neutrons such as $^{10}$C is too low for very high-dose (≥20 Gy) single-fraction treatment. The goal of the experiment was to demonstrate for the first time the ability to use a $^{11}$C-ion radioactive beam to achieve full tumor control of a radioresistant tumor, such as osteosarcoma, proximal to an OAR, while maintaining low toxicity using online PET image guidance.

**The BARB beamline**

Fig. 2 shows the full BARB beamline prepared in Cave M at GSI along with photographs of different components. The secondary beam of $^{11}$C comes from the FRS[36] (Extended Data Fig. 1). The primary intensity of the $^{12}$C-ion beam in the 18 Tm heavy ion synchrotron (Schwerionensynchrotron; SIS18) at 300 MeV/u was $1.6 \cdot 10^{10}$ particles/spill and the intensity of $^{11}$C-ions in Cave M entrance was $2.5 \cdot 10^{6}$ particles/spill. To maximize the online PET acquisition time, we used a short spill duration of 200 ms and a relatively low duty cycle with a repetition rate of 3 s (see Extended Data Table 1 for a summary of all the parameters).

A measured pristine $^{11}$C-ion Bragg curve is shown in Extended Data Fig. 2 A-D. To cover the full CTV, the pristine Bragg peak had to be widened to produce a Spread-Out Bragg-Peak (SOBP). The SOBP was formed by a 3D-printed range modulator (Extended Data Fig. 3A) from a 2D scan of the monoenergetic pencil beam. The measured SOBP dose distribution is shown in Extended Data Fig. 2 E-H.

The distal field contour was modulated to the tumour CTV (Fig. 1B) by a 3D-printed plastic compensator collar (Extended Data Fig. 3B), also used as holder for immobilization and positioning. We measured a dose-rate around 1 Gy/min in the target volume covered by the SOBP. A total of 32 mice were irradiated with either high- (20 Gy) or low- (5 Gy) tumor dose.

**PET activity**

In BARB we use online PET imaging to monitor the RIB dose delivery. We first tested the consistency of the measurements and Monte Carlo simulations in plastic phantoms having same shape and material composition of the compensator collar (Extended Data Fig. 3B). A Monte Carlo simulation of the $^{11}$C-beam interacting with a plastic phantom placed inside the SIRMIO PET scanner is shown in Extended Fig. 4, along with the measured PET image, both in transversal (A-C) and lateral (D-F) view. The Monte Carlo simulations include the full experimental setup, the beam model and the PET signal formation in the detector, to account for the imaging process through the same reconstruction as for measured data. PET images are overlaid on the µCT scan of the irradiated phantom. In Extended Data Fig. 4G we show the simulated dose, activity, and the measured PET activity profiles along the z-axis direction, integrated on the beam's eye view (BEV) aperture (±1 mm) in the x-y plane transversing the beam direction. We observed a good agreement between the simulations and experimental data, particularly in the peak region, where the PET activity peak aligns with the 80% SOBP dose fall-off. This supports the feasibility of the system for the in vivo experimental campaign. Differences in the measured and simulated activity profiles can be attributed to factors that degrade the measured PET signal, such as statistics, secondary radiation background, and detector sensitivity, as well as to uncertainties in the Monte Carlo model and in the µCT calibration. These factors contribute to a higher intensity of the measured PET signal at the target entrance and a wider distribution compared to simulations.

The PET activity in a mouse bearing the LM8 tumor is shown in Fig. 3 in sagittal view. We show the Monte Carlo simulation of the expected $^{11}$C-ion dose (in Gy) distribution calculated on the µCT of a mouse irradiated during the experiment (Fig. 3A) and the corresponding simulated PET activity (Fig. 3B). In Fig. 3C we show the PET image acquired during the experiment overlayed on the same pre-treatment µCT used for the simulations. All other measured PET images for the different mice irradiated with $^{11}$C-ion beam are reported in Supplementary Fig. 1. Supplementary video 1 shows the build-up of the measured PET signal over the course of irradiation, both in sagittal and transversal views.

Fig.4A shows the z-axis profiles corresponding to Fig. 3 integrated over the BEV aperture (±1 mm) in the x-y plane perpendicular to the beam direction. It can be noted that a shift of about 1 mm is observed between measured and simulated activity distributions. During the experiment, all PET images were overlaid on the corresponding pre-treatment µCT scans of each mouse, acquired in horizontal position. However, actual irradiations were performed with mice in vertical position (Fig. 2), which may induce some anatomical changes. We conducted additional experiments to compare imaging in horizontal and vertical positions using the cone-beam CT (CBCT) of the Small Animal Radiation Research Platform (SARRP) installed in Cave M. Although the resolution of the CBCT is lower than that of the µCT,

Extended Data Fig. 5 reveals small but consistent differences in spine curvature when the animals are placed in a vertical orientation. By including this anatomical shift in the simulation, the Monte Carlo calculation accurately reproduces the position of the measured activity peak (Fig. 4B). As for the plastic phantom measurements (Extended Data Fig. 4G), now the activity peak approximately aligns with the SOBP 80% dose fall-off. Residual discrepancies in the shape of the measured and simulated profiles are similar to those described above for the phantom image, plus additional uncertainties such as animal repositioning and non-homogenous target composition, particularly the dimple in the collar as well as different densities of the target (bone, hairs, fat, skin, etc.).

As the tumors were growing very close to the spinal cord, the online PET image was used especially in the first minutes of the irradiation to check that the SOPB was not covering the spine. Extended Data Fig. 6 shows a Monte Carlo simulation of the dose and corresponding predicted activity for an SOBP extending into the spinal cord, that is a case where the range is longer than expected. The simulation is superimposed on both μCT (C-F) and CBCT (B-E) images of the same mouse, in transversal (A-C) and sagittal (D-F) view. The images C-F were never observed in any animal (Supplementary Fig. 1). They would have meant doses as shown in panels A-D, and unacceptable profiles as shown in Extended Data Fig. 6G. Therefore, no range correction by adjusting the degrader thickness (Fig. 2) had to be applied, even if the tumour was almost leaning on the spine.

**Tumor control**

Tumor sizes of irradiated and control tumor-bearing animals were measured for four weeks after the day of irradiation using caliper or μCT. Results shown in Fig. 5 demonstrate complete tumor control after 20 Gy and prolonged tumor growth delay after 5 Gy, with evidence of recurrence after 2 weeks. Full tumor control was also achieved with 20 Gy X-rays (Extended Data Fig. 7). The data are compatible with a complete coverage of the tumor target for all animals in the $^{11}$C-beam treatment. Recurrence at the lower dose is expected considering the high radioresistance of the osteosarcoma.

**Toxicity**

Skin toxicity scoring in the tumor-bearing controls was complicated by the growth of the tumor that caused superficial lesions. In irradiated animals, all of them bearing small tumors post-irradiation, skin toxicity was radiation-induced as shown in Supplementary Fig. 2. No irradiated animals showed skin toxicity grade > 3.

Being the tumor very close to the cervical area of the spinal cord, the main expected toxicity was radiation-induced myelopathy. However, none of the animals exposed to $^{11}$C-ions presented severe morbidity such as forelimb paralysis or pronounced kyphosis. Extended Data Fig. 7 and Supplementary video 2 show the toxicity after X-rays, which was much more severe than for mice exposed to the particle beam. Animals suffered severe weight loss due to impaired feeding, ostensibly caused by damage to the esophagus, and grade 3 kyphosis, caused by cervical myelopathy as expected from previous experiments in mice[32–34] or rats[37–40]. This morbidity is caused by the X-ray dose that, unavoidably, was received by the organs behind the tumor in the neck region.

For animals exposed to $^{11}C$-ions, the lack of severe toxicity demonstrates that the spine was not exposed to high doses, as observed with the PET measurement (compare Fig. 3 to Extended Data Fig. 6). However, since the tumor is adjacent to the spine, some activity was inevitably observed in the spinal cord, located in the dose fall-off region (Fig. 4). We checked the impact of this residual dose on low-grade toxicity by measuring grip strength performance, a common test to assess cervical spinal injury[41] (Extended Data Fig. 8). All results of the bi-weekly grip strength performance for individual mice are reported in Supplementary Fig. 3. A wide inter-individual variability is noted in those curves. However, by pooling the data in Extended Data Figs. 9A-B we show that the strength of the mice is reduced after irradiation compared to controls, indicative of a minor deficit in neuromuscular function. Correlation of integral PET counts in the spine with individual grip performance is shown in Extended Data Figs. 9 C-D. Despite the wide scatter in the grip test data, there is a significant correlation between activity in the spine and decreased mouse forelimb strength. In this figure, the activity is measured on the μCT images of the irradiated animal. As shown in the Monte Carlo simulation in Supplementary Fig. 4, the anatomical change in repositioning (Extended Data Fig. 5) only leads to <5% difference in the activity counts, and therefore does not modify the correlation shown in Extended Data Figs. 9 C-D. We therefore demonstrate that the activity map of the radioactive therapeutic beams predicts toxicity in the OAR.

**Washout**

The activity in a plastic target decreases after irradiation because of the physical decay of the $^{11}C$ ($T_{1/2}$=20.34 min) projectile. Additional positron emitters in the target are $^{10}C$ and $^{15}O$. They are not beam contaminants but arise from nuclear fragmentation during tissue irradiation with the $^{11}C$ beam: $^{10}C$ is primarily produced by the beam's fragmentation upon interacting with the target, while $^{15}O$ is generated solely from tissue fragmentation (with soft tissue being ~65–75% oxygen). Their low production cross sections (40–70 mb)[42] yield abundances about two orders of magnitude lower than that of $^{11}C$. Moreover, the positron activity profile from tissue fragmentation drops near the Bragg peak—rendering $^{15}O$ negligible (despite its $T_{1/2}$ =2.04 min) and allowing the distinct impact of $^{10}C$ ($T_{1/2}$ =19 s) to be isolated from the $^{11}C$ signal and biological washout effects.

We have previously modelled the radioactive decay with exponential functions that include all fragments produced[25]. In this experiment, the radioactive decay is overlapped with an unknown biological decay due to the blood flow in the tumor that removes the radioactive isotopes from the site of decay. The degree of vascularization in our tumor model was estimated by perfusion to opacify microvasculature structure in μCT. Supplementary video 3 shows that our osteosarcoma in the neck is highly vascularized, so a strong biological washout is expected. The washout data for all animals are reported in Supplementary Fig. 5. Studies in Japan in a rat glioma model point to a double-exponential model for the biological washout[43], which was also applicable to our data based on the results of the Fisher's test on fitting parameters. We therefore used the following equation to fit the activity data measured after the irradiation was stopped:

$$A(t) = A_{phys} \cdot A_{bio} = A_0 \sum_i w_i e^{\frac{-ln2}{T_{1/2i}}t} \cdot \left[W_s e^{-k_s t} + (1 - W_s) e^{-k_f t}\right]$$

where $A_0$ is the activity at the end of the irradiation, $W_s$ - the relative weight of the slow component, and $k_s$ and $k_f$ are the slow and fast time constants, respectively. $T_{1/2i}$ is the half-life of the *i*-th contributing radioisotope and $w_i$ is its fraction in the total number of fragments. Based on the FLUKA simulation, we have considered 96% $^{11}C$, 3% $^{10}C$, and 0.5% $^{15}O$ ions. Fig. 6 shows the pooled analysis of the animals exposed to 5 or 20 Gy (individual curves are reported in Supplementary Fig. 5). The results clearly show a significant difference between the low- and high-dose experiments. The fast component, very well visible at 5 Gy, essentially disappears at 20 Gy. This suggests a quick vascular injury at high doses that delays the washout process in the first half an hour after the irradiation.

**Discussion**

The goal of the BARB project was to provide a first demonstration of tumor treatment with RIB with online range verification by PET. The results in Fig. 5 are in fact the first example of successful tumor control with RIB. We used a very high dose of 20 Gy in a single fraction, expecting tumor control based on our own previous X-ray experiment in the same murine osteosarcoma model. However, while X-rays caused severe toxicity at that dose (Extended data Fig. 7, Supplementary video 2), this was not the case with $^{11}C$-ions, where we observed minor toxicity only, correlated to the residual activity measured in the spine (Extended Data Fig. 9). Therefore, we conclude that image-guided particle therapy with RIB is feasible, safe and effective.

As previously observed in phantom experiments[44], even using RIB for online beam imaging the activity peak can be shifted compared to the fall-off of the SOBP, depending on the beam momentum spread. In our study (Fig. 4A), the measured activity is shifted compared to the simulation in the in vivo experiment due to anatomical changes (different horizontal/vertical position for planning/delivery), in addition to other uncertainties in the beam modeling, μCT calibration and imaging process. The anatomical changes after repositioning of the mouse from a horizontal to a vertical orientation (Extended Data Fig. 5) is particularly significant in light of ongoing efforts to implement particle therapy in an upright position[45]. Our findings reinforce the necessity of vertical CT planning and highlight the potential of online PET as a valuable tool for upright particle therapy.

Notwithstanding these uncertainties, our results show that the most distal position of the activity maximum from the Bragg peak can be used as a reliable indicator of the deepest location where a relevant amount (≥ 80%) of dose is deposited (Fig. 4B). The capability of RIB to predict the distal fall-off position around the 80% of the Bragg peak had been already demonstrated in phantoms[46], and is now confirmed in vivo. Therefore, our results provide first experimental support to the modelling predictions of the benefit of RIB in particle therapy for target margin reduction[19].

The results on washout (Fig. 6) are intriguing. One of the main tenets of radiotherapy is that tumor control can only be achieved when all cancer stem cells are killed[47]. However, already 20 years ago, it has been shown that at high doses microvascular endothelial apoptosis can contribute to tumor sterilization[48]. Later studies showed that the tumor damage at high doses induce vascular damage[49] and can be mediated by an ischemic-reperfusion mechanism[50]. This idea has been already translated in clinics with the single-dose radiotherapy (SDRT)[51], that uses single fractions of 24 Gy rather than fractionation in small malignancies, achieving excellent clinical results[52]. However, the concept is controversial. In fact, according to the classical linear-quadratic model, single fractions are much more effective than fractionated doses, and therefore the benefits of SDRT may simply be attributed to the high biological effectiveness of single doses[53,54]. The available experimental data are not conclusive[55]. Dynamic contrast-enhanced MRI in rats shows increased permeability after high doses of X-rays or C-ions[56], but those studies look at the effects weeks after irradiation, whereas our results cover the initial 30 minutes after exposure. This time frame is crucial, as other studies reported ischemic stress following SDRT within a few minutes[50] or hours[57]. Biological washout provides a direct measurement of the vascular perfusion in the tumor, and therefore we believe that this technique can clarify the vascular engagement after radiotherapy. Our results are consistent with an ischemic stress occurring very early after high doses. Although reperfusion was not observed within the measured time interval, we cannot rule out the possibility that it may occur at a later time.

**Outlook**

What are the next pre-clinical steps in RIB research? We will test short-lived isotopes such as $^{10}$C or $^{15}$O, which are expected to provide stronger signal and faster feedback, for increased temporal resolution. For $^{10}$C, it will be necessary to use the new Super-FRS[58] at FAIR, which will be able to provide much higher intensities of secondary beams.

For the washout studies, future experiments should investigate a wider range of doses and extended post irradiation time-points, alongside tumor histology to better assess vascular changes following irradiation.

Can these successful results lead to a clinical translation of RIB? The MEDICIS-Promed[59,60] project at CERN proposed an ISOL production of a $^{11}$C beam that can then be injected directly into medical synchrotrons currently used for $^{12}$C-ion therapy. Moving the isotope production to the low-energy injecting area can indeed be a feasible solution where RIB can be used at least for an initial test of the range before a full treatment course. The Open-PET scanner[61,62] developed at QST in Japan or the INSIDE in-beam PET[10,63] installed at CNAO in Italy can be excellent detectors for clinical applications of RIB. Finally, washout can also have an important prognostic value in the clinics as it should be correlated to the tumor vasculature state and eventually to the level of hypoxia[43,64], a notorious negative prognostic factor in therapy[65]. Our preclinical results show for the first time the feasibility of RIB radiotherapy and support these ongoing efforts for clinical translation.

## Methods

### RIB production

The $^{11}C$ beam was produced via in-flight separation. A 300 MeV/n $^{12}C$ primary beam from the SIS18 synchrotron impinges on a 8.045 g/cm$^2$ thick beryllium target at the FRS[36] and undergoes peripheral nuclear reactions, so that one or more nucleons are stripped-off, leading to a variety of lighter isotopes from carbon and other elements from boron down to hydrogen. Via a combined magnetic rigidity analysis and energy loss, which is induced in a so-called wedge-shaped degrader that is located at the central focal plane of the fragment separator, an isotopic clean $^{11}C$ beam is achieved. The purity of the beam is about 98% (Supplementary Fig. 6). This beam was used at the FRS for a variety of basic nuclear and atomic physics studies (such as reaction cross sections of the $^{11}C$-ions, their range and range straggling, basic PET studies etc.)[24,25,66] in preparation of the present experiment. Via the connecting beamline to the target hall[26], the isotopic clean beam is transported to the Cave M, where the present irradiation has been accomplished. The FRS and its three branches are shown in Extended Data Fig. 1 and exact parameters of the beam used in the present experiments are reported in Extended Data Table 1.

### Dosimetry

The $^{11}C$ beam from the FRS reached the experimental room with an intensity of $2.5 \cdot 10^6$ particles per spill. To minimize irradiation and allow for extended PET image acquisition, a beam cycle of 0.2 s ON and 3 s OFF was used. Once in the experimental room the beam was monitored with large parallel plate ionization chambers[67]. The beam was characterized in terms of beam spot size and 1D and 3D depth dose distributions in water by means of a PTW PEAKFINDER™ system (PTW Freiburg, Germany) and an in-house water phantom[68] equipped with an OCTAVIUS 1600 XDR (PTW Freiburg, Germany). The latter setup allows to acquire 2D dose distributions at different water depth which can be processed to generate 3D dose map distributions.

A range modulator (Extended Data Fig. 3A) was used to generate a 1.2 cm SOBP in water. This modulator was 3D-printed on a 3D Systems ProJet MJP 2500 Plus using VisiJet M2S-HT250 as the printing material and VisiJet M2 SUP as the support material. The printing material has a water-equivalent density of 1.162 g/cm³ and a physical density of 1.1819 g/cm³.

The measured pristine and SOBP curves are shown in Extended Data Fig. 2. The Bragg peak position in water for the pristine depth dose distribution was measured at 80.5 mm. By comparison with Monte Carlo simulations, it was then possible to estimate beam parameters such as the beam energy and momentum spread. All measured and estimated beam parameters, also used for the Monte Carlo simulations, are reported in the Extended Data Table 1.

As the $^{11}C$ beam spot size is much larger than standard clinical beam and the size of the target volume was comparably small, it was not possible to use standard active scanning techniques to deliver the desired dose to the CTV. A system of modulator, degraders, collimator, and compensator was then used to passively modulate and optimize the beam for animal irradiation.

A schematic of the complete setup for mice irradiation is shown in Fig. 2A, and a more detailed description of the experimental setup and beam characterization can be found in the Supplementary Material - Dosimetry. At first, to achieve the desired penetration depth in the mouse neck (~5.6 mm), the energy of the beam at the mouse position had to be reduced. This was achieved by introducing defined thicknesses of material in the beamline, in particular, 28.4 mm thick aluminum plates (60.1 mm water equivalent path length), and a range shifter equipped with remotely controlled polyethylene plates. This latter component allowed fine adjustments of the Bragg peak position and a dose delivery correction in case a range correction would have appeared necessary during the treatment. The distal edge of the SOBP was shaped to the distal contour of the generalized CTV using a specially designed compensator (Extended Data Fig. 3B), which was placed on the neck of the mice and secured on the bed. These compensators, produced using the same 3D-printer as the range modulator, also served for immobilization and mice positioning.

To laterally define the irradiation area, additionally to the mouse neck compensator, a system of brass collimators with a 15×12 mm elliptical aperture was placed right before the PET scanner (see Fig 2A and Supplementary Material – Dosimetry Figure S1). This system had also the function of shielding the detector and limit the noise signal in the scanner by blocking most of the ions which would not contribute to the dose on the target. Absolute dosimetry at the tumor site was performed using a small-volume pinpoint ionization chamber (PTW TM31023) placed within a custom-designed dosimetry holder (Supplementary Material – Dosimetry Figure S7). This setup ensured that the detector's sensitive volume corresponded to the tumor depth. The absolute dose measured by the chamber was then used to determine a setup-specific monitor unit calibration, allowing for the scaling of treatment plans to the prescribed dose. The chamber readings were recorded using a UNIDOS electrometer, applying the standard correction factors ($k_Q$ and $k_{TP}$) commonly used in particle therapy (see Supplementary Material - Dosimetry for further details).

**Animal model**

All experiments were performed using 11-12-week-old female C3H/He mice (Janvier Labs, France) according to German Federal Law under the approval of the Hessen Animal Ethics Committee (Project License DA17/2003). Mice were divided into five groups: 20 Gy $^{11}$C irradiation (17 animals, followed up for 6 months after the irradiation); 20 Gy and 5 Gy $^{11}$C irradiation followed by additional 30 minutes of PET signal acquisition (8 and 7 animals, respectively, followed up for four weeks after the irradiation); tumor-bearing sham-irradiated controls (11 animals, followed up for four weeks after the irradiation day); and tumor-free controls (8 animals, followed up for 6 months after the irradiation day). Mice were housed at GSI in a conventional animal facility (non-SPF) at 22°C, 12-hour light-dark cycle, with unrestricted access to water and a standard diet (Ssniff, Germany). Fourteen days before irradiation, $10^6$ mouse Dunn osteosarcoma LM8 cells (originating from C3H/He mice, purchased from Riken BioResource Center, Japan) were injected in 20 µl of PBS buffer solution subcutaneously in the neck area of the mouse, above the cervical area of the spine. To maintain the consistency of injections, during the procedure animals were anesthetized with 2% isoflurane, which was inhaled via a face mask. After two weeks, most tumors were palpable and measurable at least with the µCT.

### X-ray experiment

The primary goal of the X-ray experiment was to assess tumour control and the toxicity of high doses of radiation to mouse skin and to the spinal cord. We used the in-house X-ray generator (Isovolt DS1, Seifert, Ahrensberg, Germany) operated at 250 kVp and 16 mA, with 7 mm beryllium, 1 mm aluminum, and 1 mm copper filtering. Animals were positioned inside an isoflurane-filled plastic container, directly under the 2 mm thick lid. The setup height was adjusted to the dose rate of 5 Gy/min at the tumor position, which was confirmed by the ionization chamber measurement (PTW, Freiburg, Germany). Brass collimators of 2.5 cm thickness with 1.5×1.2 cm oval openings were used to shield the animal bodies, only exposing the tumor-bearing neck areas. Five animals each were irradiated with 20 or 15 Gy.

### CT imaging

#### *Horizontal imaging - µCT*

To collect the data on the tumor growth, we performed µCT measurements with a VivaCT 80 scanner (SCANCO Medical AG, Switzerland). To ensure the reproducible positioning of the mice as during the $^{11}C$ irradiation, we have used the custom-made bed imitating the geometry of the SIRMIO bed. Additionally, animals were immobilized using the compensator collar, so that the scans could be utilized for the later Monte Carlo calculations. During the scan, animals were anesthetized with isoflurane (3% for the induction, 1.5-2% for maintenance during the scan). Neck regions of 31 mm were scanned for approximately 5 minutes at a tube voltage and current of 45 kVp and 177 µA, respectively, adding a 0.1 mm aluminum filter, acquiring 250 projections per 180° with 45 ms integration time. The resulting images had a voxel size of 97.1 µm. After the scan, animals were allowed to recover before being transferred to the original cage. The scans performed at 14 days after the tumor cells injection into the animals used to establish the tumor model were used to contour the generalized CTV for the treatment planning. The contouring of the visible tumor mass was done manually with the 3D Slicer software[69] for every animal, then the individual GTV contours were added up. We expected that the majority of the tumors in the $^{11}C$ groups would grow in the similar location and will not differ in size from those of the test group. Nevertheless, to increase robustness towards the biological variation, we smoothened the resulting CTV contour and made it symmetrical with respect to the spine (Figure 1C). The animals selected for $^{11}C$ irradiation were scanned one day before the irradiation (13 days after the tumor cells injection).

#### *Vertical imaging – SARRP*

To resolve remaining discrepancies between simulations and measurements (see Fig. 4A), we checked for anatomical changes arising from repositioning the mice from the horizontal µCT bed, used for imaging, onto the vertical SIRMIO bed used during the $^{11}C$ irradiation (Fig. 2). We have utilized the CBCT of the Cave M SARRP (XStrahl, Germany) to acquire vertical CT scans of mice by positioning them onto a vertical holder replicating the geometry and fixation procedure (including the compensator collar) of the SIRMIO bed.

During these tests, animals were anesthetized with isoflurane (3% for the induction, 1.5% for maintenance) and, to further imitate the conditions of the main experiment, they were left in the vertical position for several additional minutes before the scans were taken. We have acquired 250 projections per 180° at tube voltage and current of 70 kVp and 1 mA, respectively, adding a 0.1 mm copper filter. The resulting images had a voxel size of 275 µm. After the scan, animals were allowed to recover before being transferred to the original cage.

We have observed some consistent variations in the spine curvature (<1.5 mm in the neck area) between the µCT and the CBCT scans. This anatomical change reconciles the measurements with the Monte Carlo simulations (Fig. 4B).

**Tumour vascularization**

To assess the tumor vascularization, animals were sacrificed and perfused ex vivo with Vascupaint™ contrast agent (yellow colloidal bismuth suspension, MediLumine, Canada). After allowing the compound to polymerize for 24 hours, tumors were extracted and scanned at a high resolution (10 micron) with the following scanner settings: tube voltage and current of 55 kVp and 145 µA, respectively, 0.5 mm aluminum filter, 1500 projections per 180° with 600 ms integration time. The 3D reconstruction of the tumor vasculature was done using the scanner's built-in software 'Bone morphology' function following the approach of another study with a contrast agent[70].

**Online PET**

A novel spherical, high resolution PET scanner developed at LMU in the framework of the SIRMIO project was used to measure the RIB implantation in-beam during irradiation. The SIRMIO PET scanner features 56 3-layer depth-of-interaction (DOI) detectors arranged in a spherical shape with an inner diameter of 72 mm[28]. Each DOI PET detector consists of a LYSO scintillator block with a pixel size of 0.9 mm readout by an 8×8 SiPM array. A charge division circuit and a custom-made amplifier circuit board developed at the National Institutes for Quantum Science and Technology (Chiba, Japan) are used to reduce the 64 signals from the SiPM array to 4 signals. The data are then acquired by a customized DAQ software using two R5560 digitizers (CAEN, Italy). In order to enable image-guided irradiation for the BARB project, the data acquisition and reconstruction software were tailored to stream out and reconstruct the list mode data with user-defined time intervals, set in this experiment to cycles of 60 seconds. This feature enables visualizing the reconstructed stopping position of the beam online during the irradiation, along with the monitoring of the irradiation build up and decay through a graphical user interface. For this specific online application, the image reconstruction was based on an in-house developed ordered subset expectation maximization (*OSEM*) algorithm, with a reduced number of iterations and limited size of the field of view for the sake of computational speed during the experiment. The 3D activity maps and washout analyses used for the reported results were based on a more time-consuming 3D maximum likelihood expectation maximization (MLEM) with relevant corrections for sensitivity and random coincidences. Attenuation corrections were not included since a separate study confirmed that they have a negligible influence on the shape of the activity distribution for the considered small size of the irradiated target.

**Co-registration**

For accurate co-registration of the imaged activity with the pre-treatment μCT, prior to the experiment a specially designed mouse bed equipped with an insert for a $^{22}Na$ point source was positioned in the PET scanner. Multiple point-source measurements were performed at well controlled positions using precision linear stages to move the SIRMIO bed with the source at different locations in the field of view of the SIRMIO PET scanner. By knowing both the physical location of the point source in the bed and its reconstructed position in the PET image, the mouse bed—and consequently the mouse position during treatment through the reproducible positioning of the mouse collar on the bed —could be accurately aligned within the PET field of view, thus enabling to accurately overlay the reconstructed activity images with the treatment planning anatomy.

**Monte Carlo**

An extensive FLUKA Monte Carlo[71] simulation study was conducted to support both the experiment design and its data analysis. Simulations were performed using the HADRONTherapy DEFAULT card in FLUKA (v2021.2.3) with the flair GUI (v2.3-0). The mean water ionization potential was set to 78 eV[72]. We activated the COALESCEnce card for light fragment spectra and residual nuclei, and the IONSPLIT card for deuteron splitting at low-energy interactions. For simulation in the SOBP configuration a user-defined USERROUTINE was implemented to read the 2DRM geometry file[73]. The number of primary particles in the simulations was chosen to ensure that statistical uncertainties due to Monte Carlo fluctuations remained below 1%.

The simulations assisted in designing beamline components, including the 2D range modulator, collimator, and mouse compensator collar, and in developing shielding strategies to protect the SIRMIO PET scanner from radiation damage. As in our previous BARB dosimetry study FLUKA simulations were also used to verify dosimetry measurements and support beam model characterization[66] (see Supplementary Material - Dosimetry).

Expected dose and positron annihilation maps inside the body were simulated by importing the mice scans with their original resolution in a voxel FLUKA geometry. For the purpose of accurate beam model and transport, the full experimental setup was implemented into the FLUKA geometry, and a setup-specific CT number to stopping power calibration, including the mouse bed and compensator, was implemented in FLUKA.

To reproduce the detector response and imaging process, the annihilation maps simulated with FLUKA were then imported in a Geant4 dedicated simulation setup which includes a detailed model of the SIRMIO PET scanner[26]. To start propagating the annihilation photons from the FLUKA simulated annihilation maps, positrons with no kinetic energy were simulated in Geant4 to enforce annihilation at the same position as the input map. The resulting annihilation photon pairs were then transported in air through the detector setup, accounting for the geometrical detector response, but omitting attenuation in the target (as its effect on the shape of the reconstructed activity was found negligible in a separate simulation study and is also not applied as correction in the image reconstruction). The Geant4 simulation output is a list of hits in the detector crystals which is then post-processed to resemble the experimental data.

The final simulated PET image was obtained using the same reconstruction method applied to the real measurements. The correction for sensitivity used in the reconstruction is based on the same Monte Carlo simulation model, which was validated at a few positions in the relevant central part of the field of view where the maximum of activity is expected. Small remaining inconsistencies between the theoretical and real sensitivity at the edge of the field of view could affect differently the reconstruction of the simulated and measured data, introducing small mismatches of different magnitude in the entrance/tail region, which are visible in the profiles in Fig. 4. Moreover, the simulated PET images are only based on the simulated annihilation distributions and do not include the prompt radiation background generated by the RIB irradiation, nor the very minor amount of intrinsic radioactivity of the used LYSO crystals[28]. Furthermore, the simulation does not include the biological washout model, which can slightly broaden the measured activity in the animal. However, the radiation background was largely suppressed in the measured data by using very narrow coincidence energy windows, exploiting the large signal-to-noise ratio of RIB irradiation. Moreover, the washout contribution can be considered negligible in the first few minutes of irradiation, in which the output of our online monitoring was used to decide on the adequate sparing of the spine.

**Mouse follow-up**

***Tumor growth***

Starting one week after the injection, tumor dimensions were measured with a caliper twice per week for 28 days after the irradiation (Fig. 5A). Assuming the ellipsoid shape of the tumor, the volume was calculated as

$$V = \frac{4}{3}\pi abc$$

where $a$ and $b$ are the measured length and width of the tumor, respectively, and $c$ (depth) is assumed to be the average of $a$ and $b$. For more accurate and reproducible estimates[74], volumes were also measured using the μCT (Fig. 5B). After the irradiation, animals were scanned weekly for four weeks; the animals remaining after the 28-days timepoint were scanned monthly afterwards until the sacrifice timepoint.

***Toxicity assays***

Skin toxicity scoring. The toxicity of the skin in the irradiated area was scored using a simplified grading system of the Gesellschaft für Versuchstierkunde (GV-SOLAS) guidelines, divided by five grades (0: no effect; 1: redness; 2: dry skin and desquamation; 3: closed, healing wound; 4: open wound, not healing; 5: necrosis). Grade 5 was never observed during the experiment, and the termination criteria was reached when the animals showed a grade 4 and did not heal after the application of a topical treatment (Bepanthen®, Bayer). The treatment healed successfully the animals with a grade 3 after irradiation.

Grip test. The grip test was performed to measure the strength of the animals´ forelimbs after irradiation. Animals were acclimatized to refined handling techniques to reduce stress and optimize the data collection. We used the Grip Strength Meter-47200 (Ugo Basile®) equipped with a T-bar. The animals were lifted by the tail and suspended over the bar, then lowered to reach a horizontal position and gently pulled back until the grasp was released. Upon release, the peak force (in newton) was recorded. To get consistent data and avoid habituation to the task, the first three measurements in which the animal successfully grabbed the bar with both forelimbs were recorded and averaged. The procedure is illustrated in Extended Data Fig. 8.

Kyphosis scoring. To score the overall appearance and health status of the animals, videos were recorded in a house-made setup consisting of a starting box, a transparent-walled corridor, and a loop structure at the end, where the animals could enter and go back to the corridor. The animals were observed for spontaneous walking, grooming behavior and posture during stationary and movement phases. To evaluate the kyphosis, we used the scoring system from Yerger *et al.*[75] where in grade 0 there is no persistent kyphosis and the mouse can always straighten the spine; 1: mild kyphosis exhibited during stationary phase, but the spine is straightened during locomotion; 2: persistent mild kyphosis even during movement, the spine cannot be straightened completely; 3: the kyphosis is always maintained and well pronounced.

## References

1. Durante, M. & Paganetti, H. Nuclear physics in particle therapy : a review. *Reports Prog. Phys.* **79**, 096702 (2016).

2. PTCOG. Particle therapy facilities in operation. (2025). Available at: https://www.ptcog.site/. (Accessed: 20th March 2025)

3. Durante, M., Debus, J. & Loeffler, J. S. Physics and biomedical challenges of cancer therapy with accelerated heavy ions. *Nat. Rev. Phys.* **3**, 777–790 (2021).

4. Lomax, A. J. Myths and realities of range uncertainty. *Br. J. Radiol.* **93**, 20190582 (2020).

5. Paganetti, H. Range uncertainties in proton therapy and the role of Monte Carlo simulations. *Phys. Med. Biol.* **57**, R99–R117 (2012).

6. Parodi, K. & Polf, J. C. In vivo range verification in particle therapy. *Med. Phys.* **45**, e1036–e1050 (2018).

7. Parodi, K., Yamaya, T. & Moskal, P. Experience and new prospects of PET imaging for ion beam therapy monitoring. *Z. Med. Phys.* **33**, 22–34 (2023).

8. Enghardt, W. *et al.* Charged hadron tumour therapy monitoring by means of PET. *Nucl. Instruments Methods Phys. Res. Sect. A Accel. Spectrometers, Detect. Assoc. Equip.* **525**, 284–288 (2004).

9. Bauer, J. *et al.* Implementation and initial clinical experience of offline PET/CT-based verification of scanned carbon ion treatment. *Radiother. Oncol.* **107**, 218–226 (2013).

10. Kraan, A. C. *et al.* In-beam PET treatment monitoring of carbon therapy patients: Results of a clinical trial at CNAO. *Phys. Medica* **125**, 104493 (2024).

11. Kong, J. *et al.* A dedicated in-beam PET system with a modular dual-head for radiotherapy imaging in HIMM. *J. Instrum.* **19**, P04021 (2024).

12. Jenkins, D. G. Recent advances in nuclear physics through on-line isotope separation. *Nat. Phys.* **10**, 909–913 (2014).

13. Blumenfeld, Y., Nilsson, T. & Van Duppen, P. Facilities and methods for radioactive ion beam production. *Phys. Scr.* **T152**, 014023 (2013).

14. Bertulani, C. A. & Gade, A. Nuclear astrophysics with radioactive beams. *Phys. Rep.* **485**, 195–259 (2010).

15. Chacon, A. *et al.* Experimental investigation of the characteristics of radioactive beams for heavy ion therapy. *Med. Phys.* **47**, 3123–3132 (2020).

16. Chacon, A. *et al.* Monte Carlo investigation of the characteristics of radioactive beams for heavy ion therapy. *Sci. Rep.* **9**, 6537 (2019).

17. Boscolo, D. *et al.* Radioactive Beams for Image-Guided Particle Therapy: The BARB Experiment at GSI. *Front. Oncol.* **11**, (2021).

18. Mohammadi, A. *et al.* Range verification of radioactive ion beams of 11 C and 15 O using in-beam PET imaging. *Phys. Med. Biol.* **64**, 145014 (2019).

19. Sokol, O. *et al.* Potential benefits of using radioactive ion beams for range margin

reduction in carbon ion therapy. *Sci. Rep.* **12**, 21792 (2022).

20. Chatterjee, A., Alpen, E. L., Tobias, C. A., Llacer, J. & Alonso, J. High energy beams of radioactive nuclei and their biomedical applications. *Int. J. Radiat. Oncol.* **7**, 503–507 (1981).

21. Durante, M. & Parodi, K. Radioactive Beams in Particle Therapy: Past, Present, and Future. *Front. Phys.* **8**, 00326 (2020).

22. Durante, M., Golubev, A., Park, W.-Y. & Trautmann, C. Applied nuclear physics at the new high-energy particle accelerator facilities. *Phys. Rep.* **800**, 1–37 (2019).

23. Durante, M. *et al.* All the fun of the FAIR: fundamental physics at the facility for antiproton and ion research. *Phys. Scr.* **94**, 033001 (2019).

24. Kostyleva, D. *et al.* Precision of the PET activity range during irradiation with 10 C, 11 C, and 12 C beams. *Phys. Med. Biol.* **68**, 015003 (2023).

25. Purushothaman, S. *et al.* Quasi-real-time range monitoring by in-beam PET: a case for 15O. *Sci. Rep.* **13**, 18788 (2023).

26. Haettner, E. *et al.* Production and separation of positron emitters for hadron therapy at FRS-Cave M. *Nucl. Instruments Methods Phys. Res. Sect. B Beam Interact. with Mater. Atoms* **541**, 114–116 (2023).

27. Gerlach, S. *et al.* Beam characterization and feasibility study for a small animal irradiation platform at clinical proton therapy facilities. *Phys. Med. Biol.* **65**, 245045 (2020).

28. Lovatti, G. *et al.* Design study of a novel geometrical arrangement for an in-beam small animal positron emission tomography scanner. *Phys. Med. Biol.* **68**, 235005 (2023).

29. Asai, T. *et al.* Establishment and characterization of a murine osteosarcoma cell line (LM8) with high metastatic potential to the lung. *Int. J. Cancer* **76**, 418–422 (1998).

30. Prudowsky, Z. D. & Yustein, J. T. Recent Insights into Therapy Resistance in Osteosarcoma. *Cancers (Basel).* **13**, 83 (2020).

31. Dong, M. *et al.* Efficacy and safety of carbon ion radiotherapy for bone sarcomas: a systematic review and meta-analysis. *Radiat. Oncol.* **17**, 172 (2022).

32. Lo, Y.-C., McBride, W. H. & Rodney Withers, H. The effect of single doses of radiation on mouse spinal cord. *Int. J. Radiat. Oncol.* **22**, 57–63 (1992).

33. Denbeigh, J. M. *et al.* Characterizing Proton-Induced Biological Effects in a Mouse Spinal Cord Model: A Comparison of Bragg Peak and Entrance Beam Response in Single and Fractionated Exposures. *Int. J. Radiat. Oncol.* **119**, 924–935 (2024).

34. Yokogawa, N. *et al.* Effects of Radiation on Spinal Dura Mater and Surrounding Tissue in Mice. *PLoS One* **10**, e0133806 (2015).

35. Ong, W. L. *et al.* Radiation myelopathy following stereotactic body radiation therapy for spine metastases. *J. Neurooncol.* **159**, 23–31 (2022).

36. Geissel, H. *et al.* The GSI projectile fragment separator (FRS): a versatile magnetic system for relativistic heavy ions. *Nucl. Instruments Methods Phys. Res. Sect. B Beam*

*Interact. with Mater. Atoms* **70**, 286–297 (1992).

37. Saager, M. *et al.* Carbon Ion Irradiation of the Rat Spinal Cord: Dependence of the Relative Biological Effectiveness on Linear Energy Transfer. *Int. J. Radiat. Oncol.* **90**, 63–70 (2014).

38. Saager, M. *et al.* Late normal tissue response in the rat spinal cord after carbon ion irradiation. *Radiat. Oncol.* **13**, 5 (2018).

39. Saager, M. *et al.* Fractionated carbon ion irradiations of the rat spinal cord: comparison of the relative biological effectiveness with predictions of the local effect model. *Radiat. Oncol.* **15**, 6 (2020).

40. Sun, Y. *et al.* Transplantation of Oligodendrocyte Precursor Cells Improves Locomotion Deficits in Rats with Spinal Cord Irradiation Injury. *PLoS One* **8**, e57534 (2013).

41. Anderson, K. D., Abdul, M. & Steward, O. Quantitative assessment of deficits and recovery of forelimb motor function after cervical spinal cord injury in mice. *Exp. Neurol.* **190**, 184–191 (2004).

42. Jin, S. Y. *et al.* Fragmentation of the positron-emitting nucleus 11C on a carbon target at 248 MeV/nucleon. *Radiat. Meas.* **171**, 107066 (2024).

43. Toramatsu, C. *et al.* Measurement of biological washout rates depending on tumor vascular status in 15 O in-beam rat-PET. *Phys. Med. Biol.* **67**, 125006 (2022).

44. Mohammadi, A. *et al.* Influence of momentum acceptance on range monitoring of 11 C and 15 O ion beams using in-beam PET. *Phys. Med. Biol.* **65**, 125006 (2020).

45. Volz, L. *et al.* Opportunities and challenges of upright patient positioning in radiotherapy. *Phys. Med. Biol.* **69**, 18TR02 (2024).

46. Hamato, A. *et al.* Dose estimation using in-beam positron emission tomography: Demonstration for 11C and 15O ion beams. *Nucl. Instruments Methods Phys. Res. Sect. A Accel. Spectrometers, Detect. Assoc. Equip.* **1066**, 169643 (2024).

47. Baumann, M., Krause, M. & Hill, R. Exploring the role of cancer stem cells in radioresistance. *Nat Rev Cancer* **8**, 545–554 (2008).

48. Garcia-Barros, M. *et al.* Tumor response to radiotherapy regulated by endothelial cell apoptosis. *Science* **300**, 1155–9 (2003).

49. Fuks, Z. & Kolesnick, R. Engaging the vascular component of the tumor response. *Cancer Cell* **8**, 89–91 (2005).

50. Bodo, S. *et al.* Single-dose radiotherapy disables tumor cell homologous recombination via ischemia/reperfusion injury. *J. Clin. Invest.* **129**, 786–801 (2019).

51. Greco, C. *et al.* Safety and Efficacy of Virtual Prostatectomy With Single-Dose Radiotherapy in Patients With Intermediate-Risk Prostate Cancer. *JAMA Oncol.* **7**, 700 (2021).

52. Zelefsky, M. J. *et al.* Phase 3 Multi-Center, Prospective, Randomized Trial Comparing Single-Dose 24 Gy Radiation Therapy to a 3-Fraction SBRT Regimen in the Treatment of Oligometastatic Cancer. *Int. J. Radiat. Oncol.* **110**, 672–679 (2021).

53. Shuryak, I., Carlson, D. J., Brown, J. M. & Brenner, D. J. High-dose and fractionation effects in stereotactic radiation therapy: Analysis of tumor control data from 2965 patients. *Radiother. Oncol.* **115**, 327–334 (2015).

54. Brown, J. M., Brenner, D. J. & Carlson, D. J. Dose escalation, not 'new biology,' can account for the efficacy of stereotactic body radiation therapy with non-small cell lung cancer. *Int. J. Radiat. Oncol. Biol. Phys.* **85**, 1159–1160 (2013).

55. Park, H. J., Griffin, R. J., Hui, S., Levitt, S. H. & Song, C. W. Radiation-Induced Vascular Damage in Tumors: Implications of Vascular Damage in Ablative Hypofractionated Radiotherapy (SBRT and SRS). *Radiat. Res.* **177**, 311–327 (2012).

56. Bendinger, A. L. *et al.* High Doses of Photons and Carbon Ions Comparably Increase Vascular Permeability in R3327-HI Prostate Tumors: A Dynamic Contrast-Enhanced MRI Study. *Radiat. Res.* **194**, (2020).

57. Jani, A. *et al.* High-Dose, Single-Fraction Irradiation Rapidly Reduces Tumor Vasculature and Perfusion in a Xenograft Model of Neuroblastoma. *Int. J. Radiat. Oncol.* **94**, 1173–1180 (2016).

58. Kalantar-Nayestanaki, N. & Scheidenberger, C. Experiments at the Interface of Nuclear, Atomic, and Hadron Physics with FRS at GSI and Super-FRS at FAIR. *Nucl. Phys. News* **34**, 21–26 (2024).

59. Penescu, L. *et al.* Technical Design Report for a Carbon-11 Treatment Facility. *Front. Med.* **8**, (2022).

60. Augusto, R. S. *et al.* New developments of 11C post-accelerated beams for hadron therapy and imaging. *Nucl. Instruments Methods Phys. Res. Sect. B Beam Interact. with Mater. Atoms* **376**, 374–378 (2016).

61. Tashima, H. *et al.* A single-ring OpenPET enabling PET imaging during radiotherapy. *Phys. Med. Biol.* **57**, 4705–4718 (2012).

62. Tashima, H. *et al.* Development of a Multiuse Human-Scale Single-Ring OpenPET System. *IEEE Trans. Radiat. Plasma Med. Sci.* **5**, 807–816 (2021).

63. Moglioni, M. *et al.* In-vivo range verification analysis with in-beam PET data for patients treated with proton therapy at CNAO. *Front. Oncol.* **12**, (2022).

64. Toramatsu, C. *et al.* Tumour status prediction by means of carbon-ion beam irradiation: comparison of washout rates between in-beam PET and DCE-MRI in rats. *Phys. Med. Biol.* **68**, 195005 (2023).

65. Bertout, J. A., Patel, S. A. & Simon, M. C. The impact of O2 availability on human cancer. *Nat. Rev. Cancer* **8**, 967–75 (2008).

66. Boscolo, D. *et al.* Depth dose measurements in water for 11C and 10C beams with therapy relevant energies. *Nucl. Instruments Methods Phys. Res. Sect. A Accel. Spectrometers, Detect. Assoc. Equip.* **1043**, 167464 (2022).

67. Luoni, F. *et al.* Beam Monitor Calibration for Radiobiological Experiments With Scanned High Energy Heavy Ion Beams at FAIR. *Front. Phys.* **8**, (2020).

68. Schuy, C., Simeonov, Y., Durante, M., Zink, K. & Weber, U. Technical note: Vendor-agnostic water phantom for 3D dosimetry of complex fields in particle therapy. *J.*

*Appl. Clin. Med. Phys.* **21**, 227–232 (2020).

69. Kikinis, R., Pieper, S. D. & Vosburgh, K. G. 3D Slicer: A Platform for Subject-Specific Image Analysis, Visualization, and Clinical Support. in *Intraoperative Imaging and Image-Guided Therapy* 277–289 (Springer New York, 2014). doi:10.1007/978-1-4614-7657-3_19

70. Downey, C. M. *et al.* Quantitative Ex-Vivo Micro-Computed Tomographic Imaging of Blood Vessels and Necrotic Regions within Tumors. *PLoS One* **7**, e41685 (2012).

71. Böhlen, T. T. *et al.* The FLUKA Code: Developments and Challenges for High Energy and Medical Applications. *Nucl. Data Sheets* **120**, 211–214 (2014).

72. Seltzer, S. M. *et al.* Key data for ionizing-radiation dosimetry: measurement standards and applications, ICRU Report 90. *J. ICRU* **14**, 5–8 (2014).

73. Charuchinda, W. *et al.* 3D range-modulators for proton therapy: near field simulations with FLUKA and comparison with film measurements. *J. Phys. Conf. Ser.* **2431**, 012081 (2023).

74. Jensen, M. M., Jørgensen, J. T., Binderup, T. & Kjær, A. Tumor volume in subcutaneous mouse xenografts measured by microCT is more accurate and reproducible than determined by 18F-FDG-microPET or external caliper. *BMC Med. Imaging* **8**, 16 (2008).

75. Yerger, J. *et al.* Phenotype assessment for neurodegenerative murine models with ataxia and application to Niemann–Pick disease, type C1. *Biol. Open* **11**, (2022).

**Acknowledgements**

The BARB experiments are supported by European Research Council (ERC) Advanced Grant 883425 BARB to M.D. The construction of the SIRMIO PET scanner was partly supported by the ERC Consolidator Grant 725539 SIRMIO to K.P. The measurements described here are performed within the experiments B-22-00046-Durante at SIS18/FRS/Cave-M at the GSI Helmholtzzentrum für Schwerionenforschung, Darmstadt (Germany) in the frame of FAIR Phase-0. The authors are grateful to: the SIS18 accelerator crew for their excellent support in the beam preparation and delivery; R. Chowdhury, L. Hartig, M. Ibáñez-Moragues, J. Oppermann, A. Puspitasari-Kokko, and C. Vandevoorde for their assistance in handling the animals during irradiations and follow-up; K. Lehmann for preparing the animal experiment proposal for the animal welfare authority in Hessen; C. Galeone, R. Kumar Prajapat, G. Li, M.C. Martire and L. Volz for their assistance during the beamtime shifts; S. Kumar Singh for the settings for $^{11}$C-beam at FRS; A.L. Gera, C. Hartmann-Sauter, T. Wagner and R. Khan for the beamline setup realization and installation; E. Rocco for refurbishing the grids needed for beam alignment; A. Noto for working on the SIRMIO PET during the irradiation; B. Franczak for the optical calculations that make possible the transfer of the RIB in Cave M; K. Zink for providing the PeakFinder tool used in our dosimetry measurements; T. Yamaya, H.G. Kang, A. Zoglauer, G. Dedes and C. Gianoli for their collaboration on the SIRMIO PET scanner and its simulation and image reconstruction environment. This paper is dedicated to the memory of Prof. Dr. Hans Geissel (1950-2024), whose unwavering support, encouragement, and insightful suggestions were invaluable to this research.

**Author information**

D.B. prepared the physics protocol, the beamline setup and performed the dosimetry; G.L. is the primary responsible for PET data acquisition and analysis, and simulation of the imaging process; O.S. linked biology and physics protocols, performed all μCT and performed a large part of the data analysis; T.V. worked on the animals, tumor growth and performed all toxicity tests; M.M. performed all FLUKA simulations and prepared the longitudinal profile figures; F.E., M.N. and P.G.T. worked on the PET detector and data acquisition; E.H., S.P., and D.K. were responsible of the secondary beam production at FRS; W.T. supervised the animal model and participated in the experiments; C.G. provided essential conceptual input and participated in the experiments; U.W. and C. Schuy performed the beam dosimetry; A.B. analyzed all the CT in horizontal and vertical position and calibrated the data; J.B. prepared the animal holders for the horizontal/vertical CT scanning and irradiation; C. Scheidenberger supervised all FRS activities and is responsible for production, separation and identification of the RIB; K.P. conceived and supervised the realization of the SIRMIO PET scanner, participated in all experiments and supported the interpretation and analysis of the PET data; M.D. conceived and supervised the BARB project, participated in every phase of the experiments, analyzed part of the data and wrote the first draft of the paper; all authors read and revised the manuscript.

**Ethics declarations**

Competing interests

The authors declare no competing interests.

**Figure captions**

**Figure 1. Mouse model and μCT. A.** LM8 osteosarcoma as visible by eye or at the μCT at different times after cell inoculation. For the irradiation, a two-weeks timepoint was chosen. **B**. Two slices of the CT depicting the contours of the individual tumor GTVs (green) near the OARs (spine and trachea with lungs, marked with yellow and blue, respectively). The generalized CTV contour (purple) was applied to all animals, covering all the possible GTV locations previously identified in the tumor induction study. Not all the GTVs are depicted here, since some of them were located on the different neighboring CT slices. **C.** The CTV obtained from the contours of individual GTVs of tumors that grew in the animals used to establish and confirm the tumor model. To account for further biological variation, the resulting contour was smoothened and made symmetrical with respect to the spine. The CTV is depicted in purple; the mouse skeleton is shown in light yellow; the trachea and the lungs are shown in blue. The light grey color depicts the contours of the mouse body.

**Figure 2. Experimental beamline. A**. Drawing of the different elements along the experimental beamline. **B.** SIRMIO animal holder with the anesthesia tubes and a mouse in position. **C.** Animal holder aligned in the beamline while the SIRMIO PET is raised. **D.** The SIRMIO PET scanner is then lowered to surround the animal. **E.** Lateral view of the full beamline.

**Figure 3**. **PET imaging in mouse**. **A**. FLUKA simulation showing the expected $^{11}$C-ion dose (in Gy) distribution in the μCT of the mouse in the sagittal view. Doses are normalized to the planned target dose. **B**. Corresponding Monte Carlo simulation of the PET activity. **C**. Online SIRMIO PET image of the positron activity distribution deposited during $^{11}$C-irradiation overlaid on the same pre-treatment μCT used for the simulations. All the panels show data referring to the same animal, the generalized CTV contour (Fig. 1) is highlighted with a black line, while the spine (OAR) contour is marked in red. All the images are integrated on the BEV aperture (±1 mm) in the x (axial) plane transversing the beam direction.

**Figure 4**. **Activity profiles in mice. A**. For the same mouse in Fig. 3, we show the z-axis depth profiles of the simulated dose (normalized to the target dose; blue), the simulated PET activity (dashed blue), and the measured (solid red) PET activity profiles, normalized to their maximum and laterally integrated on the BEV aperture (±1 mm) in the x-y plane orthogonal to the beam direction. The CTV and μCT spine regions are highlighted by pink and red bands, respectively, while the collar is depicted in yellow and the dimple

(see Extended Data Fig. 3B) in light blue. **B**. Comparison between simulated dose and PET activity profiles for a mouse analyzed at the CBCT in the SARRP in vertical position. Although the mouse is not the same as in panel **A**, we observed that switching to the vertical position consistently induces the same anatomical change in all animals. Therefore, we overlaid the measured PET activity profile of the mouse from Fig. 3 in red. All profiles are integrated as in panel **A**.

**Figure 5**. **Tumor growth. A.** Volumes calculated from 2D caliper measurements of the visible tumor (see Methods). Asterisks indicate statistical significance using the ANOVA test. **B.** Measurements of the volumes using μCT. Data are more precise than caliper measurements, but they are less frequent than external measurements. **C.** Zoom of the data points for irradiated groups shows the recurrence of the tumor irradiated with 5 Gy. Bars are standard errors of the mean values of the different animals.

**Figure 6**. **Radioactive washout.** Individual activity data recorded after the end of irradiation in Supplementary Fig. 4 are grouped in the upper panel (left: 5 Gy, middle: 20 Gy, right: comparison of fit functions assuming the physics decay of the beam containing 96% $^{11}$C, 3% $^{10}$C, and 0.5% $^{15}$O ions). As a fit function, a double-exponential decay function was chosen over a single-exponential decay following the results of the F-test (ratio of the fit $\chi^2$ with 1 or 2 parameters) with number of degrees of freedom df > 100. F(5 Gy) = 1126 ($\gg$1), F(20 Gy)=63 ($\gg$1). Double-exponential decay rates and the weight of the slow component are shown in the bottom panel for 5 Gy and 20 Gy. Significance of the differences is assessed by t-test, and *p*-values are shown.

**Extended data**

**Extended data Fig. 1. Fragment separator at GSI.** A schematic view of the FRS is shown. A primary beam of $^{12}$C-ions from the SIS-18 synchrotron was incident on a beryllium target to produce $^{11}$C-ions, separated using the Bρ-ΔE-Bρ method. An achromatic degrader was placed at the mid-focal plane of the FRS, . The FRS has three experimental branches for delivering the separated radioactive ion beam. The branch directed to Cave-M is indicated in the figure. The elements between the last dipole of the FRS and Cave-M belong to the high-energy beam transport line of GSI, designed for primary beams and therefore having smaller apertures than the FRS magnetic elements, leading to reduced transmission efficiency of secondary beams to the Cave-M.

**Extended data Fig. 2. Dosimetry of the $^{11}$C beam**. **A**. PEAKFINDER$^{TM}$ measurements of the laterally integrated monoenergetic (pristine) beam depth dose profile in water. **C**. Water phantom measurement of the beam spot 2D dose distribution at the minimal water equivalent depth inside the phantom. **B** and **D.** Water phantom measurements of the horizontal and vertical 2D dose distributions in the central plane along the beam direction respectively. Figures are normalized to their maximum values. **D.** PEAKFINDER$^{TM}$ measurements of the laterally integrated SOBP depth dose profile in water. **G.** Water phantom measurement of the beam spot 2D SOBP dose distribution at the minimal water equivalent depth inside the phantom. **E.** and **F.** Water phantom measurement of the horizontal and vertical SOBP 2D dose distributions in the central plane along the beam direction respectively. Figures are normalized to their maximum values.

**Extended Data Fig. 3. Cave M beamline components.** The scheme of the beamline is shown in Fig. 2A. **A**. Range modulator used to generate the SOBP from a monoenergetic pencil beam scan. **B**. Plastic compensator to shape the lateral and distal edges of the irradiation field to the CTV. The length is 25 mm to shield the base of the skull and the rest of the animal spine. Top: inner side with an anatomical cut for fixing the animal neck area Bottom: outer side with a dimple corresponding to the distal edge of the CTV calculated in water-equivalent thickness. For a precise description of the collar material see the section "Dosimetry".

**Extended data Fig. 4. RIB imaging in phantoms. A.** FLUKA simulation showing the expected $^{11}$C-ion dose (in Gy) distribution in the µCT of the plastic phantom in the axial view. Doses are normalized to one in the target region. **B**. Corresponding Monte Carlo simulation of the PET activity. **C**. Online SIRMIO PET image of the activity distribution deposited during $^{11}$C-irradiation overlaid on the same pre-treatment µCT used for the simulations. All the 2D images are integrated over the BEV aperture (±1 mm) in the plane transversing the beam direction. **D.-E.-F**. The same as before, but in the sagittal view, with all the 2D images integrated on the same BEV aperture. **G**. Simulated dose normalized to the target dose (blue), simulated PET activity (dashed blue), and measured (solid red) PET activity depth profiles, normalized to their maximum, along the z-axis direction, integrated over the BEV aperture (±1 mm) in the x-y plane transversing the beam direction.

**Extended data Fig. 5. Mouse CT.** Comparison of a scans with the μCT (left) and the SARRP CBCT (middle) for the same mouse. The two scans are superimposed in the image on the left to highlight the small differences in the spine position.

**Extended data Fig. 6. Simulation of PET imaging of a RIB damaging the spinal cord. A.** FLUKA simulation showing a hypothetical case of a $^{11}$C-ion SOBP dose (in Gy) distribution damaging the spinal cord for a vertical CBCT SARRP scan of a mouse in the axial view. Doses are normalized to the planned target dose. **B**. Corresponding Monte Carlo simulation of the PET activity as should have been seen in the SIRMIO PET scanner, overlaid on the CBCT image. **C**. The activity simulation shown in **B** is now overlaid on the μCT scan of the same mouse to illustrate what we would have seen online in case of RIB range exceeding the target distal position. The generalized CTV contour (Fig. 1) is highlighted with a black line, while the OAR contour is marked in red. All the images are integrated over the BEV aperture in the y-plane transversing the beam direction. **D-E-F** The same as before, but in the sagittal view, with all the 2D images integrated on the BEV aperture in the x-plane transversing the beam direction. **G.** The corresponding simulated dose normalized to the target dose (green) and the simulated PET activity (dashed green) depth profiles, normalized to their maximum, along the z-axis direction, integrated over the BEV aperture in the x-y plane transversing the beam direction. The plot shows that an online image as depicted in panels **C** and **F,** never actually observed in our experiments, would have required the insertion of a range shifter to reduce the range during the irradiation.

**Extended data Fig. 7. Treatment of the mouse osteosarcoma with 20 Gy X-rays. A**. Tumor growth curve, showing full tumor control after 20 Gy (5 mice, bars are standard errors of the mean values). **B.** Image of a mouse 15 weeks after exposure to 20 Gy X-rays suffering severe weight loss and kyphosis (see also Supplementary video 2). **C.** Pooled weights of tumor-bearing animals after receiving X-rays (orange circles, 15 and 20 Gy) or $^{11}$C treatment (blue squares; 20 Gy). Lines represent the linear fit of the data from each group. The slope coefficients are 0.215±0.046 and 0.525±0.019 for the X-ray and $^{11}$C groups, respectively and they are significantly different (t-test, $p<0.0001$). According to the ethical protocol, animals losing >20% of the weight have to be sacrificed, and for this reason some data acquisition is stopped early. The weight loss and myelopathy in these mice is due to the high dose to spine and esophagus of the X-ray beam that goes through the tumor.

**Extended data Fig. 8. Grip strength test used to estimate the cervical myelopathy**. **A**. In this quantitative test the animal is pulled by the tail and the strength of the grip on the bar is measured in newton (N). **B**. Control mouse during the test (lateral view). **C**. 20 Gy irradiated mouse during the test. The area behind the head is shaved prior to the tumor inoculation, and after the irradiation only some white fur grows back.

**Extended data Figure 9**. **Toxicity vs. activity in the spinal cord**. Animals exposed to $^{11}$C-ions shows a lower forelimb strength in the grip test (Extended data Fig. 8) compared to controls. Individual data are shown in Supplementary Fig. 3. Data for irradiated animals were only considered from week 6, assuming no radiation effect in the first month post-irradiation. **A.** Median grip strength values in the control (8 animals) and irradiated (13 animals) groups. Bars are standard errors of the median values ($= 1.2533 \cdot \sigma/\sqrt{n}$). Median strength in the control group is significantly higher than in the irradiated group (Mood's median test, $p$=0.0152). **B.** Fraction of time points (Supplementary Fig. 3) where the measured peak force F was lower than 100 N. Bars are standard errors of the mean. The fraction is significantly higher in irradiated animals (Mann-Whitney test, $p$=0.00024). **C.** Correlation between median grip strength values in single irradiated animals and total PET counts in the spinal cord. **D.** Correlation between fraction of tests with F<100 N in individual irradiated animals and total PET counts in the spinal cord. Each point represents a single animal. The counts in the spinal cord are normalized to the total number of counts observed in the whole image reconstructed with the OSEM method. The grey area is the 95% CI around the regression line, *r* is the correlation coefficient and significance of the correlation was evaluated by Pearson's test both in **C** and **D** panels.

**Extended Data Table 1.** Summary of the beam parameters used in the experiments.

| | |
|---|---|
| Spill length | 200 ms |
| Duty cycle | 6.25% |
| $^{12}C$ beam energy | 300 MeV/u |
| $^{11}C$ beam energy | 209 MeV/u |
| $^{12}C$ intensity in SIS18 | $1.6 \cdot 10^{10}$ particles/spill |
| $^{11}C$ intensity in Cave M | $2.5 \cdot 10^{6}$ particles/spill |
| Beamspot shape | Horizontal direction: 2.335 cm FWHM<br>Vertical direction: 1.415 cm FWHM |
| Momentum spread | 60% Gaussian – FWHM=0.011 GeV/c·u<br>40% flat distribution: FWHM=0.018 GeV/c·u |

Fig.1 A + B

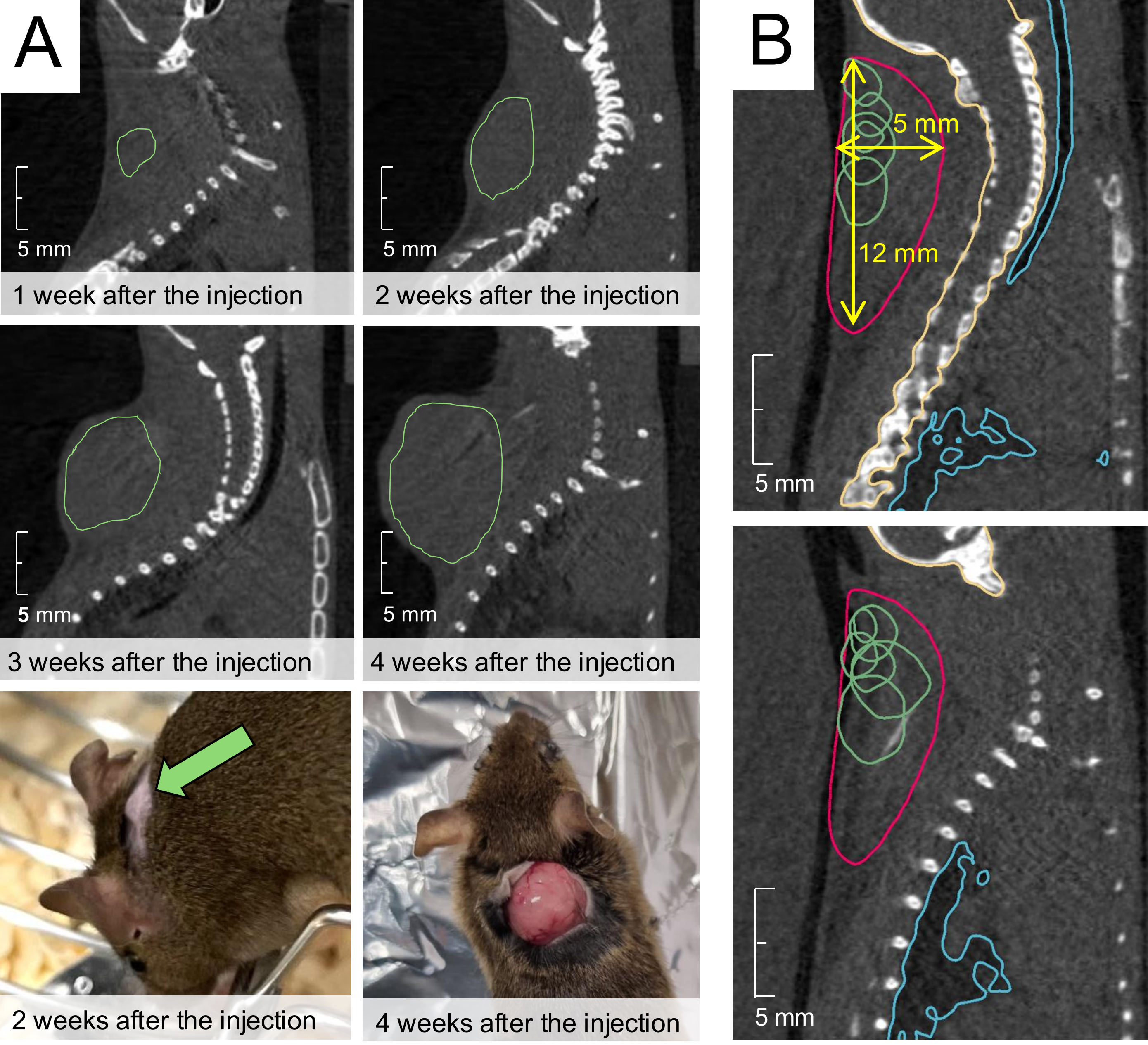

**Fig. 1C**

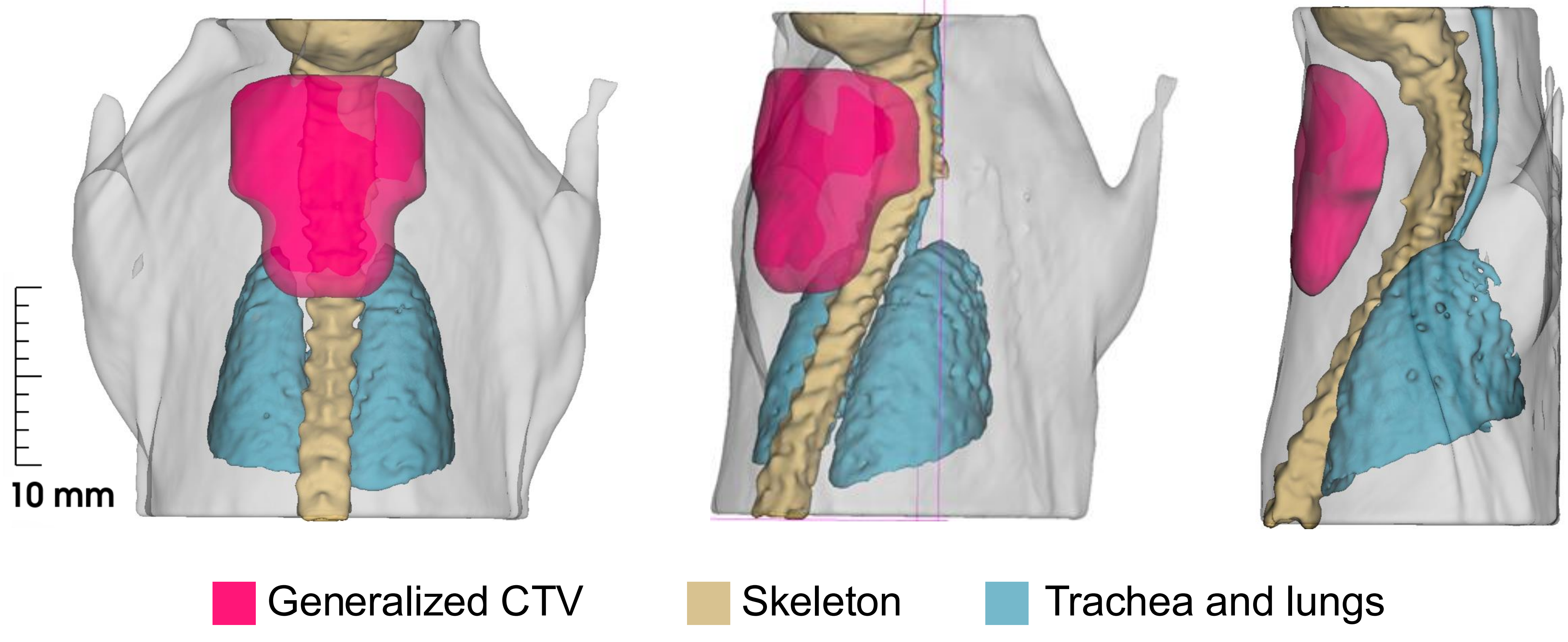

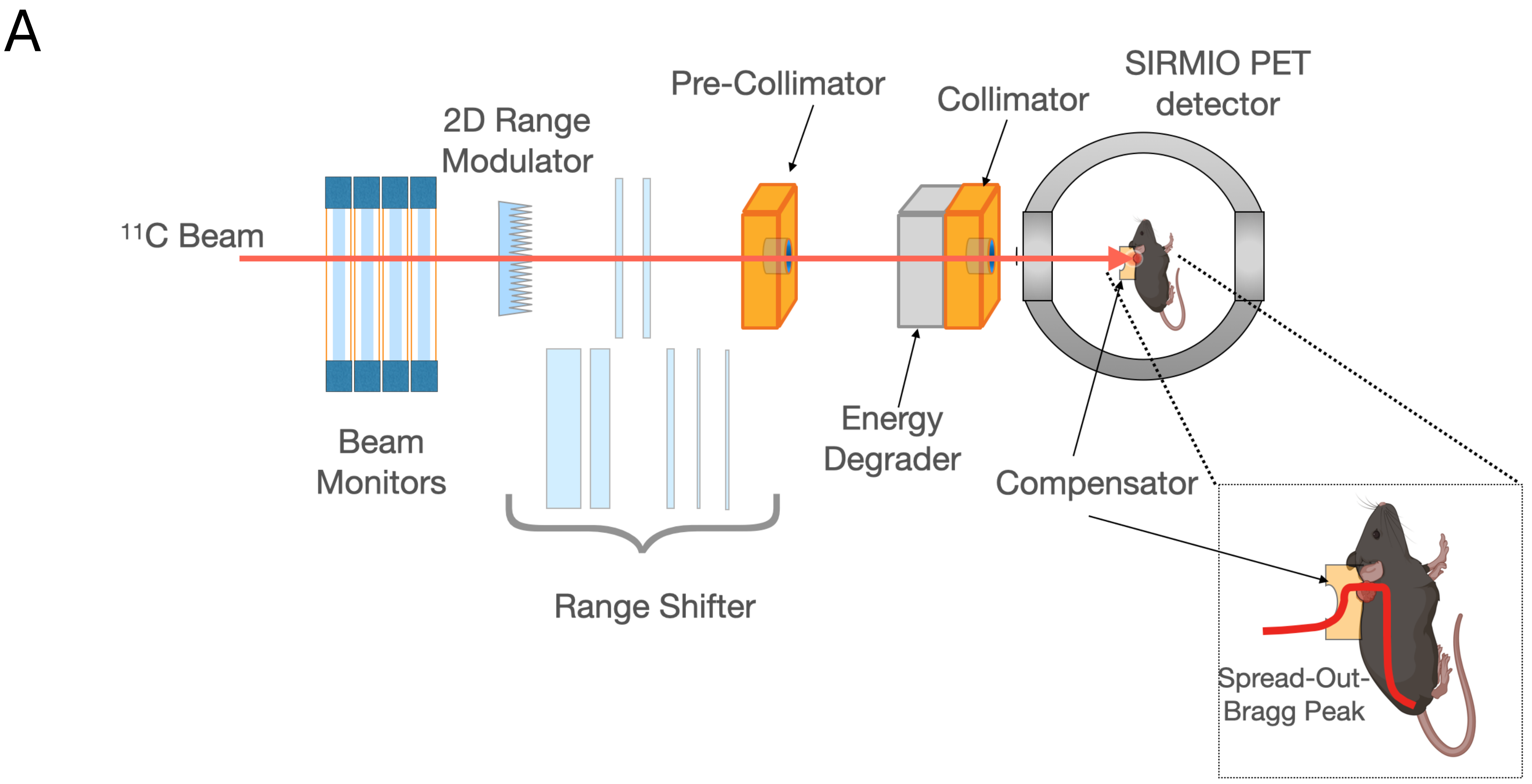

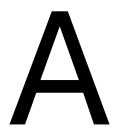

Figure 2A

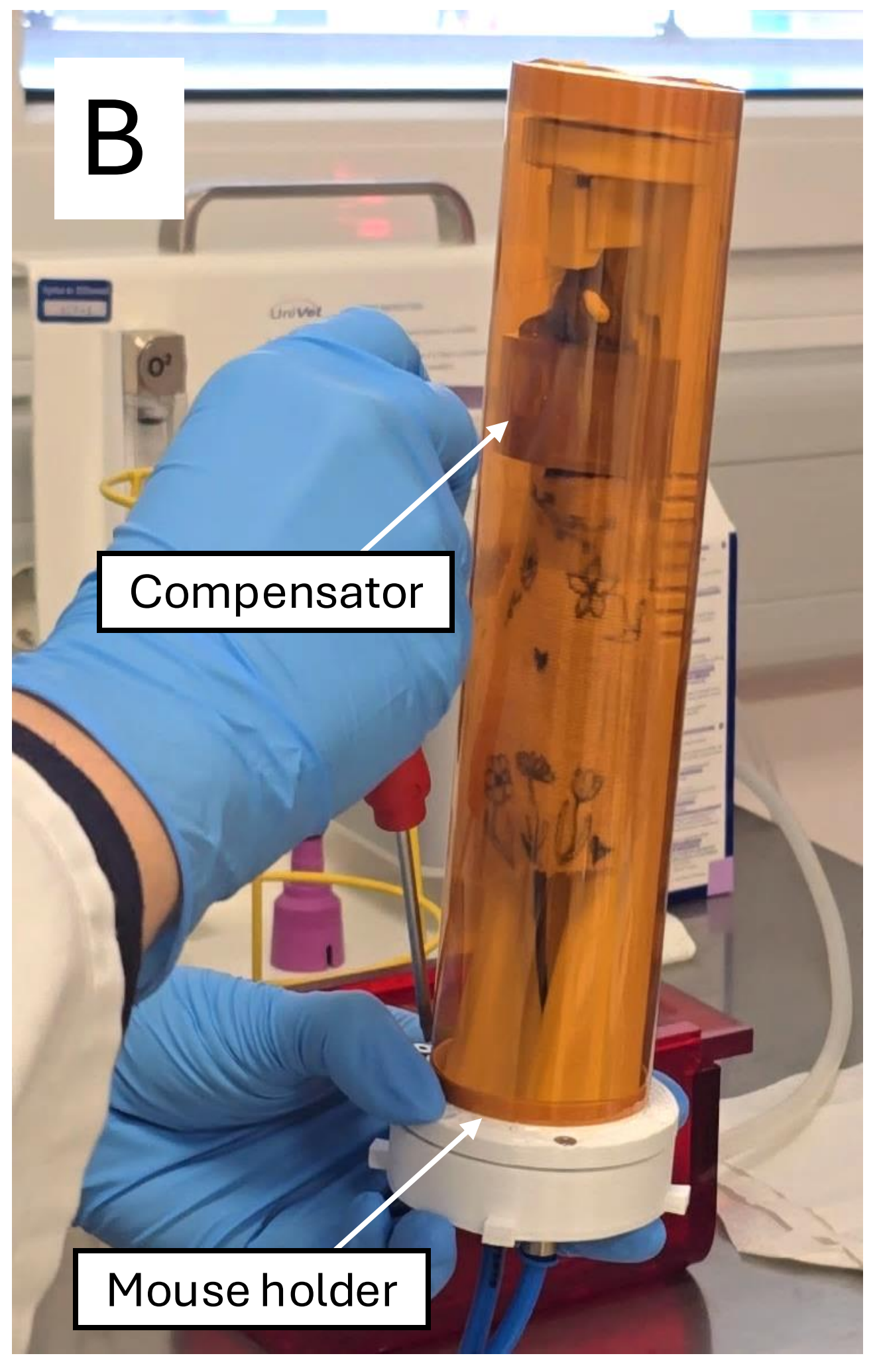

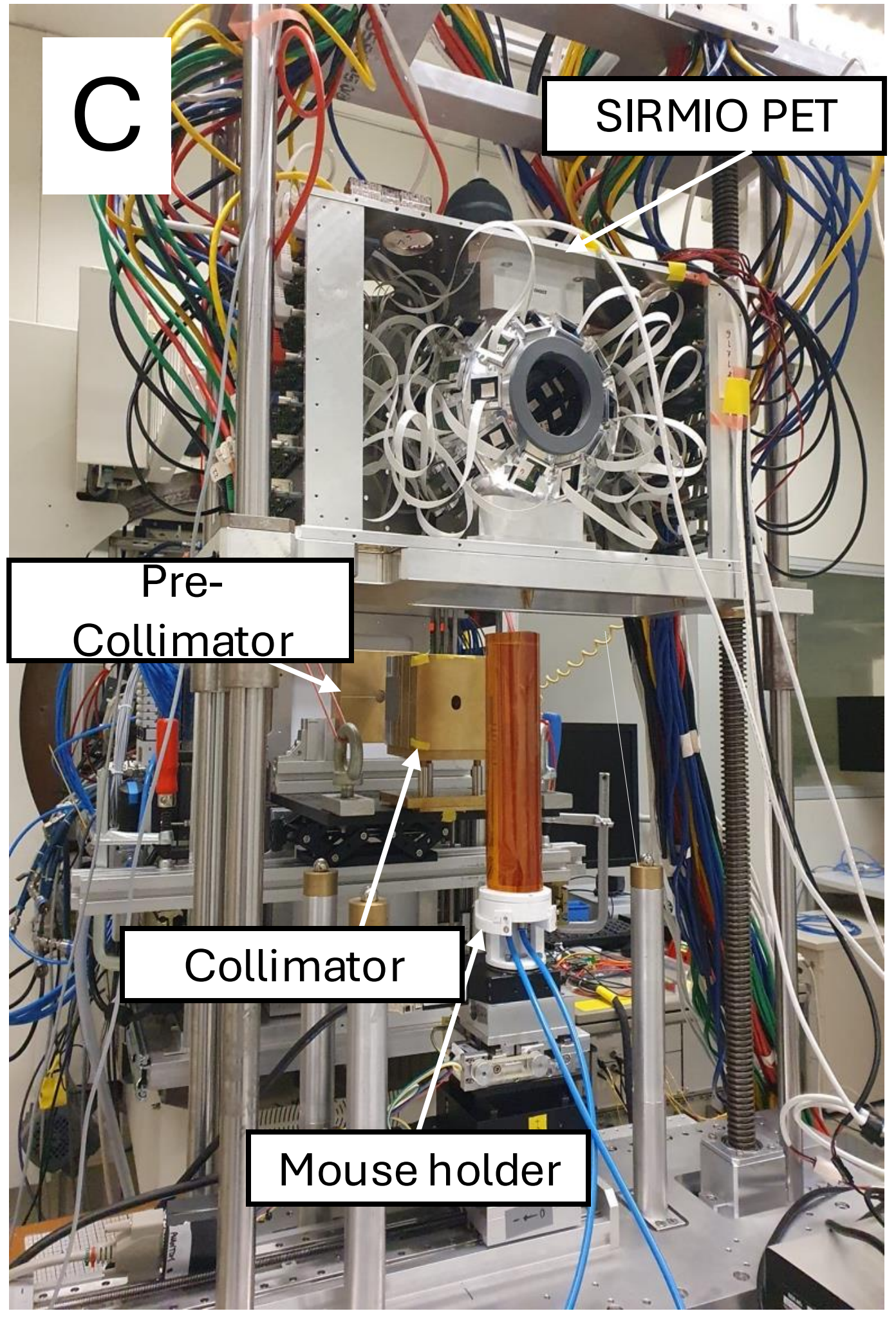

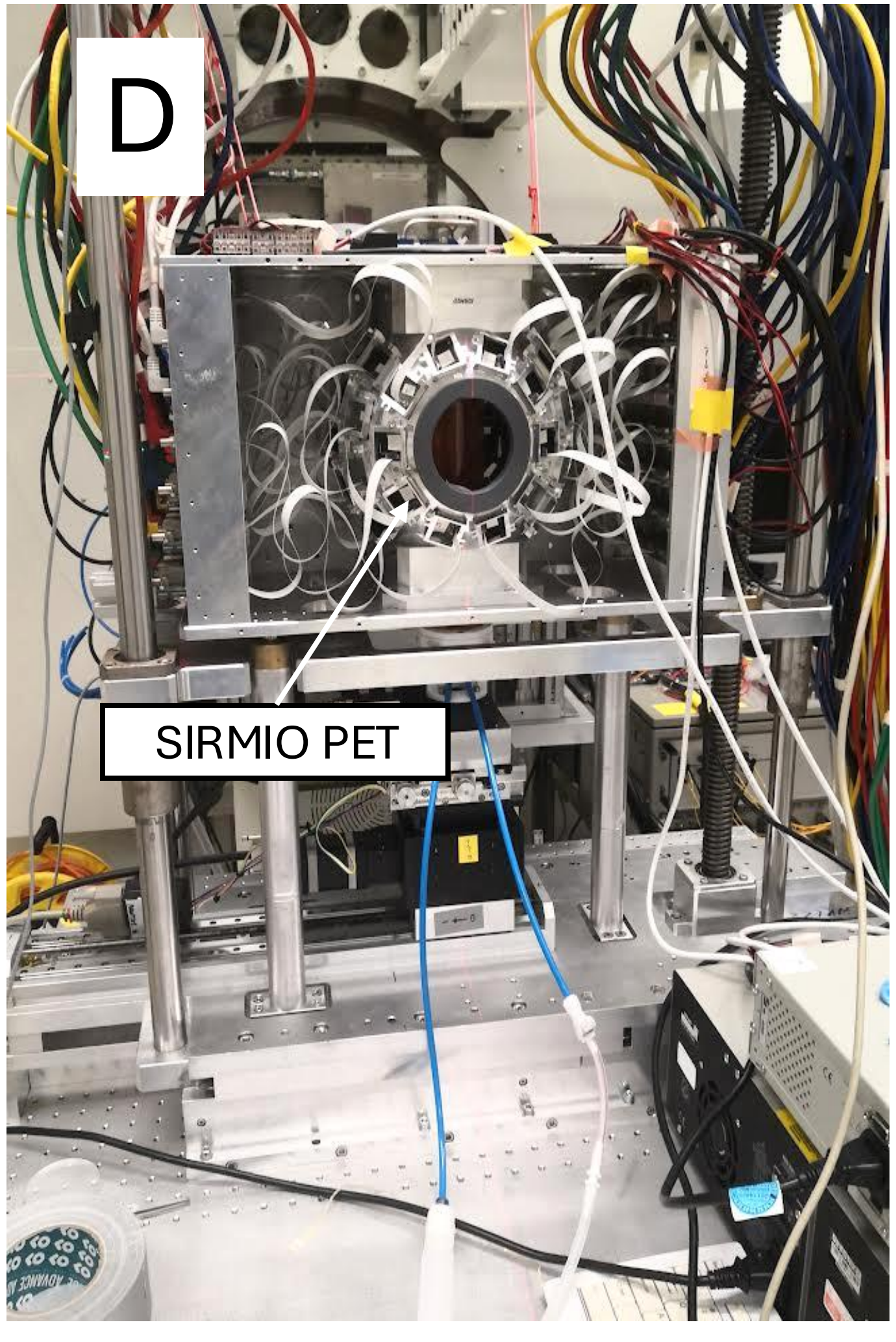

Figure 2B-D

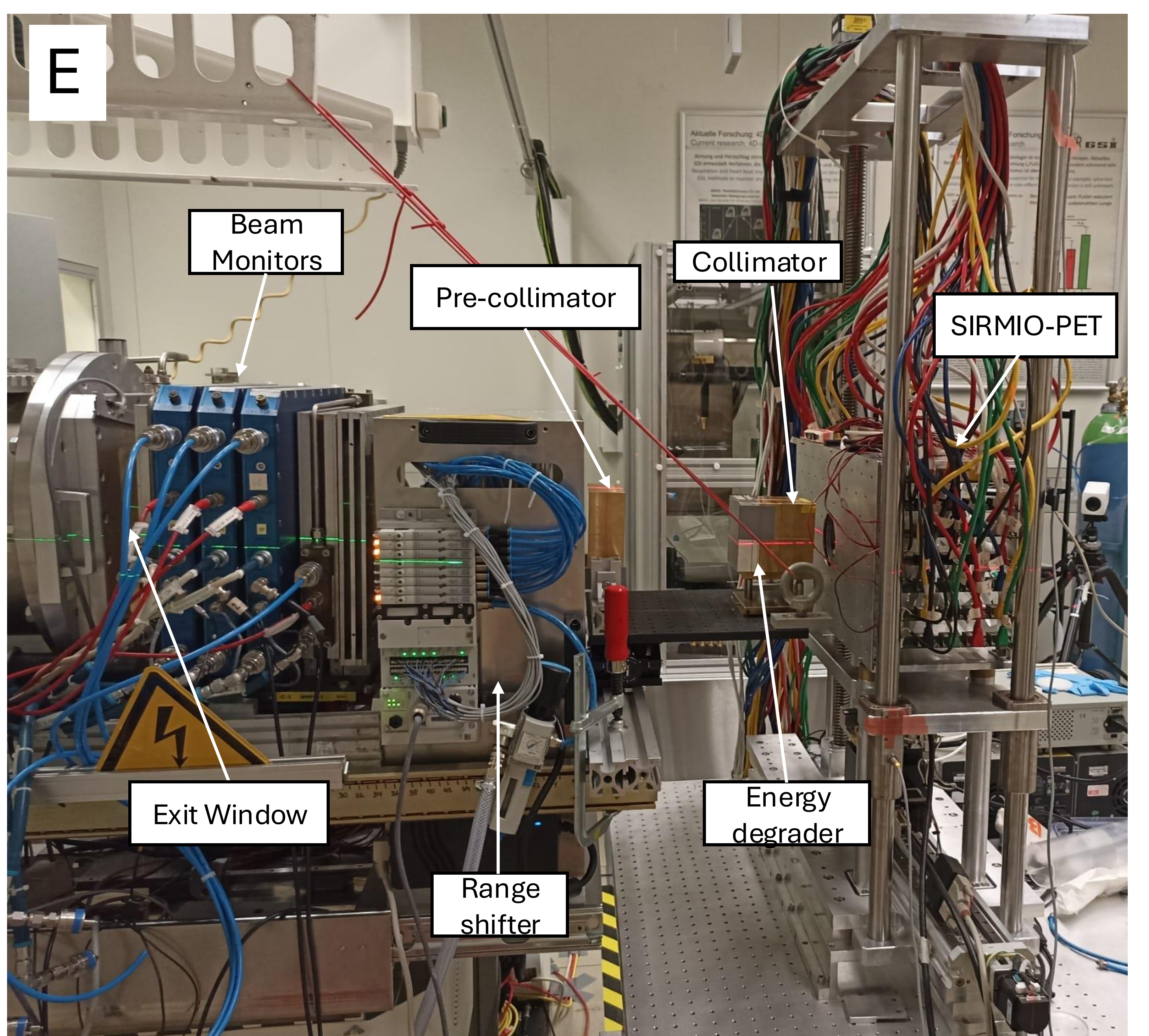

Figure 2E

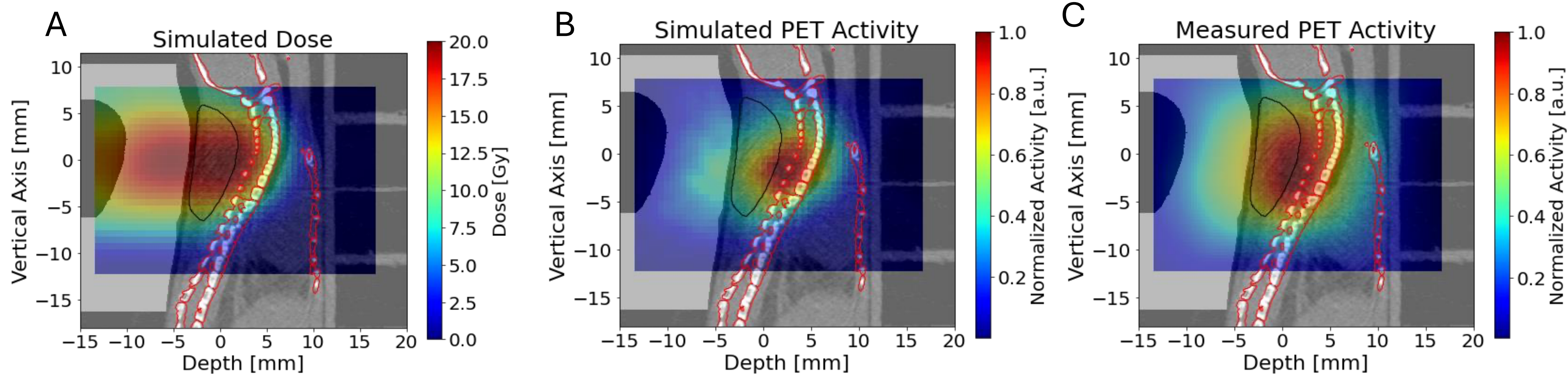

Figure 3

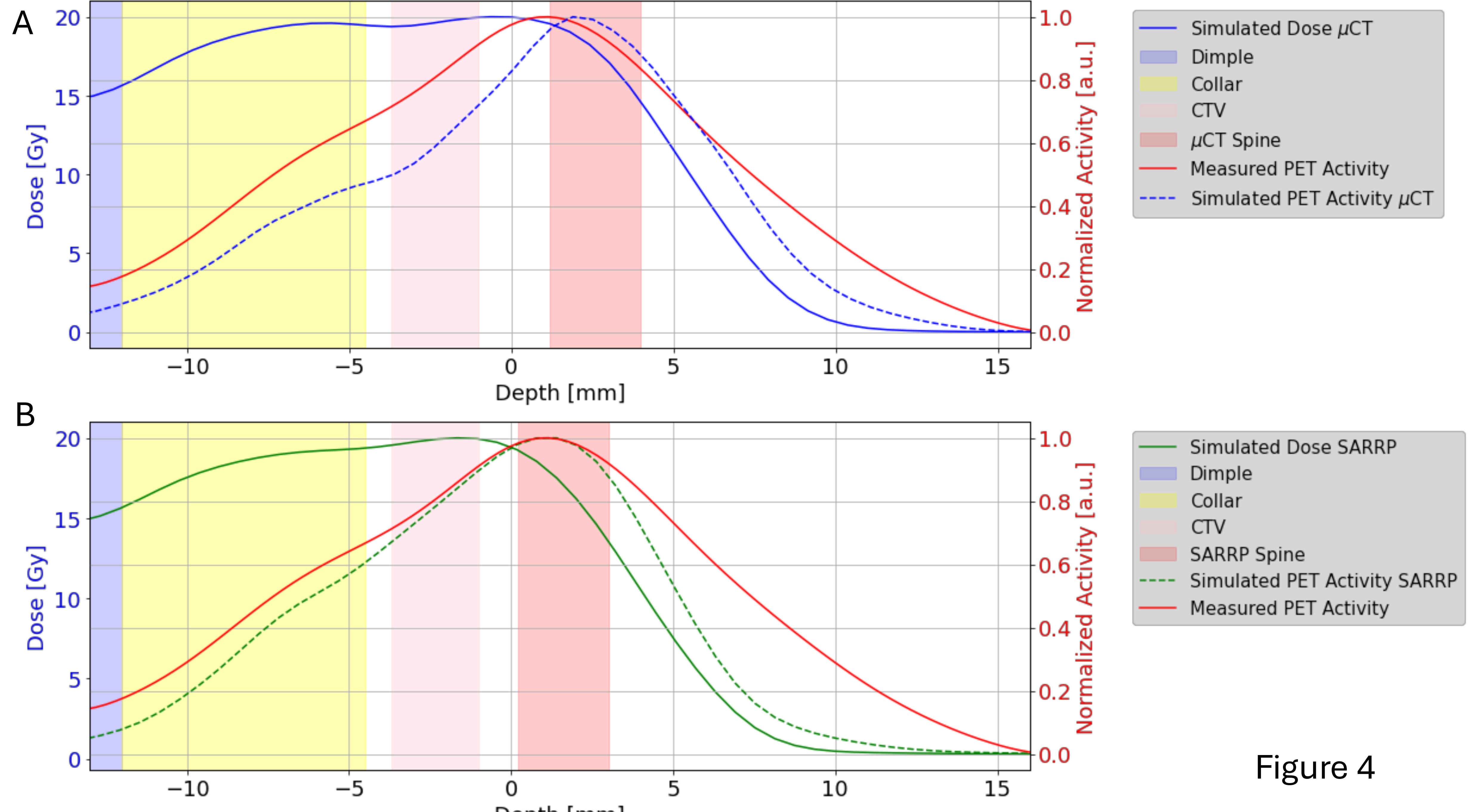

Figure 4

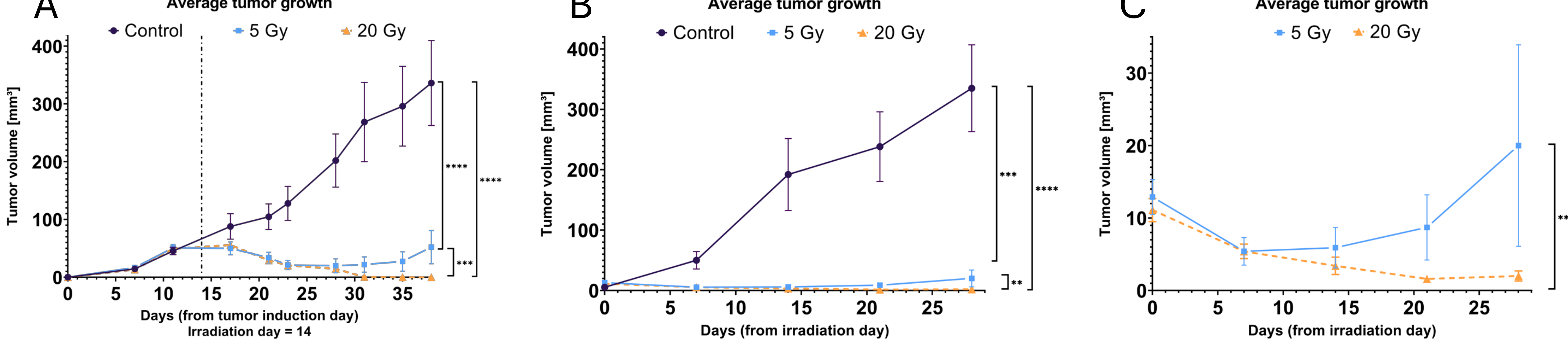

**Figure 5**

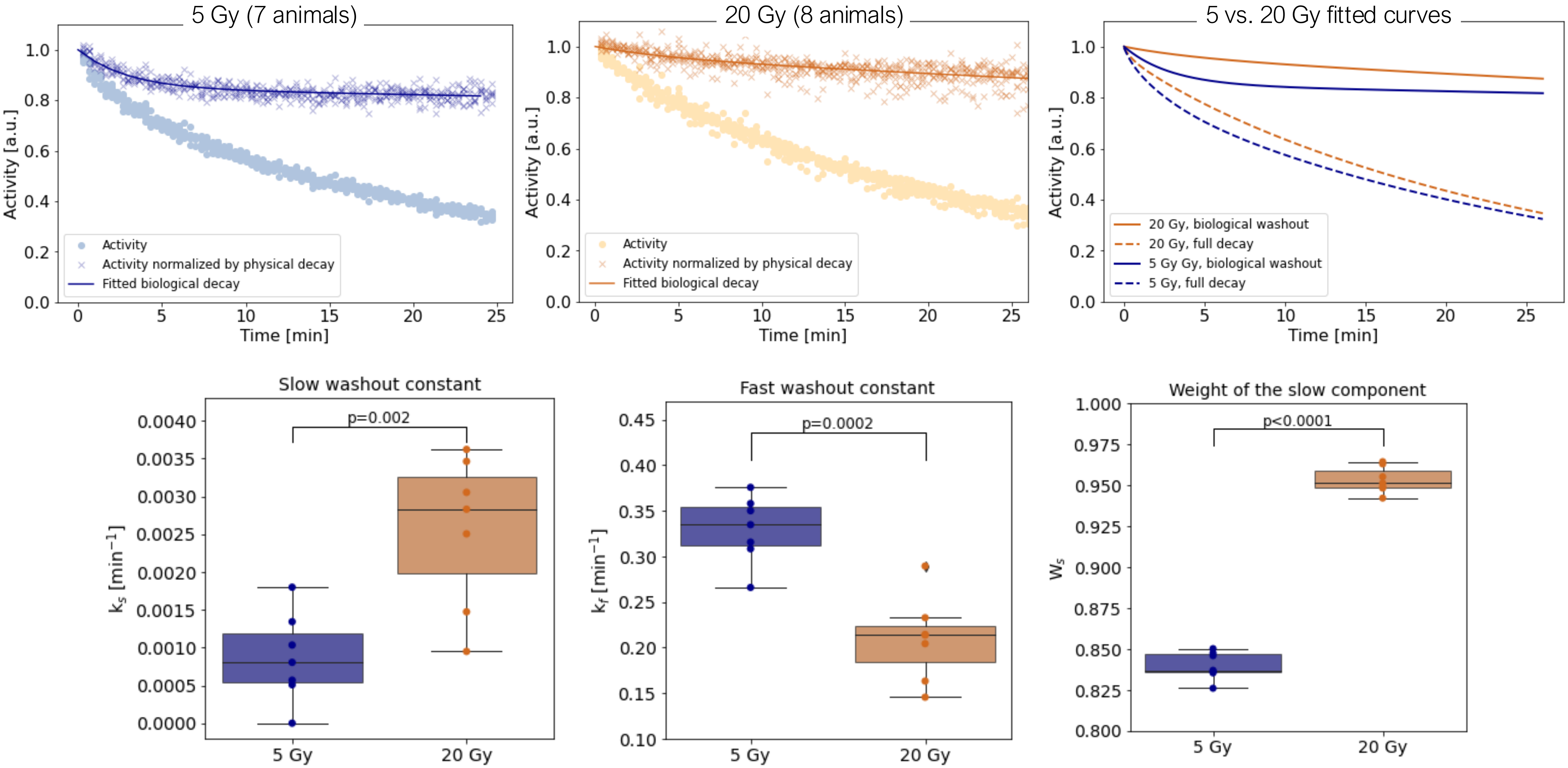

Figure 6

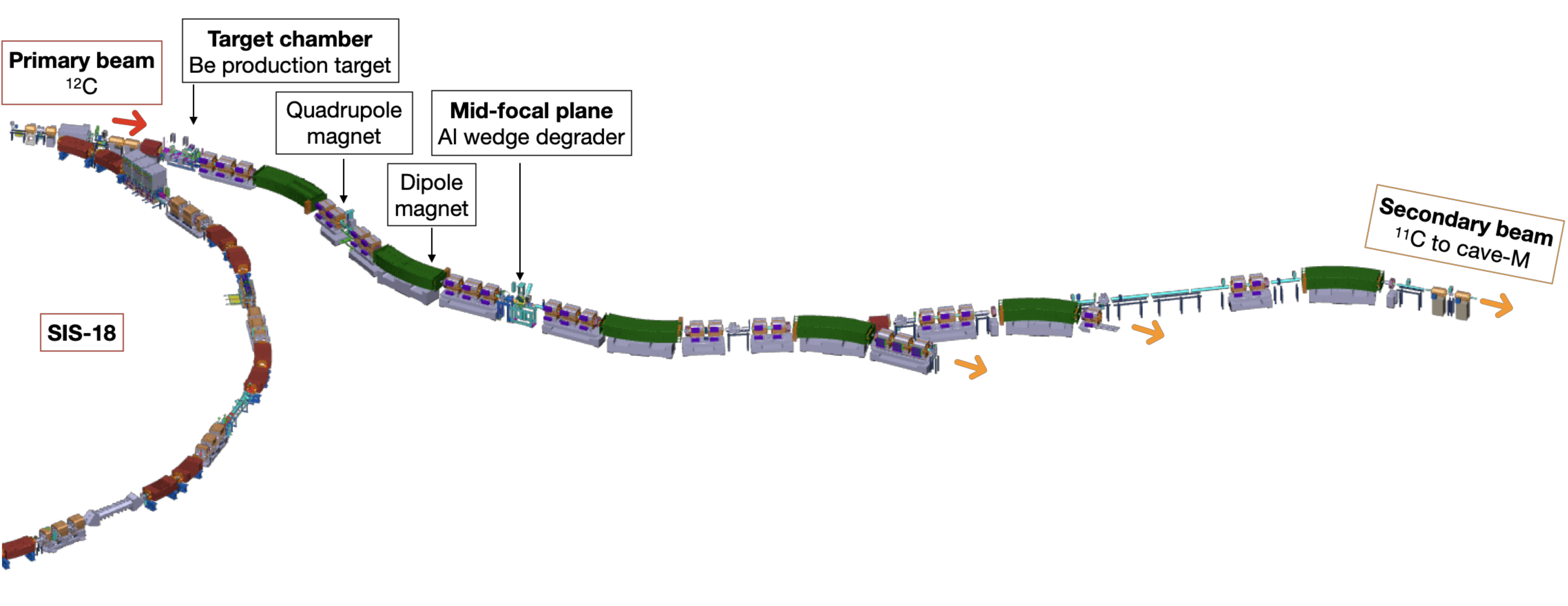

Extended Data Fig. 1

**Extended Data Fig. 2**

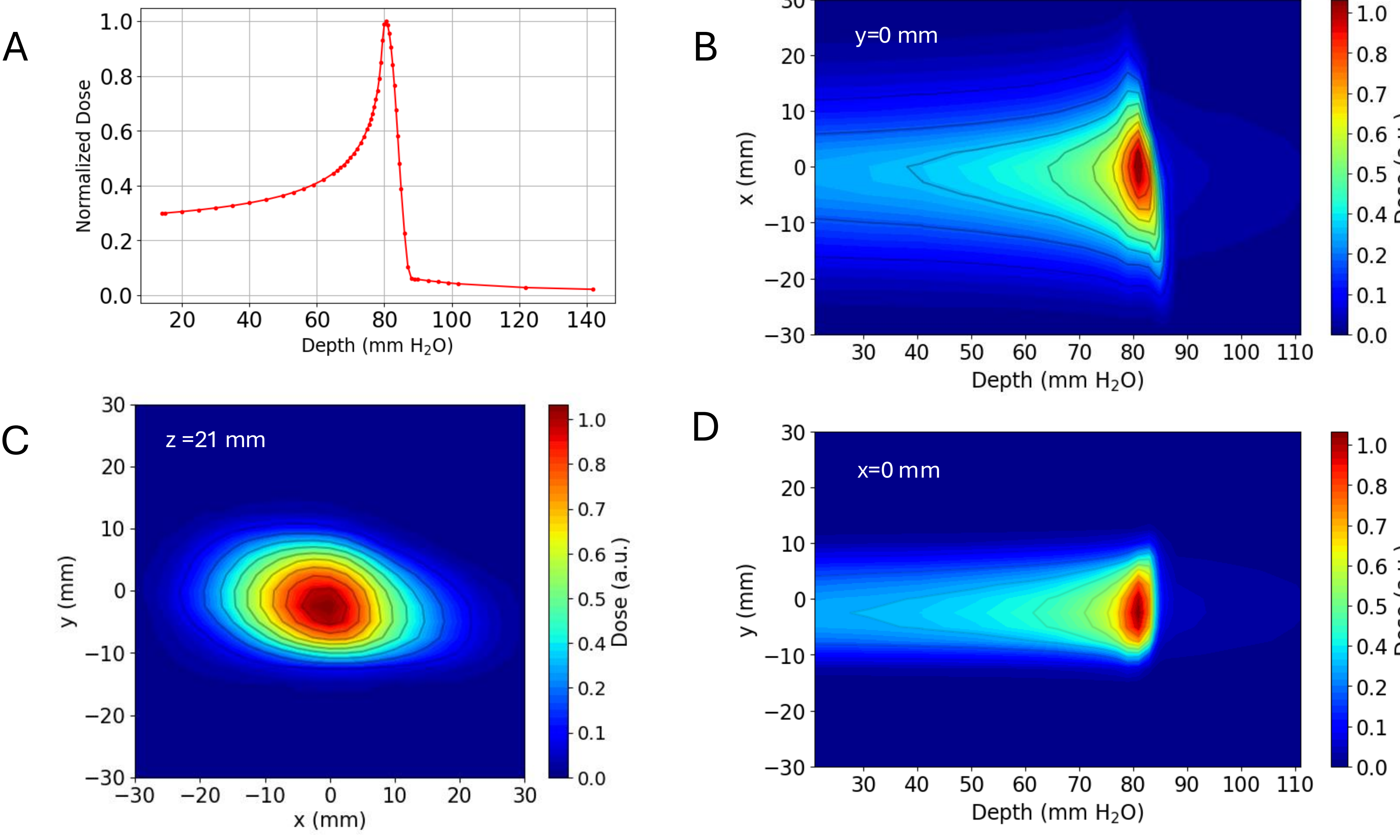

## Extended Data Fig. 2

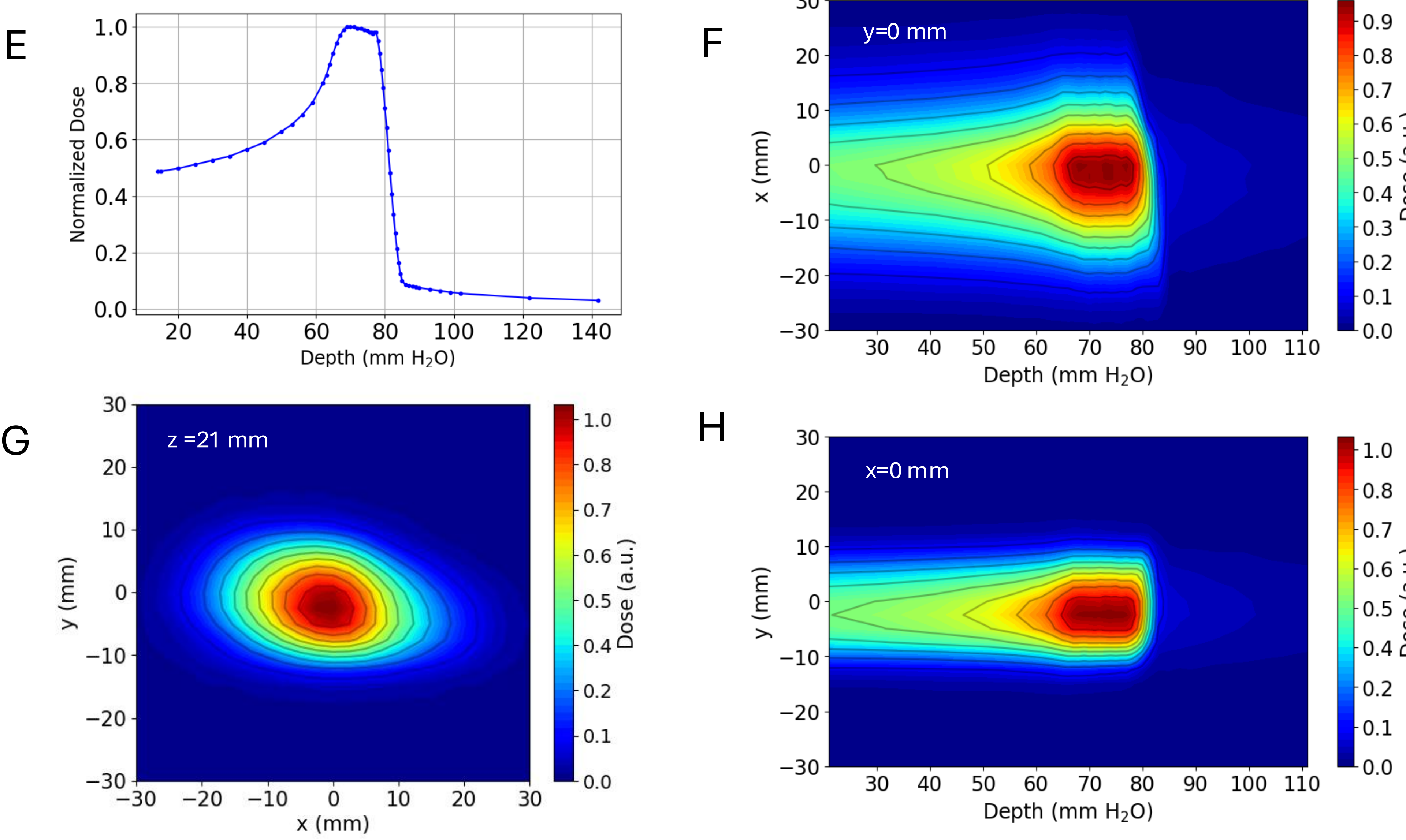

## Extended Data Fig. 3

A

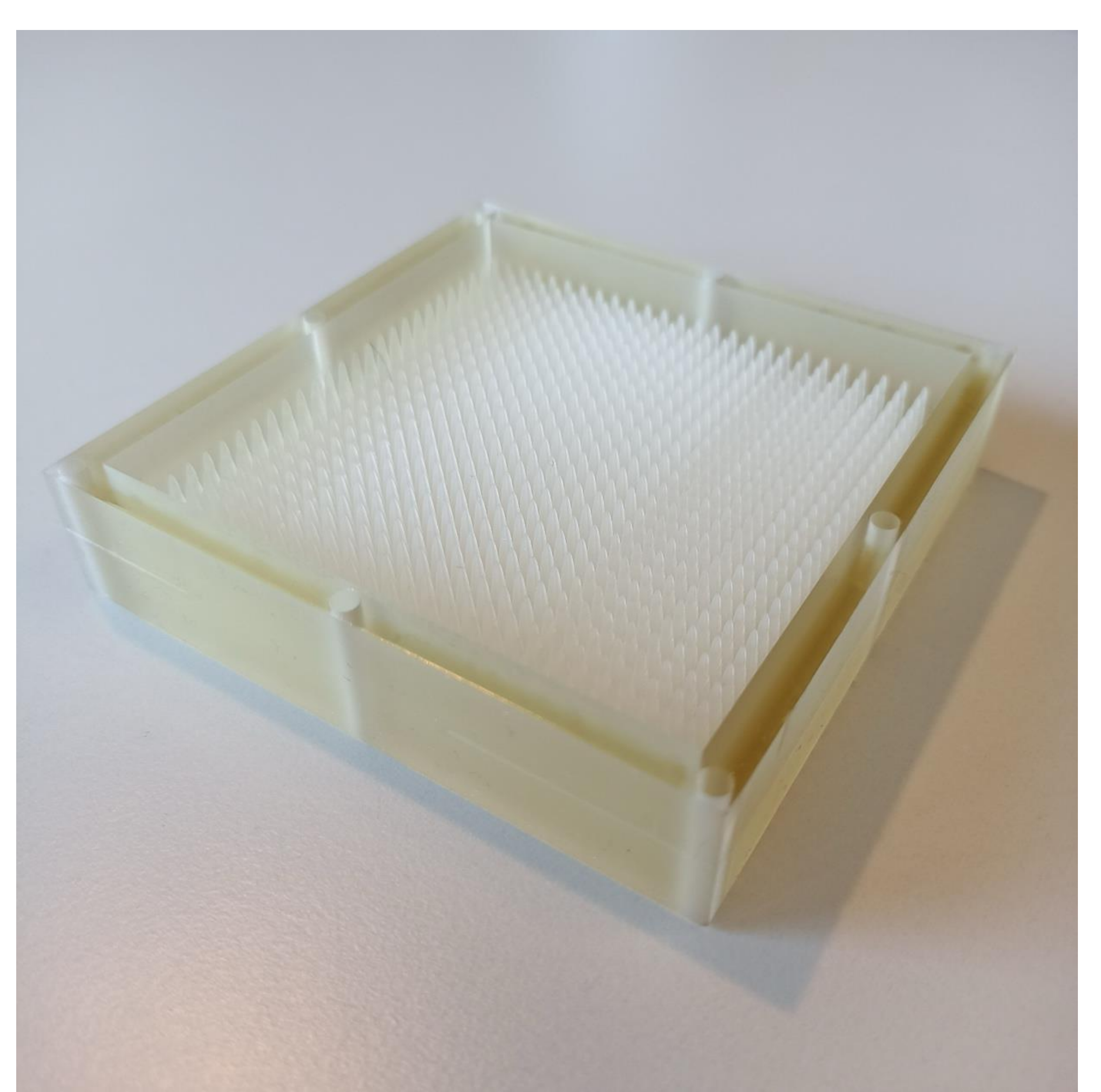

B

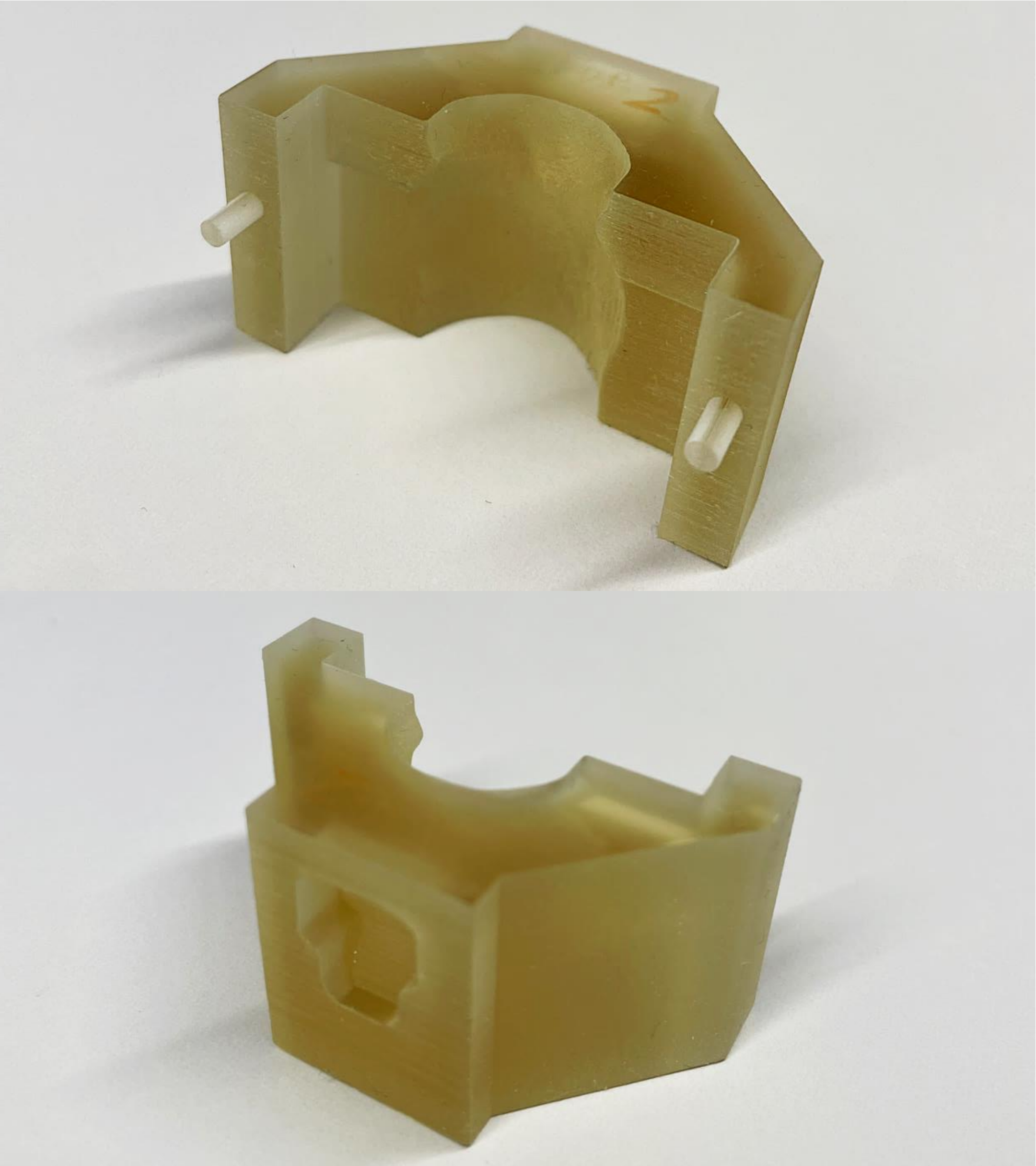

Extended Data Fig. 4

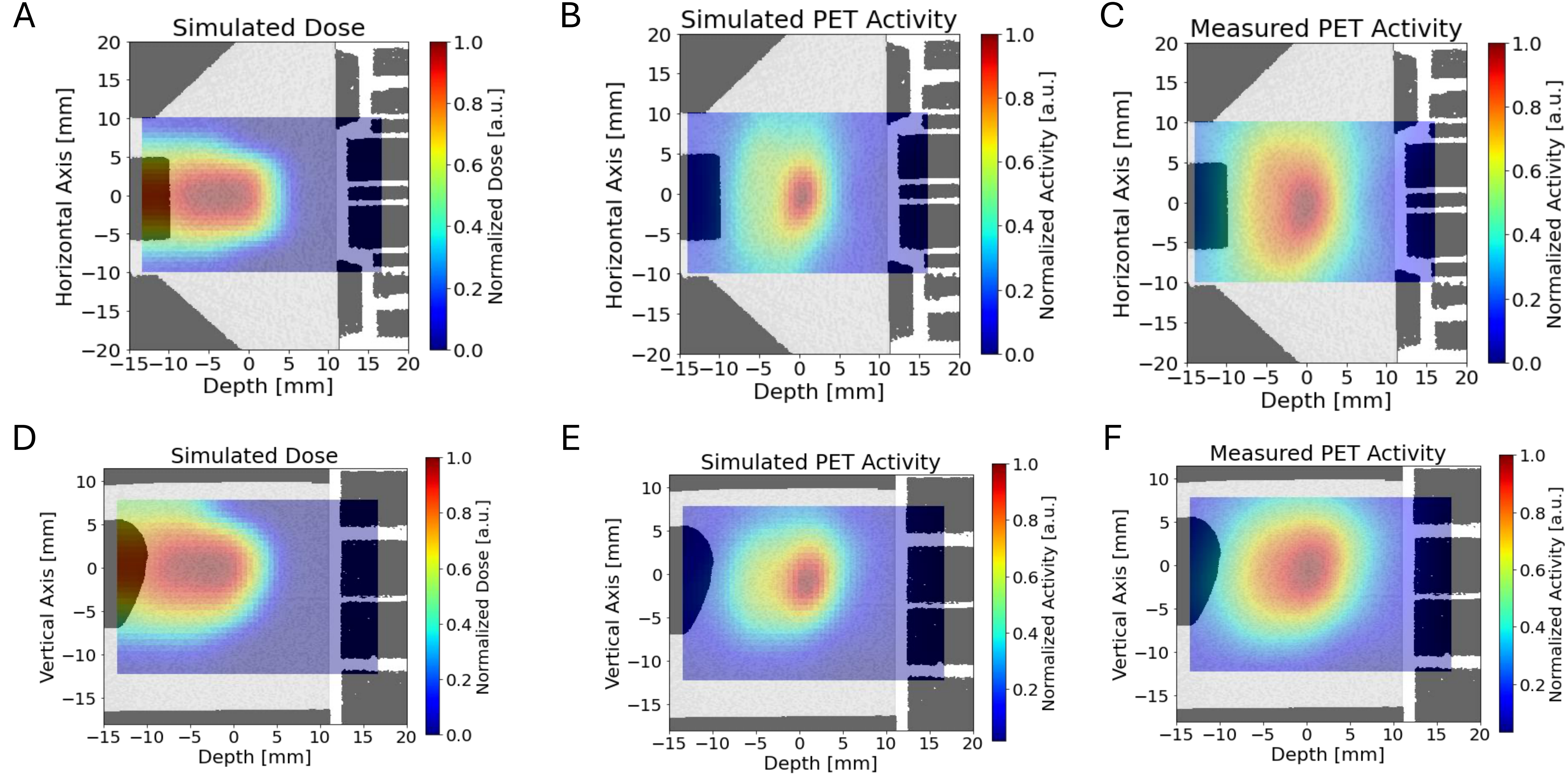

Extended Data Fig. 4 G

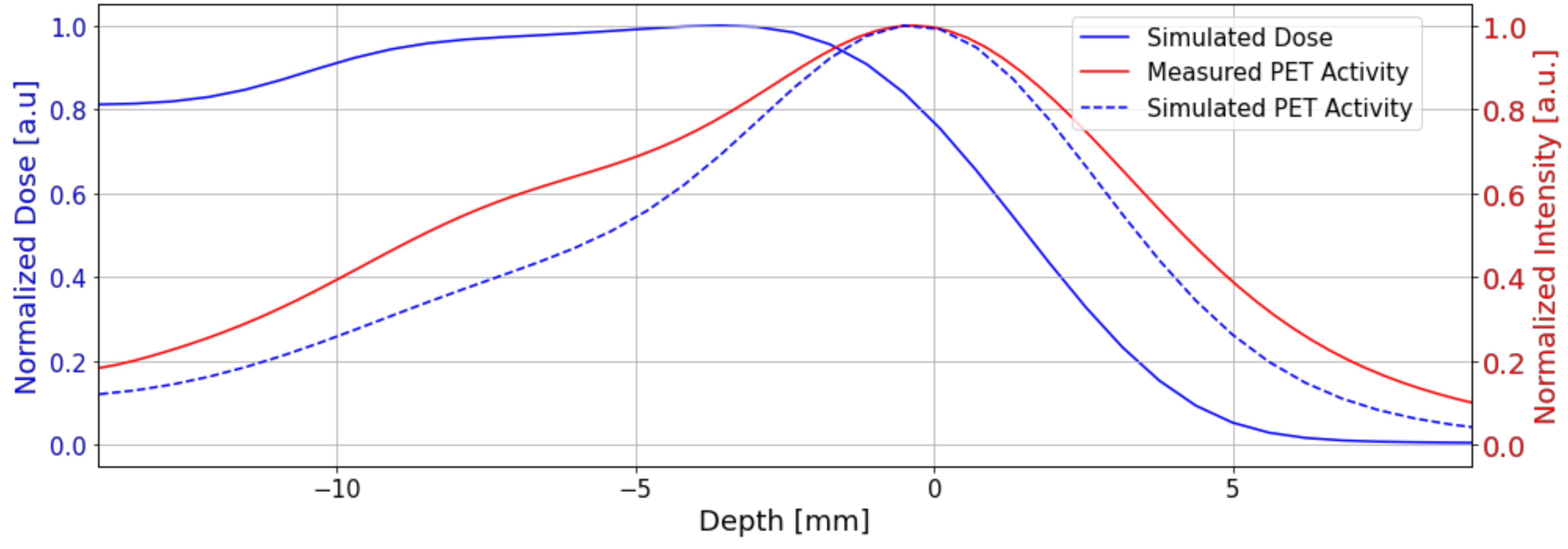

**Extended Data Fig. 5**

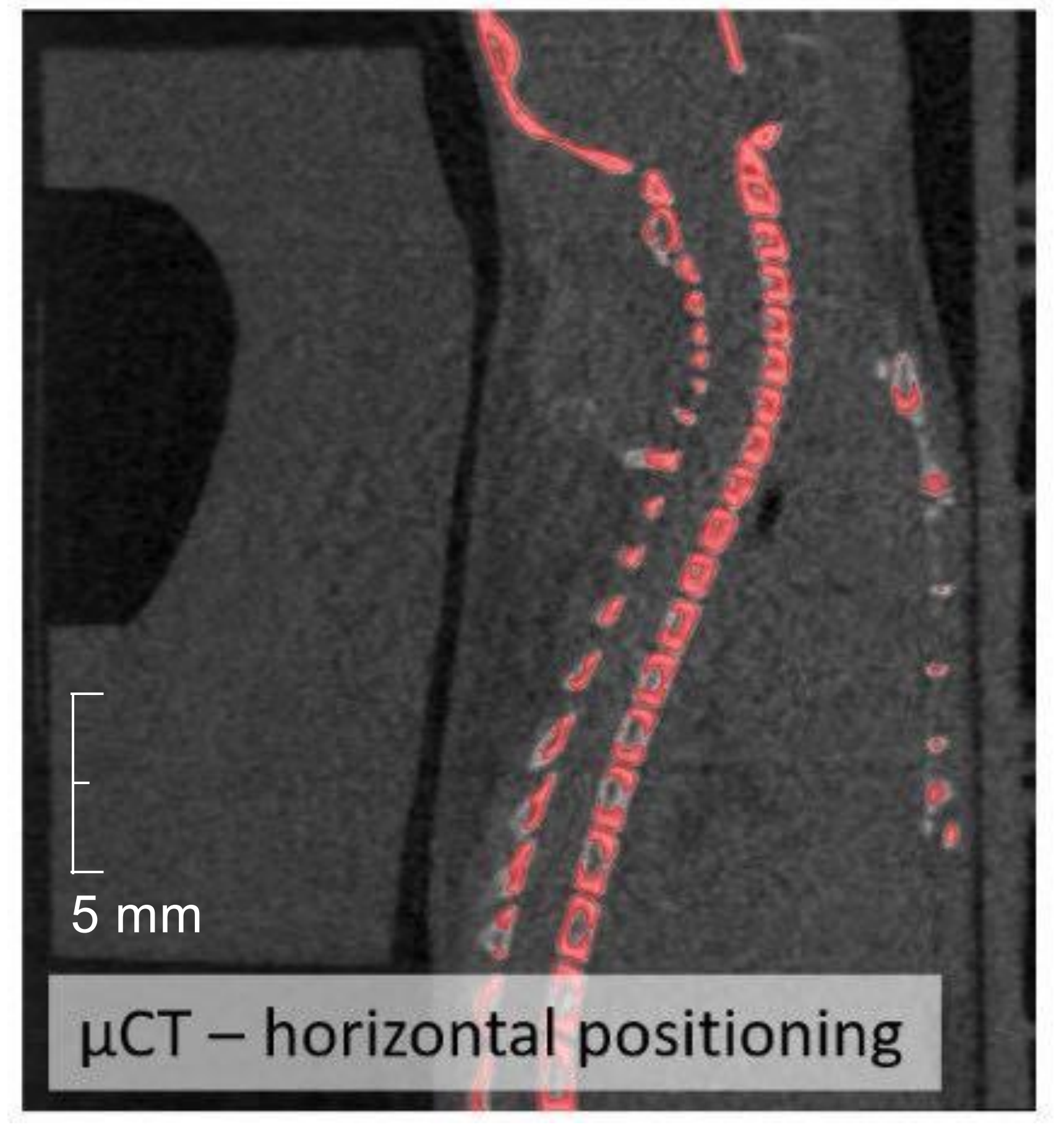

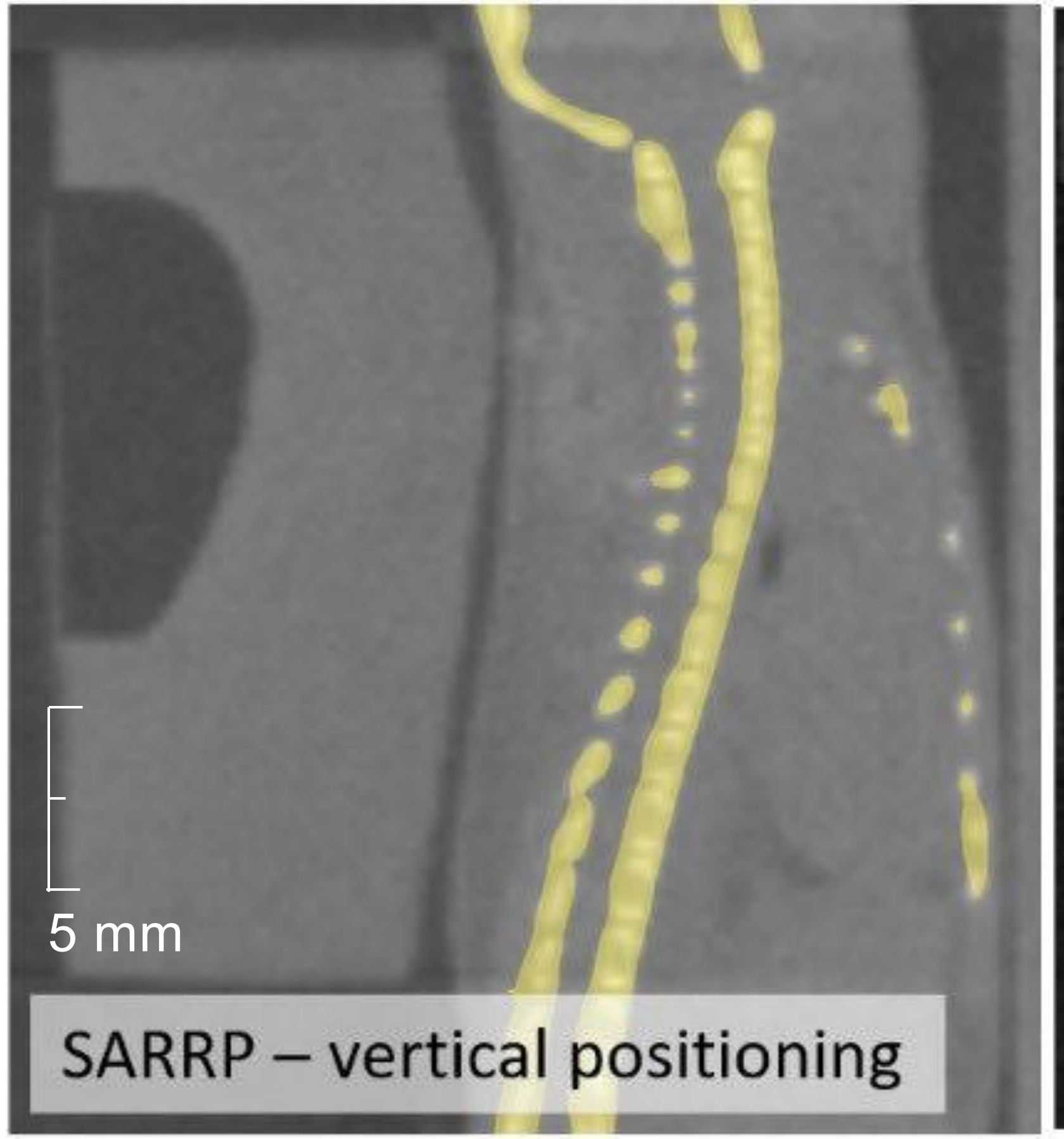

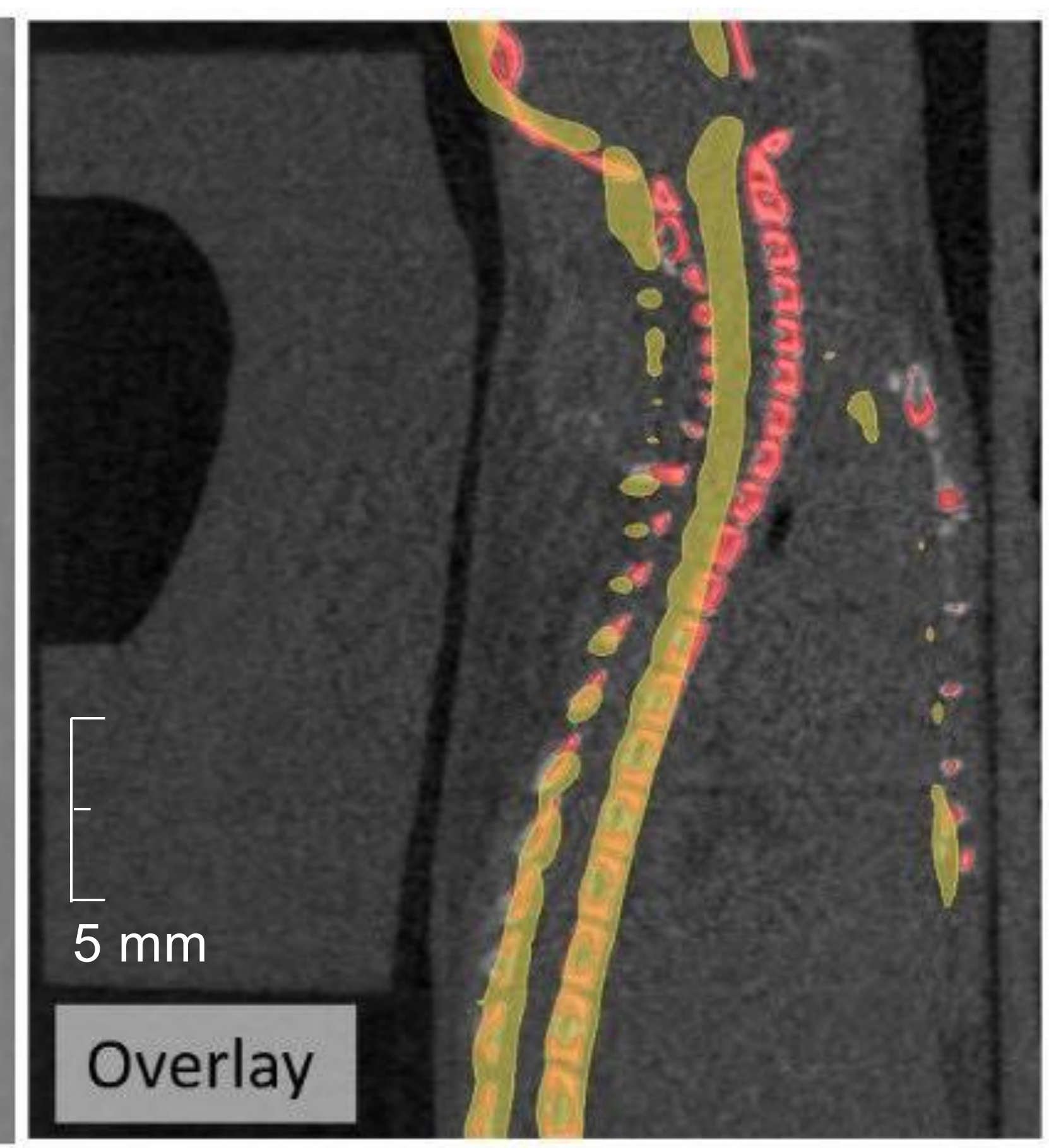

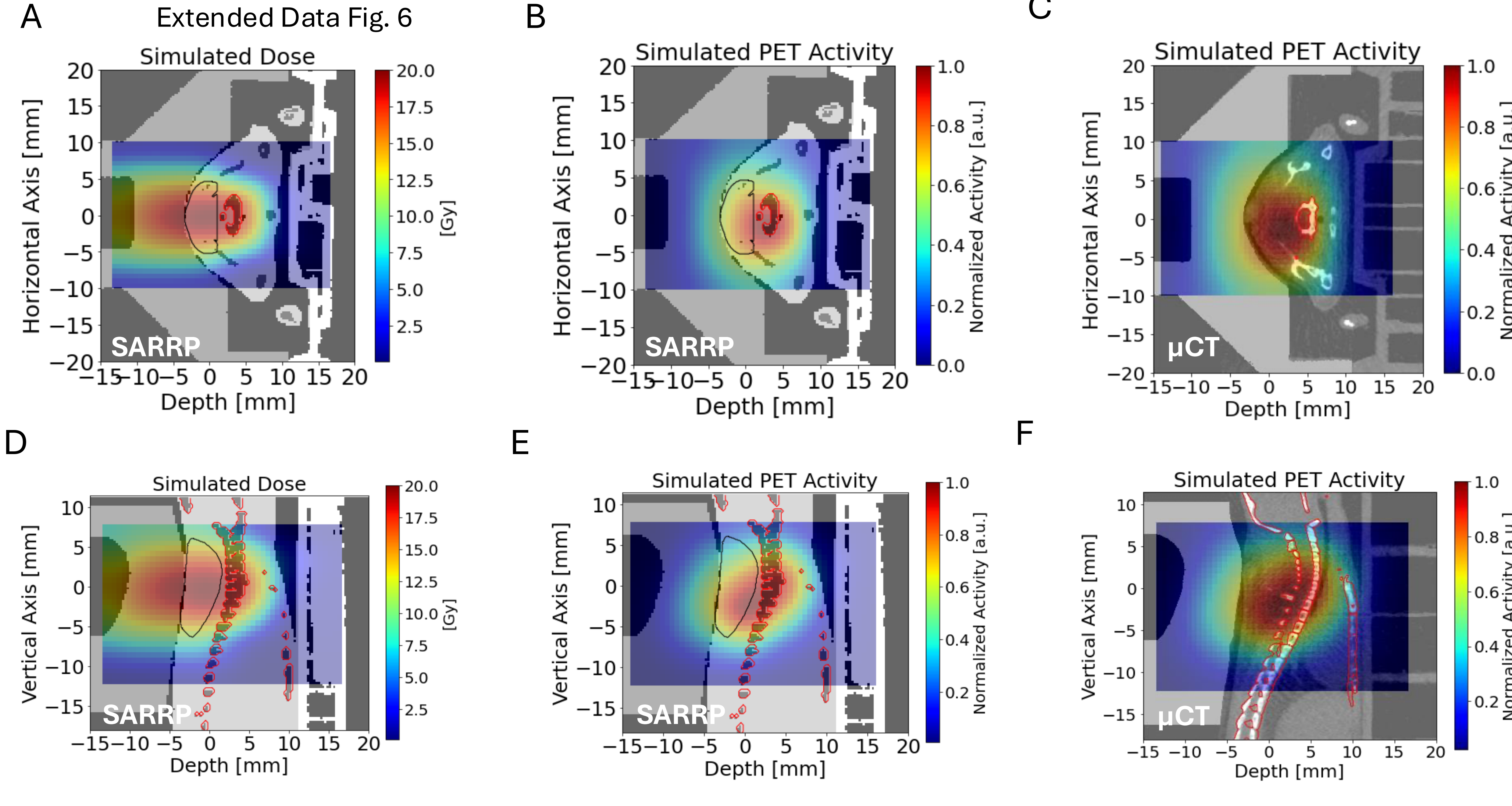

Extended Data Fig. 6G

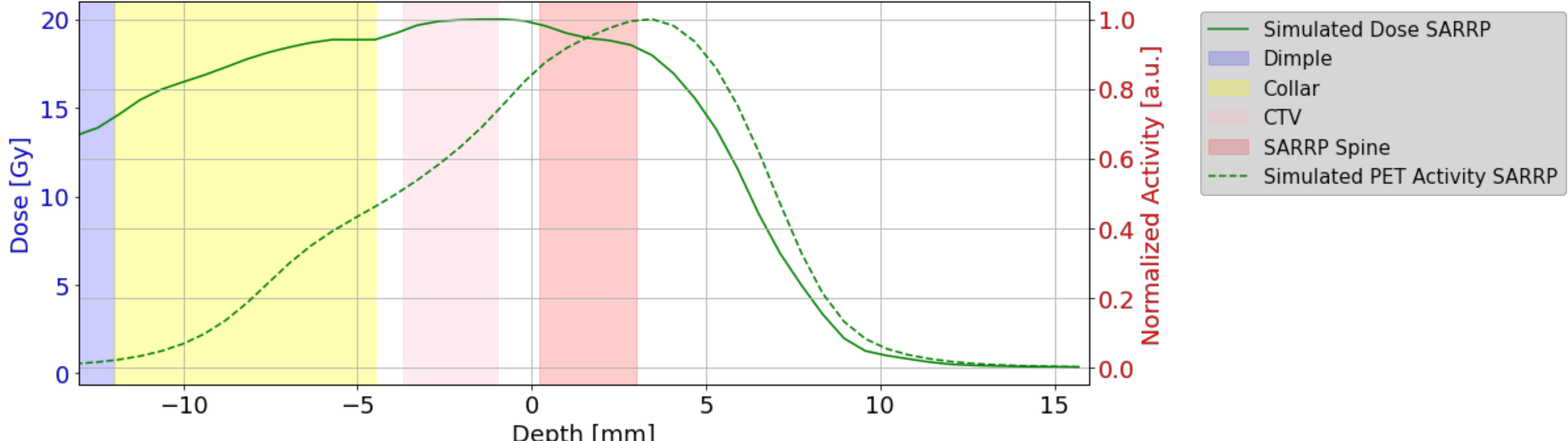

# Extended Data Fig. 7

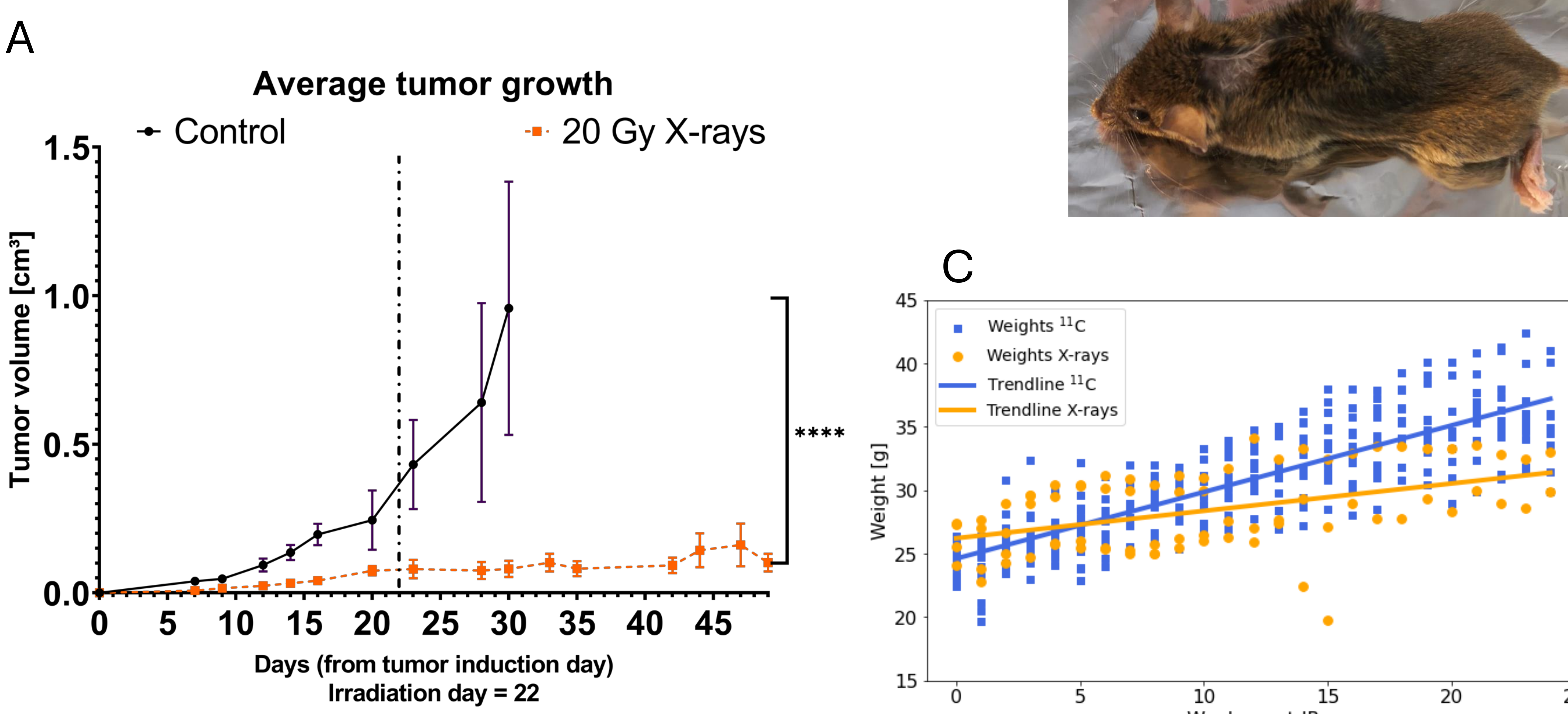

**Extended Data Fig. 8**

A

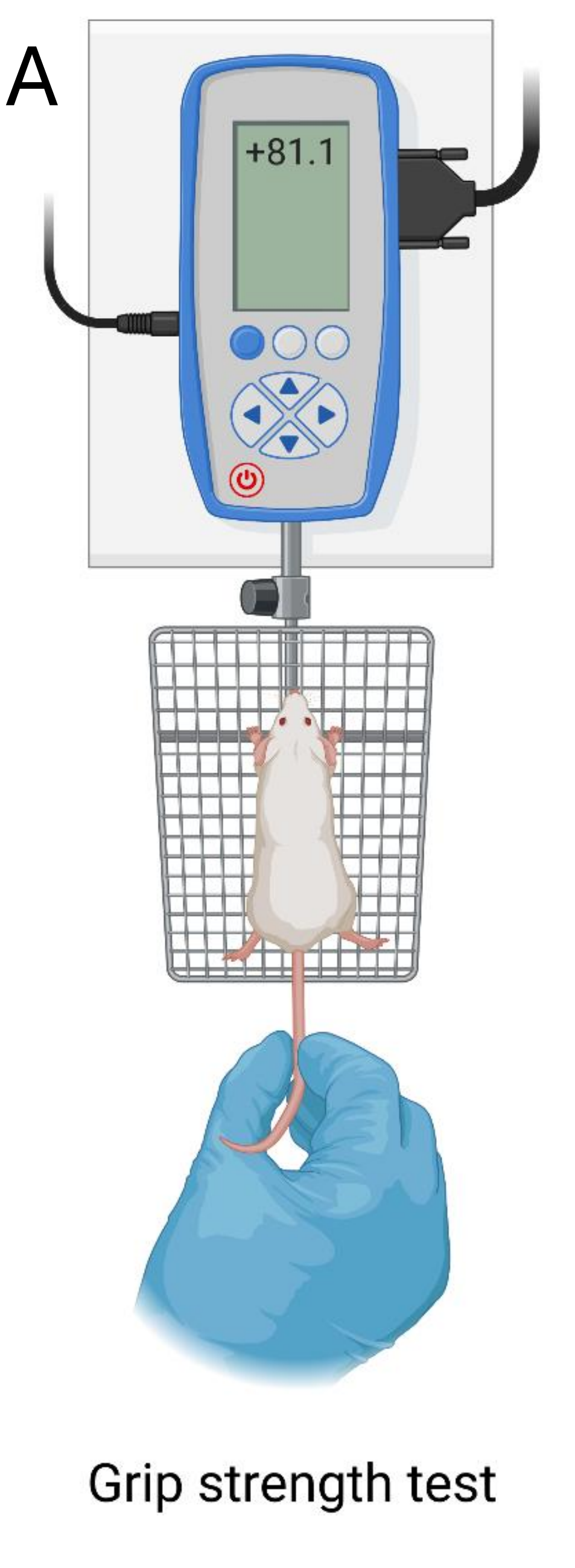

Grip strength test

B

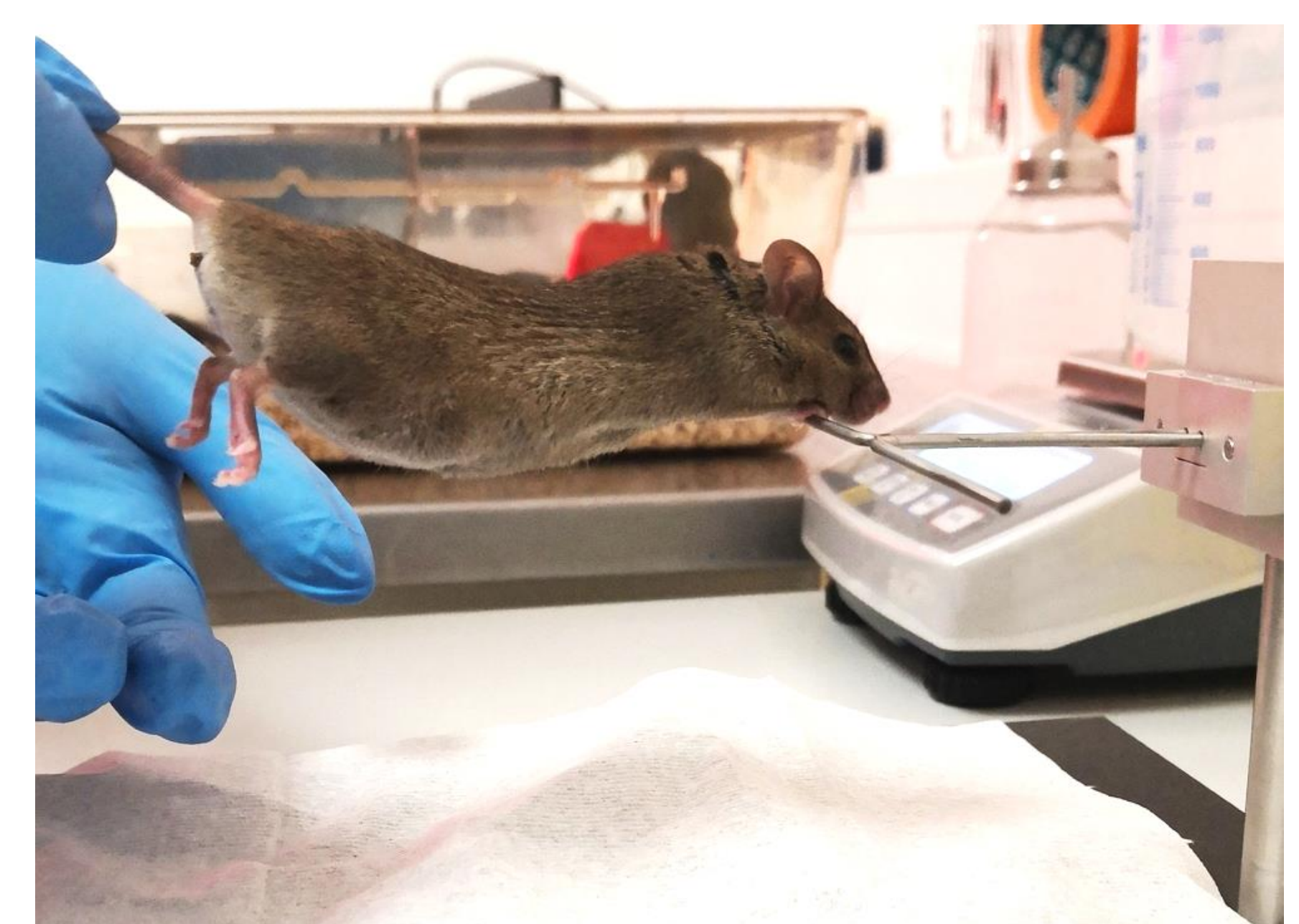

C

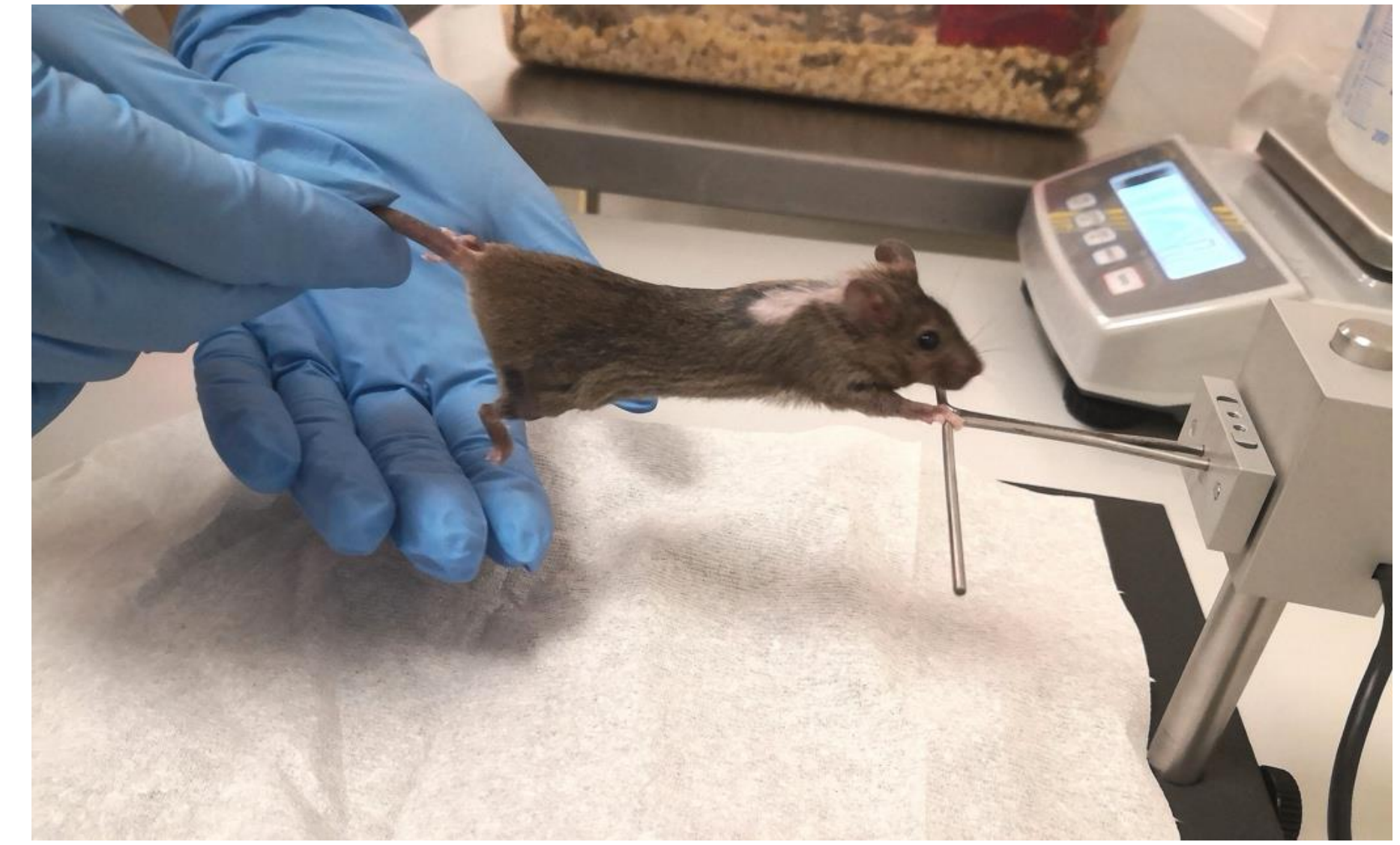

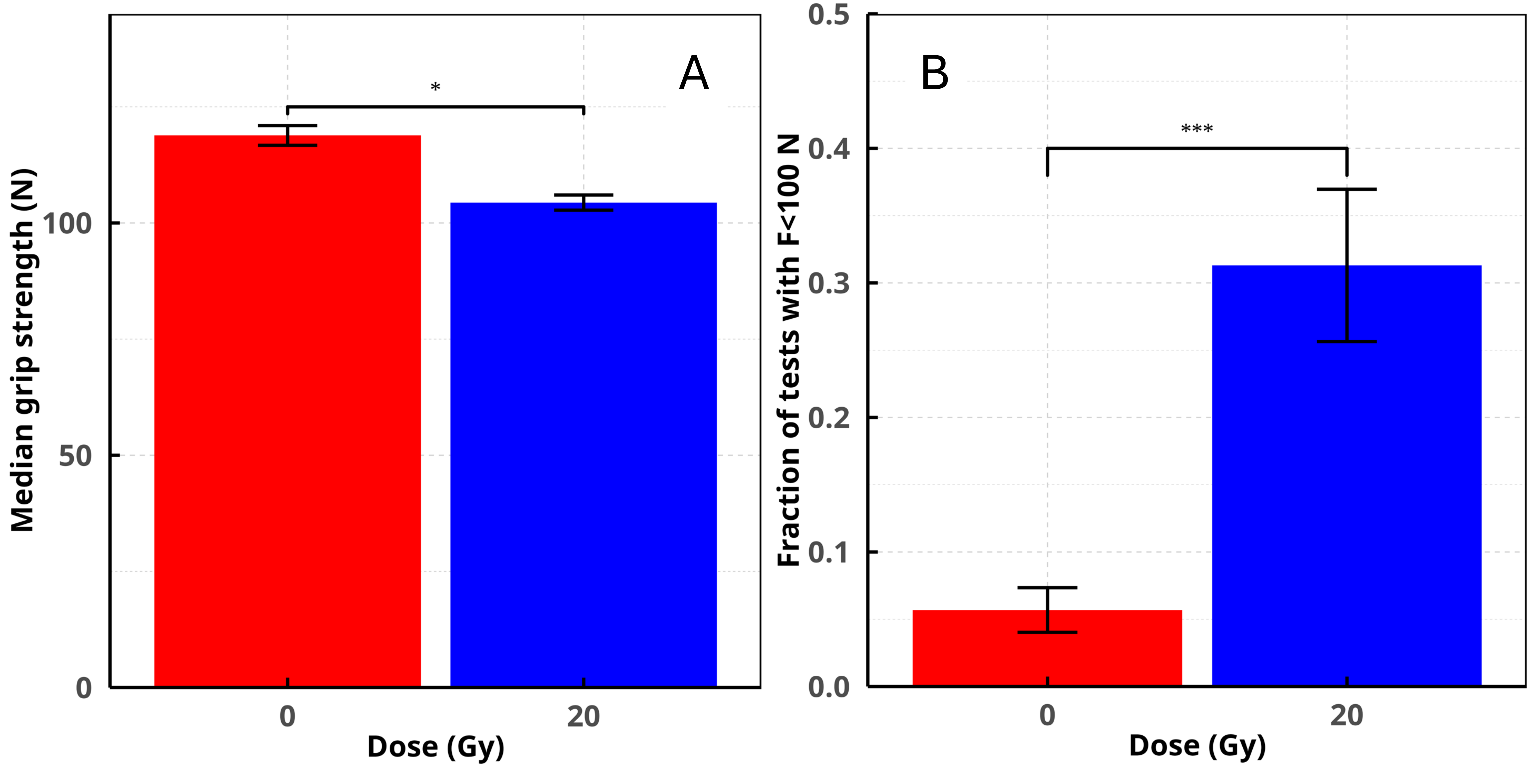

**Extended Figure 9 A-B**

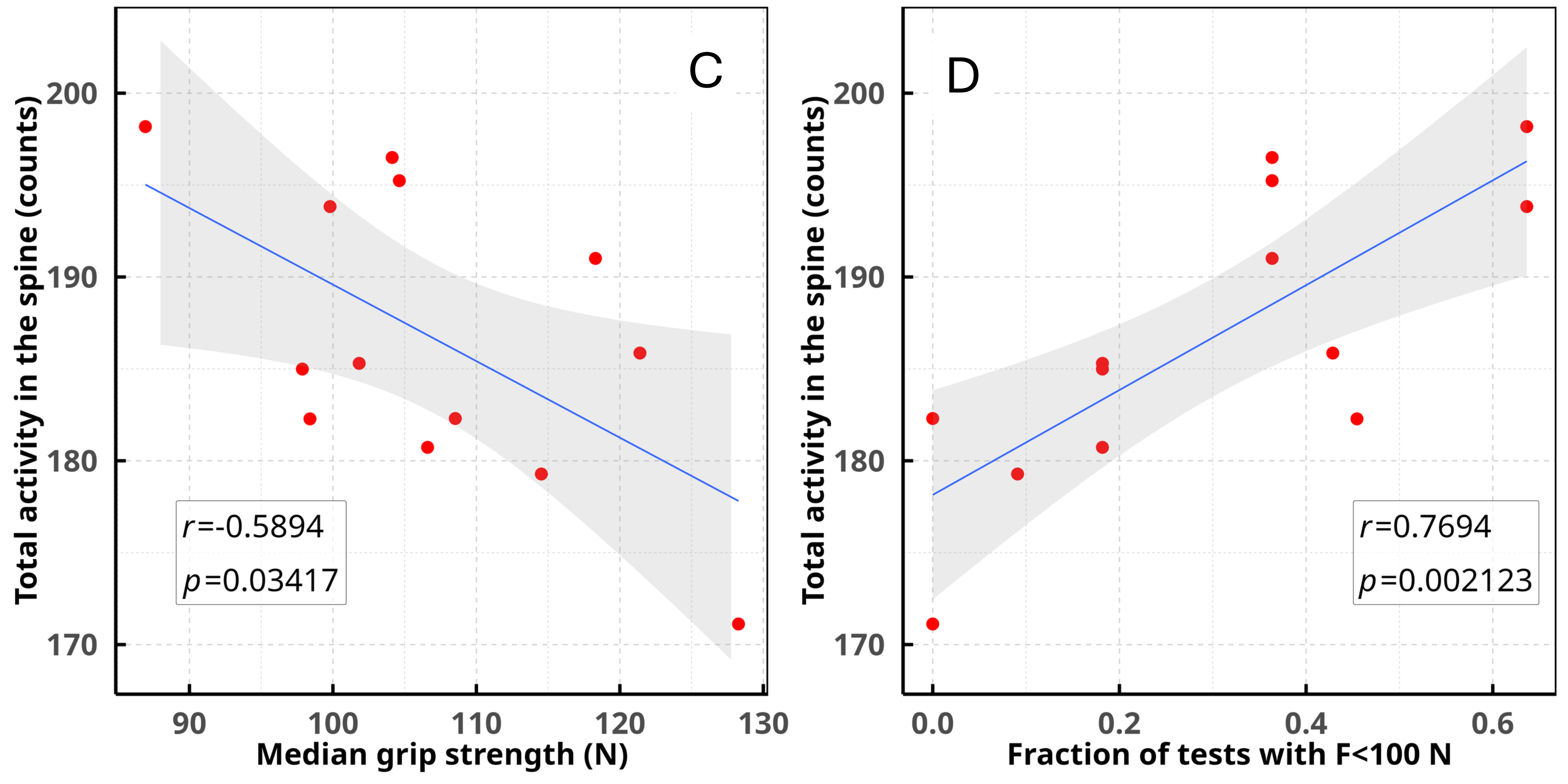

**Extended data Figure 9 C-D**

## SIPPLMENTARY MATERIAL: DOSIMETRY

### 1. Monte Carlo simulations of the beam profile along the experimental beamline

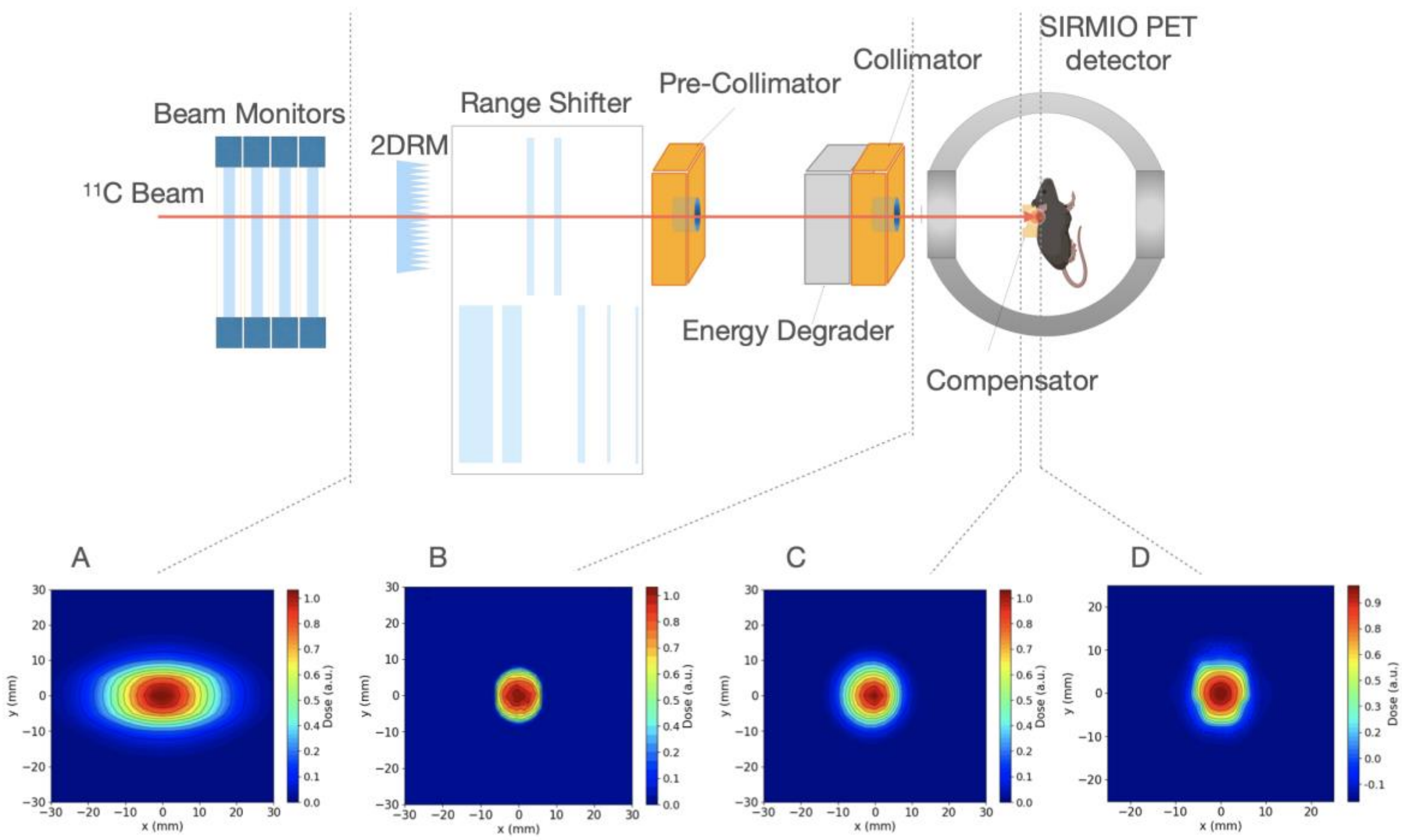

Figure S1: Schematic representation of the experimental beamline used for the animal irradiation. Panel A-D show FLUKA simulations of the lateral dose distribution at different position along the beamline, respectively: A) at the entrance in the experimental room after the beam monitor detectors, B) at the collimators, C) at the mouse collar, D) at the tumor position.

Figure S1 presents a schematic representation of the experimental setup used for animal irradiation. To illustrate the function of the different components, the simulated lateral dose distribution at various positions along the beamline is also shown. The beam parameters used in these simulations were obtained during beam characterization measurements, as described in the following sections.

From left to right, the beamline consists of the following elements:

- **Beam Monitoring:** The pristine $^{11}$C beam enters the experimental room and is monitored using standard large-plate ionization chambers installed in the nozzle of the GSI medical cave. These chambers, calibrated in terms of the number of primary particles, are integrated with the dose delivery system. Since the $^{11}$C beam is a secondary beam, its phase space is larger than that of a typical $^{12}$C pencil beam used in particle therapy, resulting in a larger beam spot size and momentum spread. Measured values are reported in Extended Data Table 1. A simulated 2D dose distribution of the beam at the entrance of the experimental room is shown in Figure S1 Panel A.
- **2DRM for SOBP Energy Modulation:** A 2D range modulator (RM) was used to shape the beam energy, creating a 1.2 cm spread-out Bragg peak (SOBP) in water. This SOBP fully covers the clinical target volume (CTV) while allowing some margin for range adjustments during treatment.

- **Range Shifter and Aluminum degraders for range adjustment:** Aluminum plates with a total thickness of 28.4 mm (equivalent to 60.1 mm of water path length) were introduced into the beamline as bolus material. These plates shifted the beam range to approximately match the tumor position. A range shifter equipped with movable PMMA plates was used to fine-tune the distal fall-off of the SOBP in the mouse, with a resolution of 0.5 mm water-equivalent thickness. The range shifter plates and aluminum thickness were chosen to achieve a residual beam range of 1.32 cm in water, corresponding to 0.56 cm in the mouse neck (after accounting for the mouse collar/compensator). The water-equivalent thicknesses of all degradation materials were characterized in the beam before the animal experiment.
- **Collimation:** A series of brass collimators was used to reduce the lateral irradiation field and blocked parts of the beam that did not contribute to the target dose. This was done to improve the signal to noise ratio and limit the amount of stray radiation on the SIRMIO PET scanner. To reduce as much as possible the effects of lateral scattering, the collimators were placed as close as possible to the aperture of the SIRMIO PET scanner. The collimator aperture consisted in an ellipse with the horizontal axis of 1.2 mm and the vertical one of 1.5 mm. To uniformly cover the collimator aperture, the final irradiation plan used two pencil beams with a 6 mm shift in the y-direction, delivered over 20 rescans to improve uniformity. A simulated lateral dose distribution of the collimated beam at the exit point from the collimator and at the surface of the mouse collar is shown in Figure S1B and Figure S1C respectively.
- **Compensator:** Finally, a plastic mouse collar, acting as a compensator, was fixed onto the mouse bed. This component was designed to partially absorb the beam outside the CTV and shape the distal edge of the SOBP to match the target contour. The expected lateral dose distribution at the center of the CTV (corresponding to the isocenter of the experimental room) is illustrated in the third panel of Figure S1D.

## 2. $^{11}$C beam characterization and depth dose measurements in water

As described in the *Materials and Methods* section, both 1D and 3D dose measurements of the pristine and range modulated $^{11}$C beam were acquired in water using two setups: a PEAKFINDER$^{TM}$ system (PTW Freiburg, Germany) for the 1D depth dose distribution and a water phantom setup equipped with a PTW OCTAVIUS 1600 XDR detector for 3D dose maps. A schematic representation of the water phantom setup is shown in Figure S2. The entrance window of the water phantom was positioned at the isocenter, and no attenuator or beam-shaping device—other than the 2DRM for SOBP generation—was introduced into the beamline.

This setup allowed for the measurement of 2D dose maps at different positions in water, starting from a minimum water-equivalent thickness of $2.10 \pm 0.05$ cm. These maps were then merged to create a full 3D dose distribution with an accuracy on the relative dose better than 2%. Figures S3 and S4 present a comparison of these measurements with Monte Carlo simulations for both the pristine and SOBP configurations. Additionally, lateral dose profiles at selected water depths are compared.

The measured data were essential for extracting key beam parameters, including beam range in water, beam energy, beam spot size, and momentum spread. This information was subsequently used to determine the exact absorber thickness required to completely stop the beam at the distal end of the mouse CTV and to define the beam model for Monte Carlo simulations (see Extended Data Table 1).

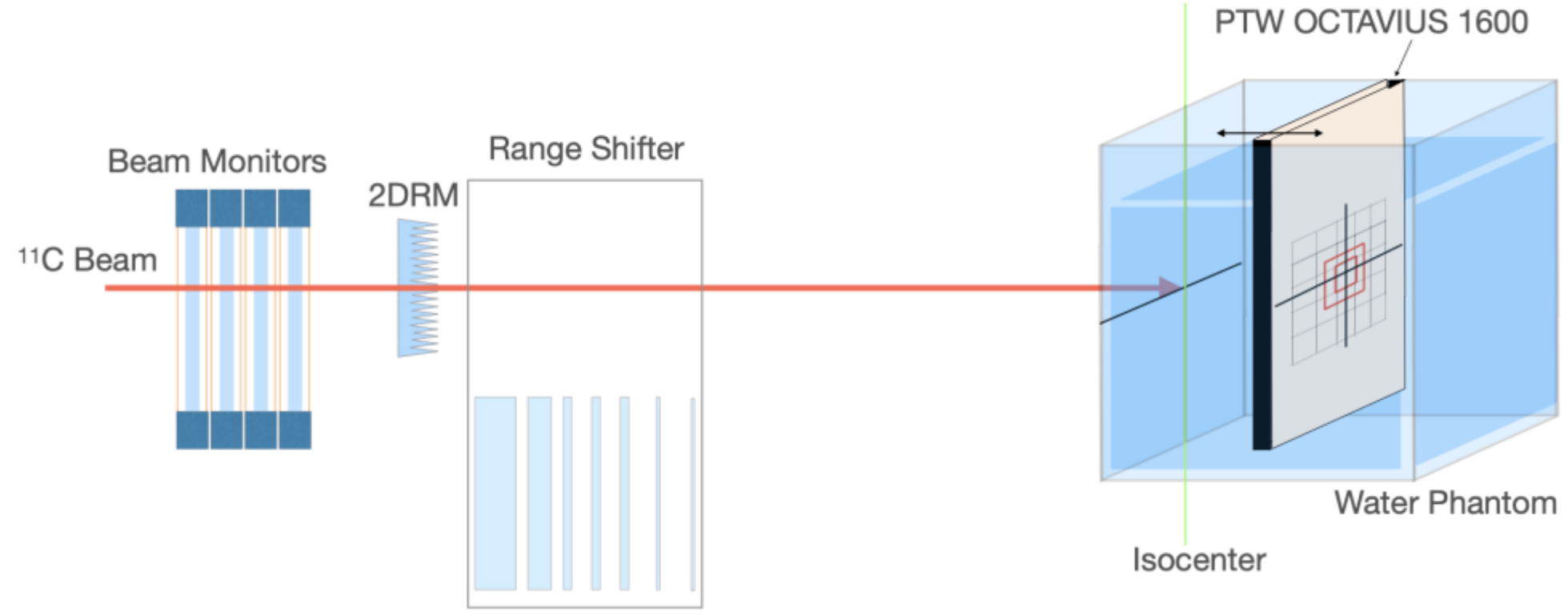

**Figure S2: Schematic representation of the water phantom experimental setup used for 3D dose measurements and beam characterization.**

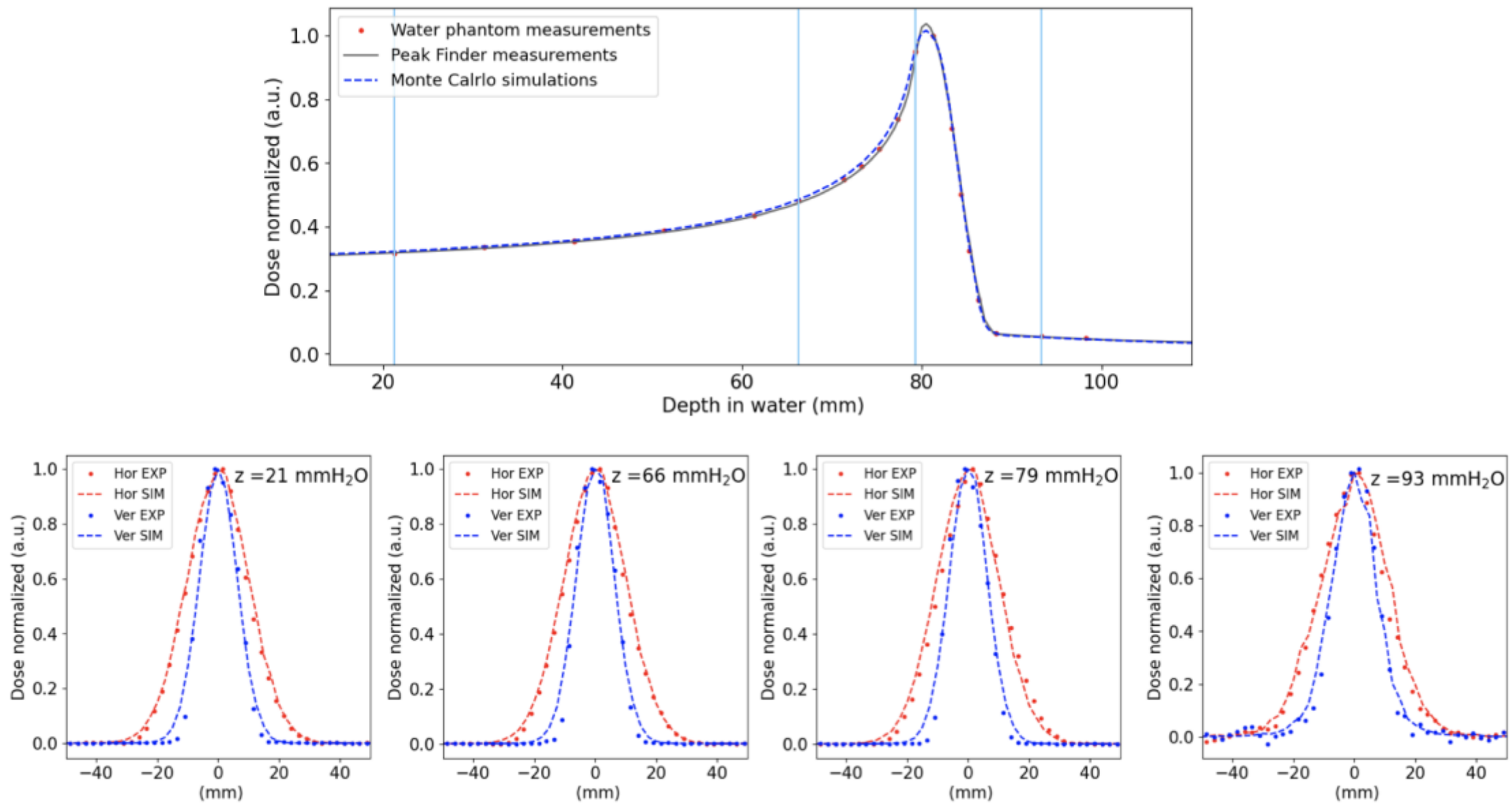

**Figure S3: Measured laterally integrated depth dose distribution for the pristine beam and respective Monte Carlo simulations (upper panel). In the lower panel comparison of the measured lateral dose distribution with the corresponding Monte Carlo simulation at selected depth in water (represented with light blue lines on the depth dose profile).**

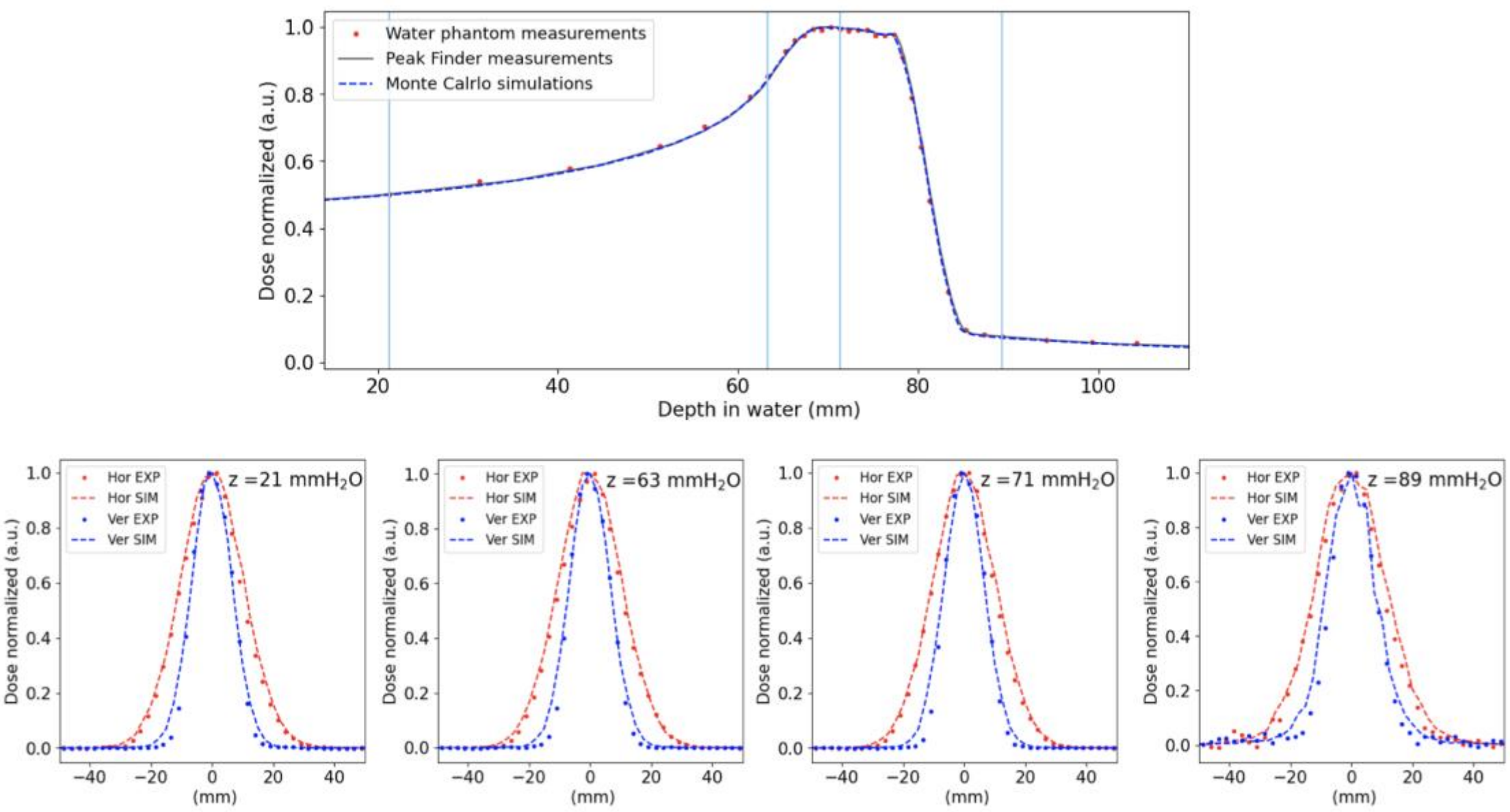

**Figure S4: Measured laterally integrated depth dose distribution for the SOBP beam configuration and respective Monte Carlo simulations (upper panel). In the lower panel comparison of the measured lateral dose distribution with the corresponding Monte Carlo simulation at selected depth in water (represented with light blue lines on the depth dose profile).**

## 3. Beam characterization and dosimetry at the mouse position

### 3.1. Lateral dose distribution measurements at the target positions

Once the beamline for the animal irradiation was defined and built, the 2D dose distribution of the beam arriving at the target position was measured with OCTAVIUS 1600 XDR detector free in air (Figure S5). In this case it was not possible to use the water phantom, since the residual range in water at this stage was smaller than the minimal measurable water thickness. Figure S6 shows the measured 2D dose distribution and a comparison of the lateral dose profiles along the central lateral and horizontal axis with the corresponding Monte Carlo simulations.

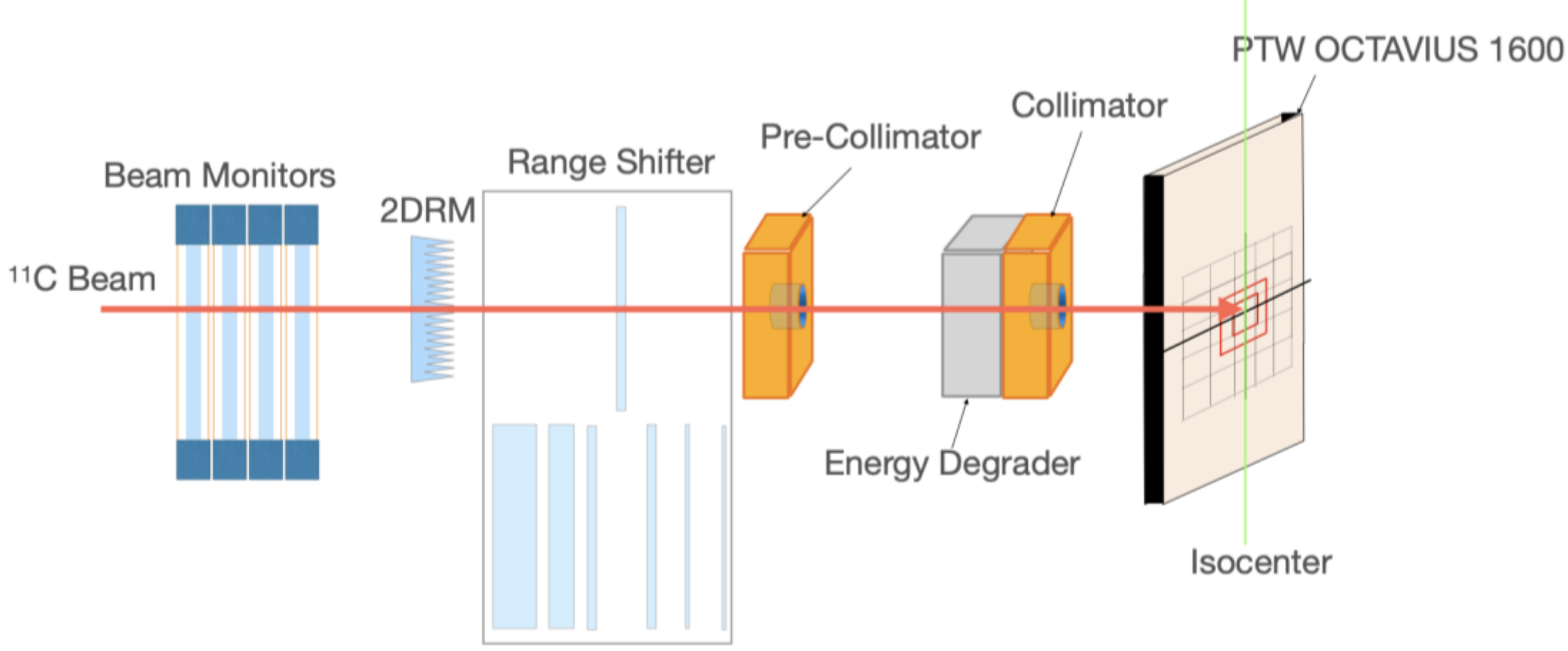

**Figure S5: Schematic representation of the experimental set up used to verify the dose distribution at the target position with the set up for the mice irradiation on the beamline.**

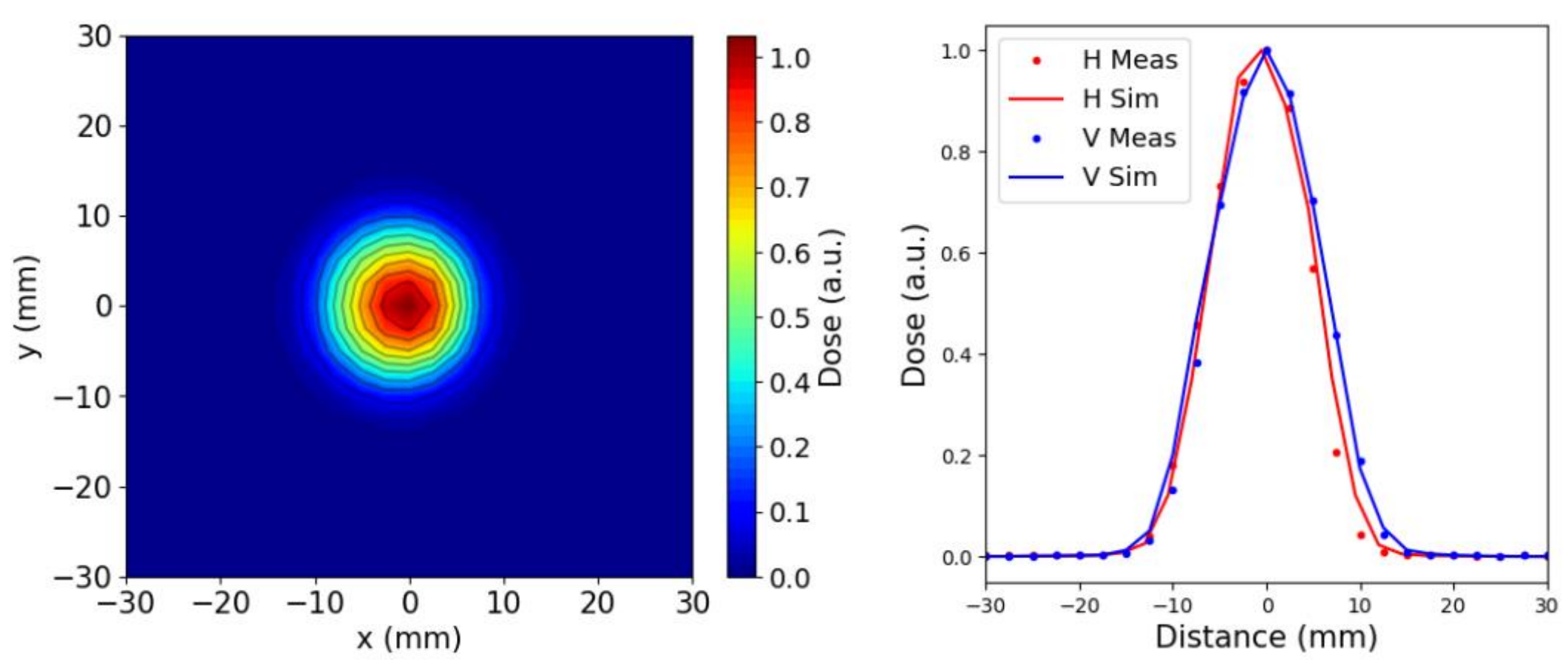

**Figure S6: Panel A, measured 2D dose distribution after beam energy degradation and collimation as it was then adopted for the animal experiment. Panel B, a comparison of measured and simulated lateral dose profiles at the target position after the beamline to be used for the animal irradiation was set up.**

### 3.2. Dosimetric calibration at the tumor position

A final dosimetric calibration was performed at the tumor position using a volume ionization chamber (PTW TM31023), as shown in Figure S7. A custom-designed compensator, shaped like a mouse collar, was used to secure the chamber in a fixed position at the center of the CTV (Figure S8). This measurement enabled us to establish a monitor unit calibration relative to the dose at the target position, which was then used to adjust the irradiation plan and ensure the accurate delivery of the desired dose to the target. The chamber readings were recorded using a UNIDOS electrometer, applying the standard correction factors commonly used in particle therapy. Specifically, a $k_Q$ factor of 1.03 was used, along with a $k_{TP}$ correction based on the measured temperature and pressure inside the irradiation room during the experiment. The $k_S$ factor was neglected due to the low dose rates.

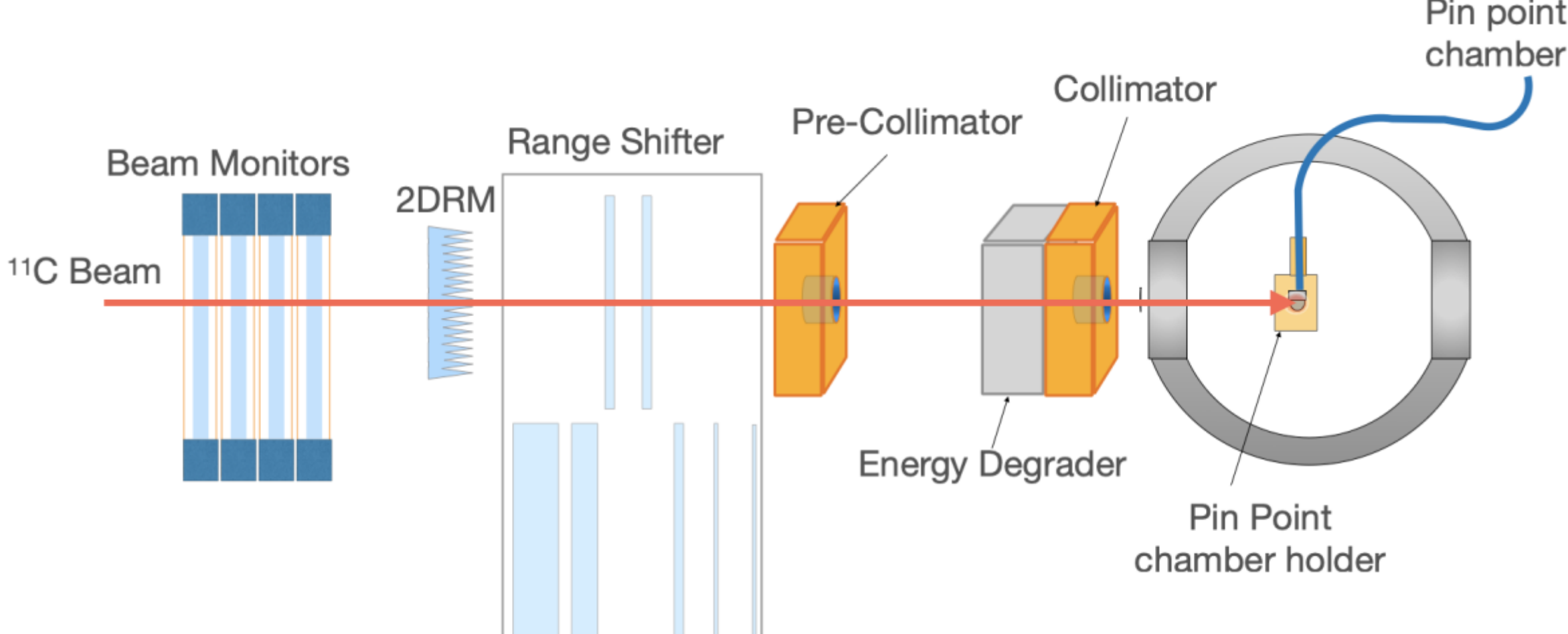

Figure S7: Schematic representation of the experimental set up used for absolute dosimetry at the tumor position

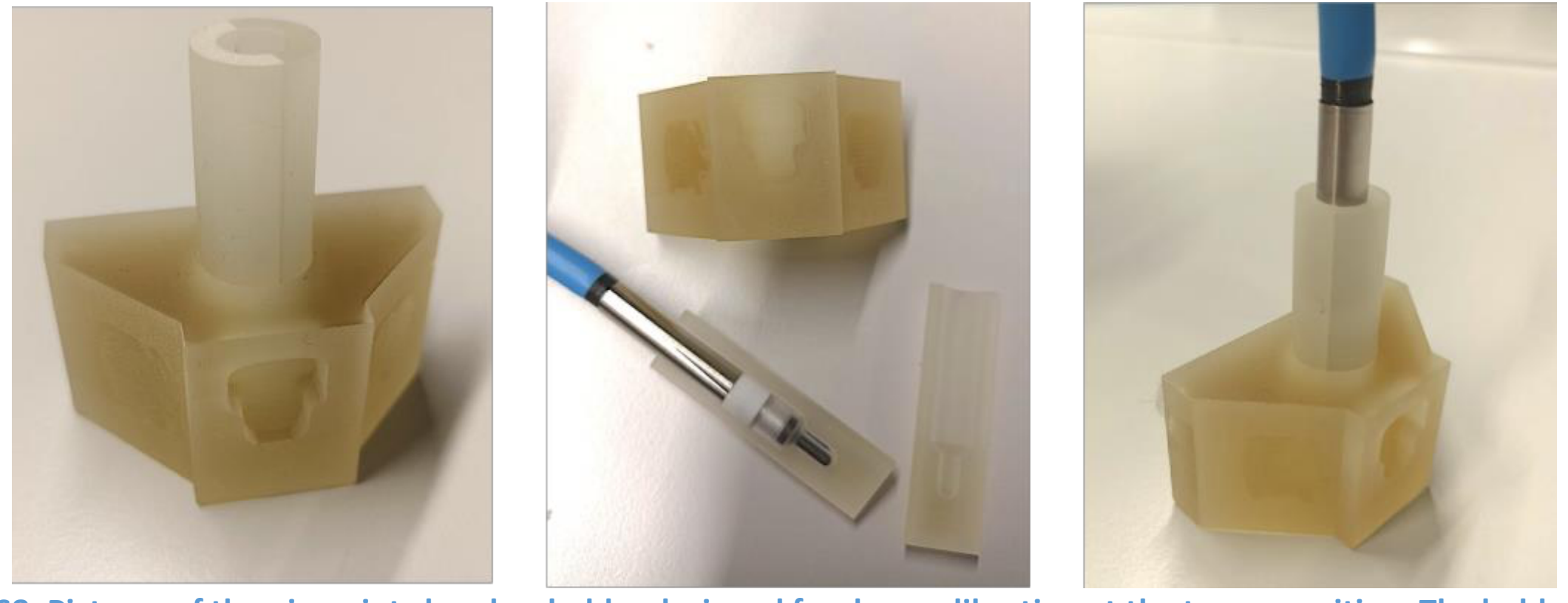

Figure S8: Pictures of the pin point chamber holder designed for dose calibration at the tumor position. The holder shape resemble the mouse compensator and has an insert for a PTW pin point chamber holding the chamber at a position and a water equivalent depth corresponding to the center of the CTV. The holder was attached to the mouse bed in the same position as a the mouse compensator.

## Supplementary Fig. 1

All PET images

# Supplementary Fig. 2

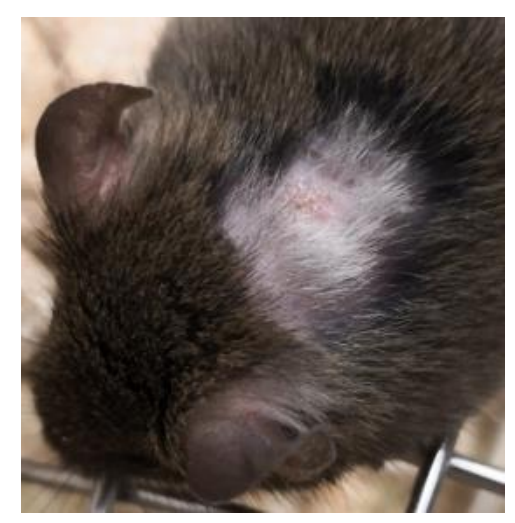
Grade 1

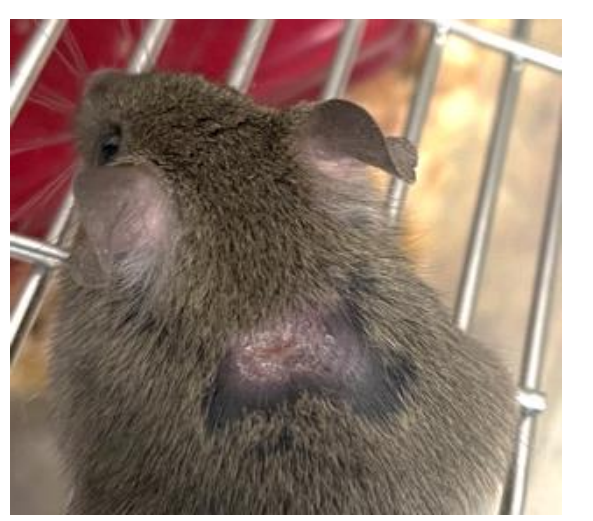
Grade 2

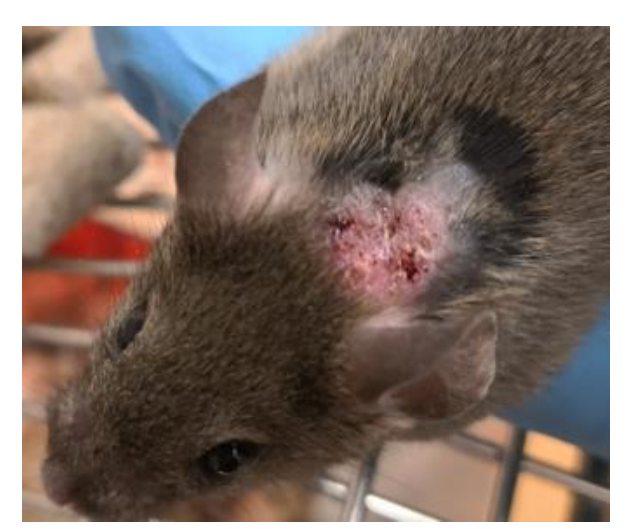
Grade 3

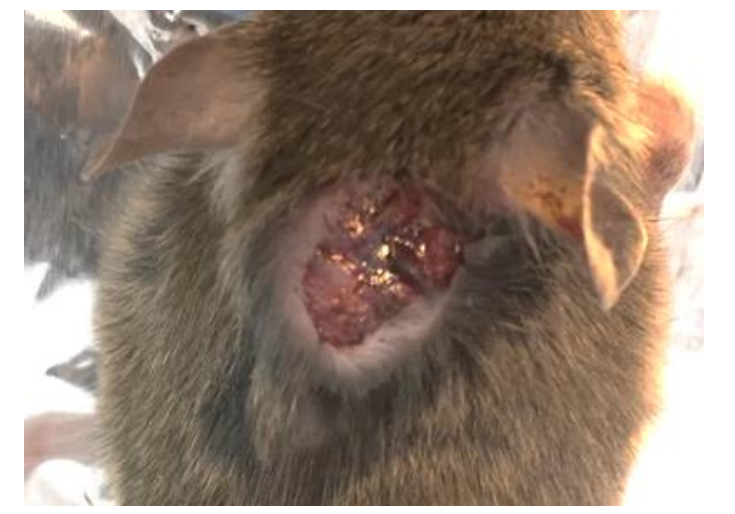
Grade 4

| Grade 5 | Necrosis |
| --- | --- |
| Grade 4 | Open wound not healing, leading to sacrifice of the animal |
| Grade 3 | Closed wound or small scratch |
| Grade 2 | Dry skin, desquamation |
| Grade 1 | Redness |
| Grade 0 | No effect |

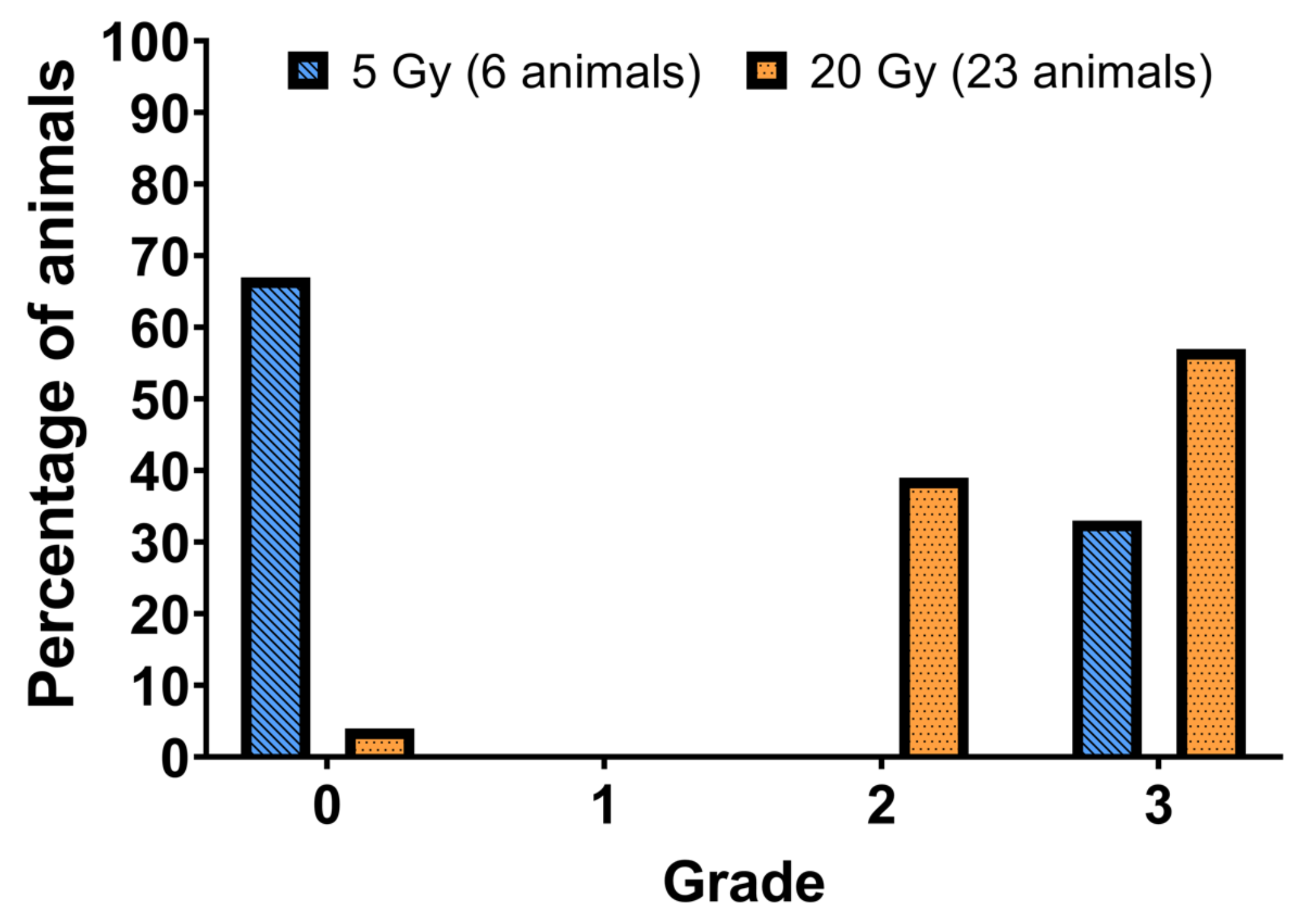

# Supplementary Fig. 3A

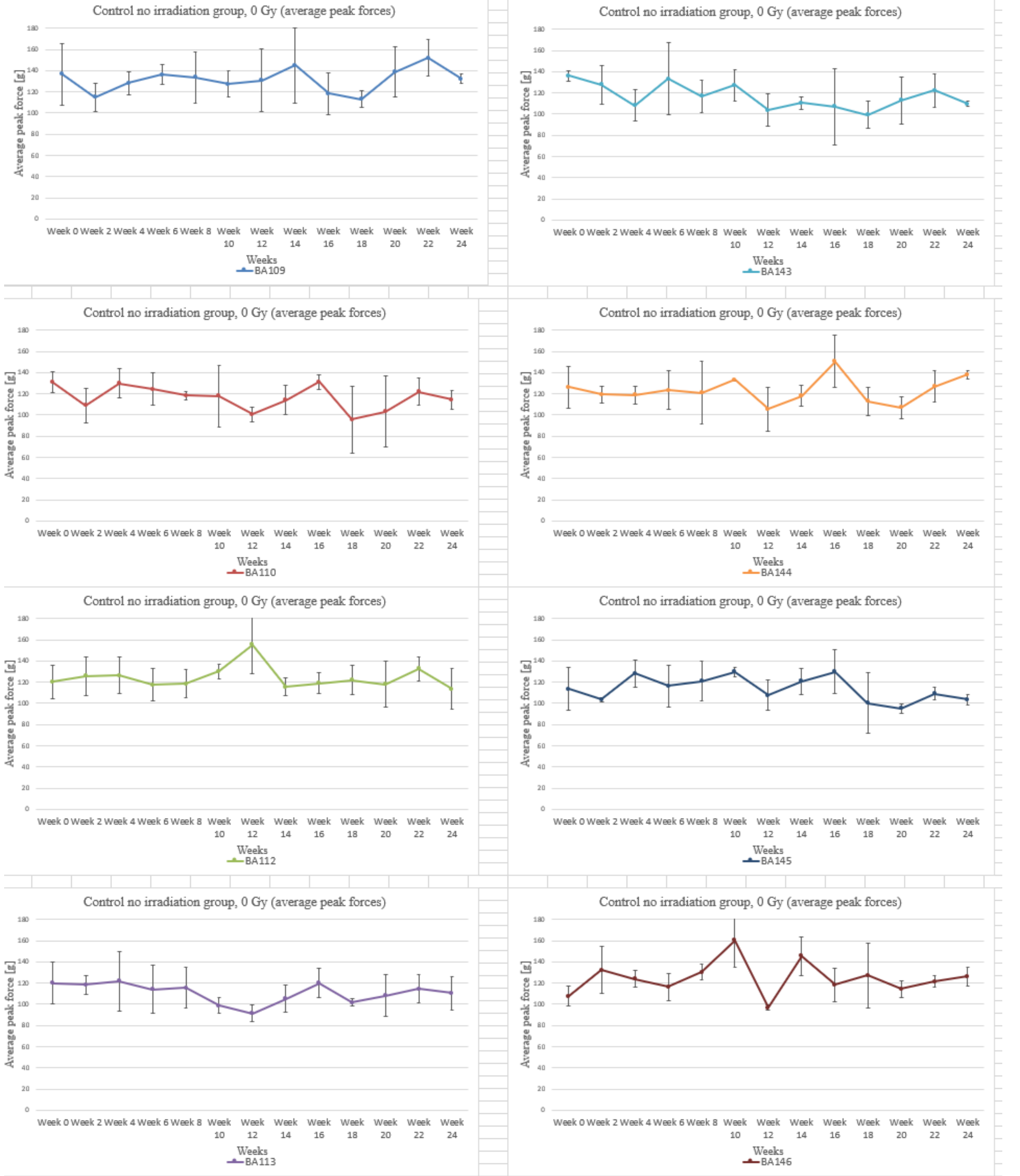

Single mouse plots
0 Gy

# Supp. Fig. 3B

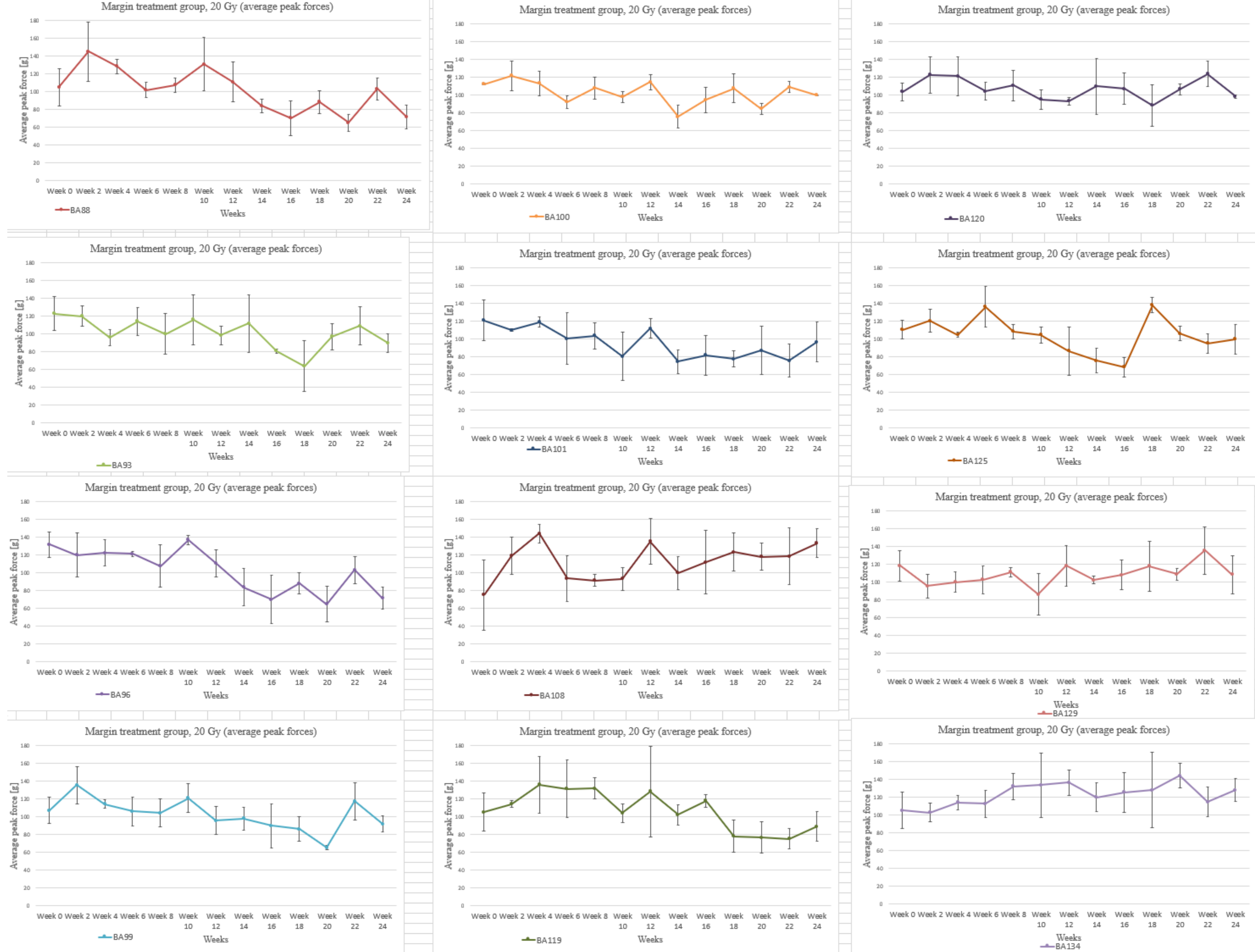

Single mouse plots
20Gy

## Supplementary Fig. 4

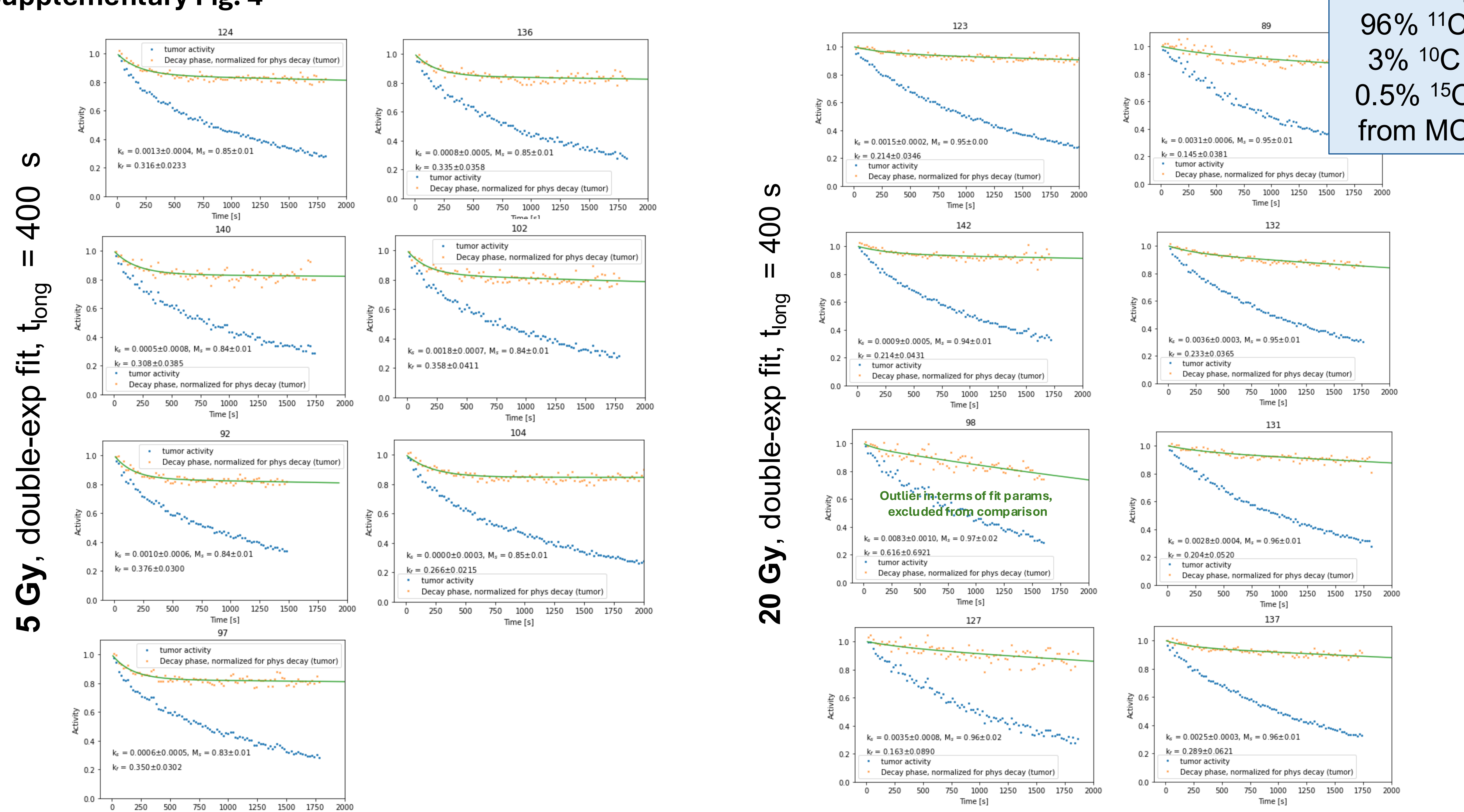

**Supplementary Fig. 5**

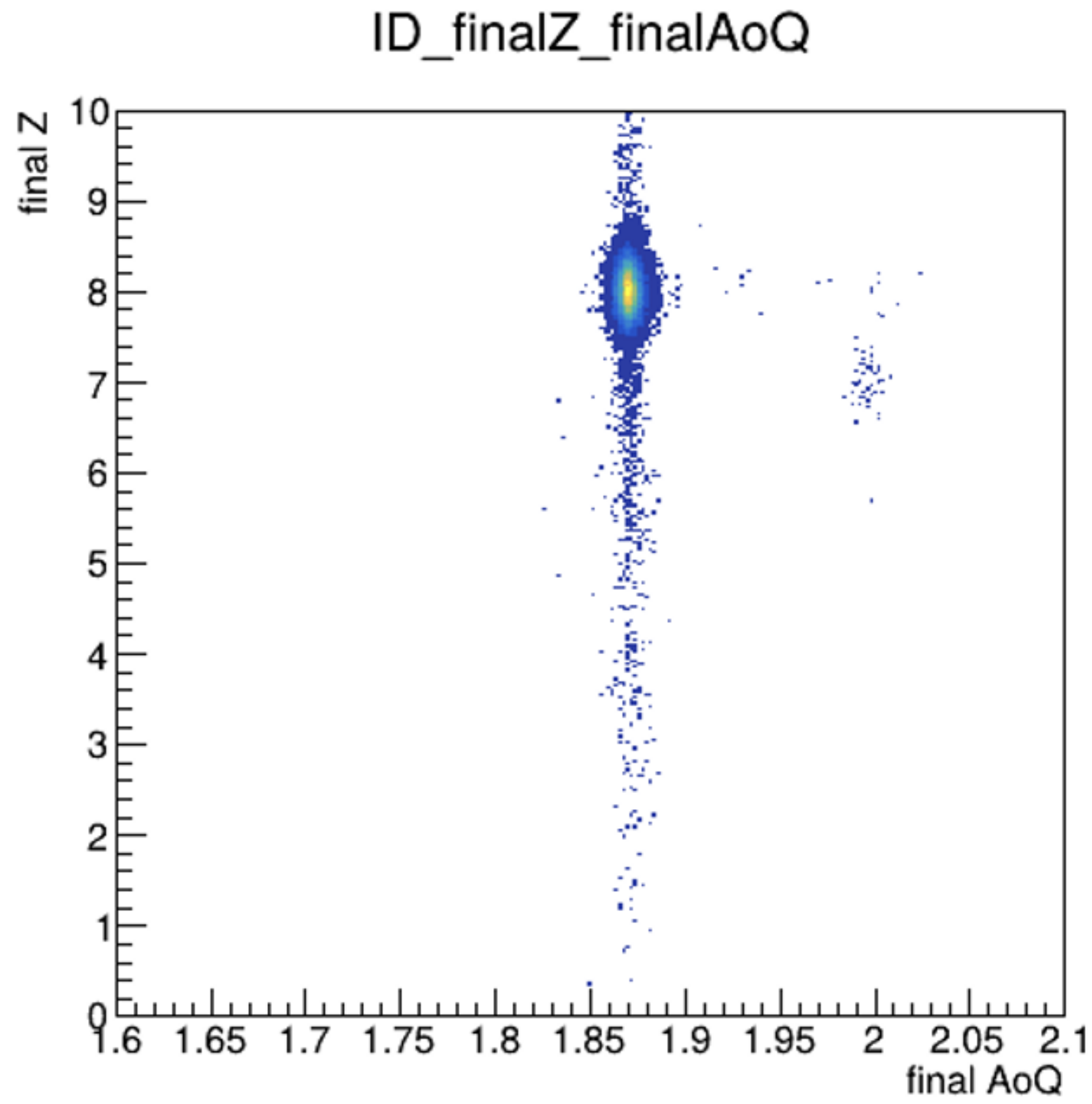

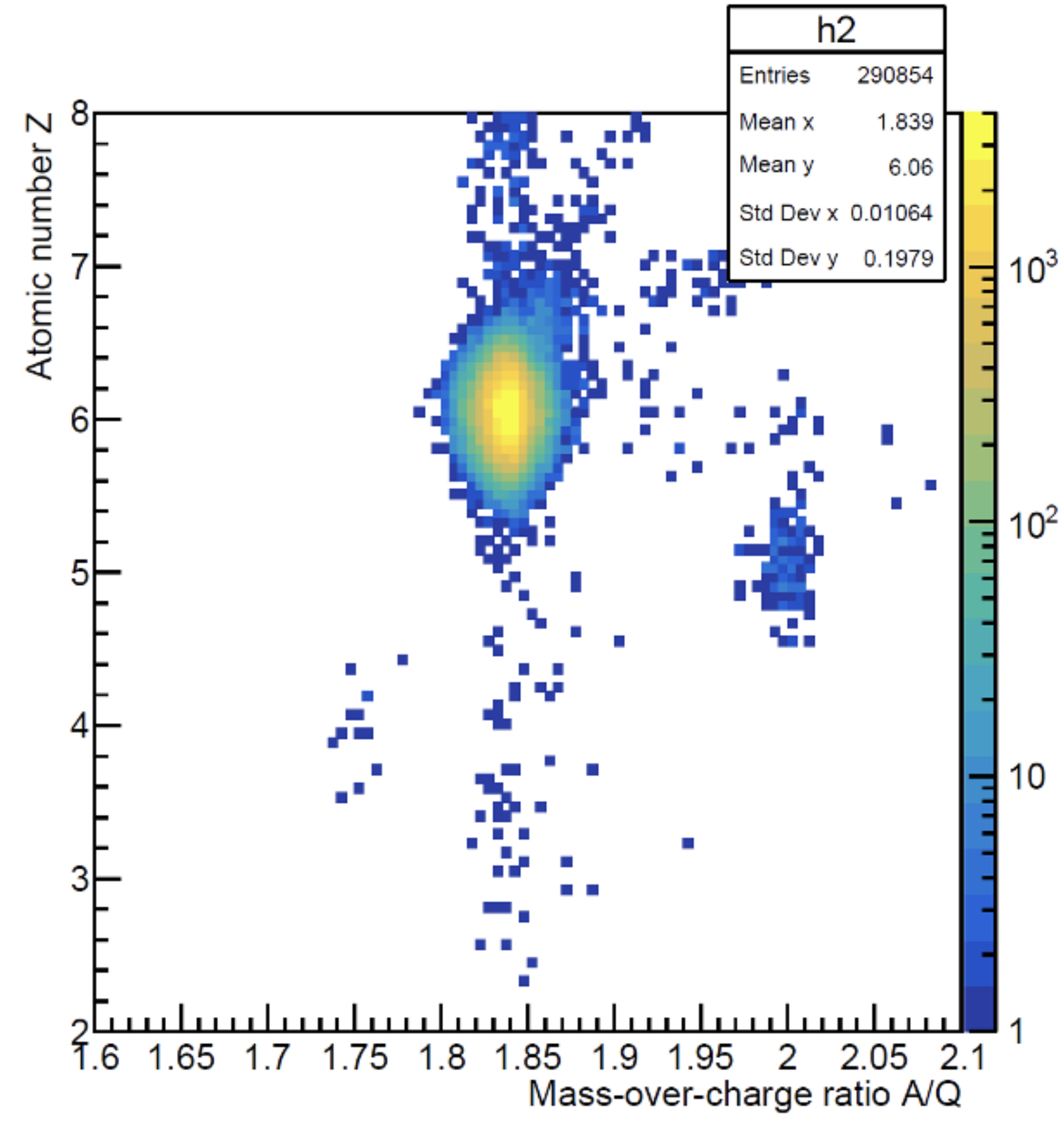